\documentclass[a4paper,fleqn,usenatbib]{mnras}
\usepackage{newtxtext,newtxmath}
\usepackage[T1]{fontenc}
\usepackage{ae,aecompl}
\hypersetup{draft}

\usepackage{graphicx}	
\usepackage{amsmath}	
\usepackage{amssymb}	
\usepackage{array}	

\newcommand{\nugc}{{\scriptsize $\nu^{2} GC$}}
\newcommand{\LX}{{$\log(L_X /\mathrm{erg~s^{-1}})$}}
\newcommand{\ii}{I\hspace{-.1em}\nobreak I}
\defcitealias{nu2gc}{M16}
\defcitealias{KH00}{KH00}

\title[AGNs and SMBHs in $\nu^2$GC]{The New Numerical Galaxy Catalogue ($\nu^2 GC$): Properties of Active Galactic Nuclei and Their Host Galaxies}

\author[H.~Shirakata et al.]
  {Hikari~Shirakata,$^1$\thanks{E-mail: shirakata@astro1.sci.hokudai.ac.jp} Takashi~Okamoto,$^{1}$ Toshihiro~Kawaguchi,$^2$
    \newauthor Masahiro~Nagashima,$^3$ Tomoaki~Ishiyama,$^{4}$ Ryu~Makiya,$^{5,6} $ Masakazu~A.~R.~Kobayashi,$^7$
    \newauthor Motohiro~Enoki,$^8$Taira~Oogi,$^5$ and Katsuya~Okoshi$^9$ \\
  $^1$Department of Cosmosciences, Graduate School of Science, Hokkaido University, N10 W8, Kitaku, Sapporo, 060-0810, Japan\\
  $^2$Department of Economics, Management and Information Science, Onomichi City University, 1600-2, Hisayamada, Onomichi, Hiroshima, 722-8506, Japan\\
  $^3$Faculty of Education, Bunkyo University, 3337, Minami-ogishima, Koshigaya, Saitama 343-8511, Japan\\
  $^4$Institute of Management and Information Technologies, Chiba University, 1-33, Yayoi-cho, Inage-ku, Chiba, 263-8522, Japan\\
  $^5$Kavli Institute for the Physics and Mathematics of the Universe (WPI), The University of Tokyo Institutes for Advanced Study, \\
     ~~~ The University of Tokyo, , 5-1-5, Kashiwa, Chiba, 277-8583, Japan\\
  $^6$Max-Planck-Institut fur Astrophysik, Karl-Schwarzschild Str. 1, D-85741 Garching, Germany\\
  $^7$ Faculty of Natural Sciences, National Institute of Technology, Kure College, 2-2-11, Agaminami, Kure, Hiroshima, 737-8506, Japan\\
  $^8$ Faculty of Business Administration, Tokyo Keizai University, 1-7-34, Minami-cho, Kokubunji, Tokyo, 185-8502, Japan\\
  $^9$ Tokyo University of Science, 102-1 Tomino, Oshamambe-cho, Yamakoshi-gun, Hokkaido, 049-3514, Japan\\
  }

\date{Accepted XXX. Received YYY; in original form ZZZ}

\pubyear{2017}

\begin{document}
  \label{firstpage}
  \pagerange{\pageref{firstpage}--\pageref{lastpage}}
  \maketitle

  \begin{abstract}
    We present the latest results of a semi-analytic galaxy formation model, ``\textit{New Numerical Galaxy Catalogue}'',
    which utilises large cosmological $N$-body simulations.
    This model can reproduce many statistical properties of galaxies at $z \lesssim 6$.
    We focus on the properties of active galactic nuclei (AGNs) and supermassive black holes,
    especially on the accretion timescale onto black holes.
    We find that the number density of AGNs at $z < 1.5$ and at hard $X$-ray luminosity $< 10^{44}$ erg/s
    is underestimated compared with recent observational estimates
    when we assume the exponentially decreasing accretion rate
    and the accretion timescale which is proportional to the dynamical time of the host halo or the bulge,
    as is often assumed in semi-analytic models.
    We show that to solve this discrepancy, the accretion timescale of such less luminous AGNs
    should be a function of the black hole mass and the accreted gas mass.
    This timescale can be obtained from a phenomenological modelling of the gas angular momentum loss
    in the circumnuclear torus and/or the accretion disc.
    Such models predict a longer accretion timescale for less luminous AGNs at $z < 1.0$
    than luminous QSOs whose accretion timescale would be $10^{7-8}$ yr.
    With this newly introduced accretion timescale, our model can explain the observed luminosity functions of AGNs at $z < 6.0$.
  \end{abstract}




  \section{Introduction}
    Galaxies are one of the main components of the Universe.
    Understanding galaxy formation and evolution is thus
    one of the main goals of astrophysics.
    Almost all galaxies have a super massive black hole
    (SMBH) at their centre and the mass of SMBHs correlates with
    properties of their host galaxies, such as the mass and velocity dispersion
    of the bulges \citep[e.g.][]{Magorrian98, FM00, HR04, MM13}.
    SMBHs and their host galaxies would thus have co-evolved with each other.
    The gas can lose its momentum via the growing processes of the bulge and/or galactic bars,
    and a part of the gas would get accreted onto the SMBH
    \citep[e.g.][]{Mihos1994August,WH95}, which could be observed as
    active galactic nuclei (AGNs).
    After that, AGN radiation, jets, and outflow inject the energy and/or angular momentum
    to the surrounding gas, which would cause the increase/decrease of the star formation rate (SFR) of
    their host galaxies \citep[e.g.][for a review]{Wagner16}.
    This ``co-evolution'' is a standing question in astrophysics,
    and has been the subject of theoretical and observational studies over three decades.
    Such work has focused on the mechanism of black hole (BH) feeding and the energetic feedback related with BH growth
    in the context of galaxy formation \citep[see, however, ][]{JM11}.

    Understanding the growth mechanisms and evolution of SMBHs is challenging
    because they cannot be directly observed.
    AGNs are the main sources to obtain information
    on SMBHs observationally,
    which emit light when material is accreted onto the SMBHs.
    To overcome the difficulty in investigating growth mechanisms and evolution of SMBHs,
    we need close comparisons between model predictions and observations of both galaxies and AGNs.

    Semi-analytic models of galaxy formation (hereafter SA models)
    are powerful tools for making theoretical predictions
    that can be directly compared with observations.
    In SA models, merging histories of dark matter (DM) haloes
    are obtained from $N$-body simulations \citep[e.g.][]{Roukema97,ON03,nugc,DeLucia10,nu2gc,Guo16}
    or analytic algorithms based on the extended Press-Schecter formalism \citep[e.g.][]{PS74,LC93,NY04,Menci05,Valiante11}.
    The evolution of baryonic components such as gaseous haloes, galaxies, and SMBHs is
    followed by phenomenological modellings to diminish the computational cost
    and to enlarge the sample size.
    Therefore, SA models are an excellent approach for statistical studies of galaxies and SMBHs
    and particularly useful for theoretical studies of rare objects, such as AGNs.

    There are a large number of previous studies using SA models
    aimed at revealing the evolution of
    SMBHs within their host galaxies.
    The evolution of the AGN luminosity function (LF),
    which has been well known to imply the ``anti-hierarchical
    trend'' of SMBH growth, is regarded as one of the main observational constraints on the models.
    In the earlier studies, they tested the
    ``merger-driven AGN scenario'' by comparing their QSO LFs
    with observational ones in optical bands
    \citep[e.g.][]{KH00, Enoki03}.
    Other triggering mechanisms of AGN activities
    such as disc instabilities \citep[e.g.][]{Lagos08,Fanidakis11,Hirschmann12},
    direct accretion from its hot gas halo \citep[e.g.][]{Fanidakis12, Griffin18},
    and galaxy-galaxy interactions \citep[e.g.][]{Menci14}
    are also studied.
    Some authors investigated the SMBH and AGN growth over cosmic time
    \citep[e.g.][]{Fontanot06,Monaco07,Marulli08,Fanidakis12,Enoki14,Menci14},
    the effect of BH spins \citep[e.g.][]{Lagos09,Fanidakis11,Griffin18},
    the effect of the seed BH mass \citep[e.g.][]{VN09,Shirakata16},
    and AGN clustering \citep[e.g.][]{Fanidakis13,Oogi16,Oogi17}.

    There are, however, uncertainties related with phenomenological modellings of the SMBH evolution,
    e.g. the triggers and the duration of gas accretion, the relation between the accretion rate
    and the AGN luminosity, the dust attenuation, Compton absorption, BH seeding, and AGN feedback.
    Unfortunately, seveal physical processes degenerate.
    Different combinations of phenomenological modellings and free parameters in a model
    could equally well explain observational properties of AGNs.
    Therefore, it is important to understand the effect of each phenomenological modelling on properties of SMBHs and AGNs.

    In this paper, we focus on the accretion timescale onto SMBHs.
    Estimation of this timescale is important as it reveals the co-evolution between
    SMBHs and their host galaxies.
    If all galaxies have undergone the AGN phase, the duration of this phase
    should be short to explain the observed AGN LFs.
    In contrast, AGNs should be long-lived if a small fraction of galaxies
    have experienced this phase \citep[e.g.][]{Soltan82}.

    There are some constraints on the accretion timescales obtained from previous studies
    \citep[see][for more details]{Martini04}.
    \cite{YT02} estimate the timescale by comparing present-day mass density of BHs with
    the integrated accreted mass density in luminous AGN phases obtained from
    optical AGN LFs at various redshifts.
    They suggest that the average ``AGN lifetime'' is $3-13\times 10^7$ years
    for $10^{8-9} M_\odot$ BHs if the radiation efficiency, $\epsilon$, is $0.1-0.3$.
    On the theoretical side, \citet[hereafter KH00]{KH00} estimate the AGN lifetime by using an SA model.
    They assume a constant radiation efficiency for AGNs,
    which are triggered only by major mergers of galaxies.
    They derive the average AGN lifetime to explain observed AGN LFs
    with $M_B\lesssim -23$ (where $M_B$ is the $B-$ band absolute magnitude).
    They suggest that the lifetime is $~\sim~3\times10^7$ yr at $z = 0$
    and that the timescale would scale with the dynamical time of the halo;
    $\propto~(1+z)^{-1.5}$.

    In these studies, the AGN lifetime is assumed to be the timescale within which SMBHs are observed as optical AGNs.
    This timescale is not necessarily equal to the accretion timescale onto SMBHs.
    \cite{Hopkins05S625} estimate not the AGN lifetime but
    the ``total'' accretion timescale considering the obscured accretion phases
    by using hydrodynamic simulations.
    They suggest that the accretion onto an SMBH is not visible at first because gas and dust components
    are surrounding the nuclear region.
    After blowing out these components by AGN winds, AGNs can be observed as optical sources.
    The AGN lifetime is then $\sim 20$ Myr and the total accretion timescale is $\sim~100$ Myr
    for AGNs with $M_B~<~-22$.

    There are still two uncertainties about the accretion timescale.
    One is the physical processes that govern the timescale.
    Several authors have proposed different mechanisms that
    determine the accretion timescale.
    \citetalias{KH00} suggest
    it is proportional to the dynamical time of the host halo.
    \cite{NS88} propose that the gas accretion continues during a starburst in its host galaxy,
    because they assume that the gas fueling to an SMBH is promoted by the
    mass loss from large star clusters.
    \cite{Granato04} and \cite{Fontanot06}
    assume the accretion rate to be determined by the viscosity of the accretion disc.
    The effect of these different assumptions on statistical properties of AGNs and SMBHs remains unclear.

    It is also unclear whether the timescale of less luminous AGNs is the same order
    as that of luminous ones.
    Previous work has focused on the timescale of optical AGNs with $M_B~<~-22$ (hard $X$-ray (2-10 keV) luminosity,
    $L_X$, corresponds to $\sim 5~\times~10^{43}$ erg/s) whose SMBH mass is larger than
    $\sim 10^8 M_\odot$.
    Less luminous AGNs with $L_X \lesssim 10^{44}$ erg/s
    would have wide range of SMBH masses.
    The accretion timescale of such less luminous AGNs is not necessarily in the same order as luminous AGNs.

    There is a well-known problem of SMBH growth scenario.
    Assuming that AGN activities are triggered only by mergers of galaxies and
    that the accretion timescale is $\sim 10^{7-8}$ yr,
    the number density of less luminous AGNs are underestimated in SA models.
    This implies that to explain the observed ``anti-hierarchical trend'' of SMBH growth,
    we need to consider other triggering mechanisms of SMBHs and/or to reconsider the accretion timescale.
    \footnote{Even when we consider AGNs triggered only by major mergers, the ``anti-hierarchical trend'' of
    optical QSOs (i.e. luminous AGNs) can be explained \citep{Enoki14}.}
    As an example, \cite{Hirschmann12} assume that AGNs are triggered solely by galaxy mergers,
    and set the accretion timescale is proportional to the Salpeter timescale,
    \begin{equation}
      \frac{\epsilon \sigma_T c}{4\pi G m_p} \frac{L_\mathrm{Edd}}{L} \sim 4.5\times10^7 \left(\frac{\epsilon}{0.1}\right) \left(\frac{L_\mathrm{Edd}}{L}\right) \mathrm{yr},
    \end{equation}
    for all AGNs, where $\epsilon = L/\dot{M} c^2$ is the radiation efficiency, $L$ is the bolometric luminosity
    of AGNs, $L_\mathrm{Edd} = 4\pi G m_p c M_\mathrm{BH} /\sigma_T$ is the Eddington luminosity,
    $M_\mathrm{BH}$ is the SMBH mass,
    $\sigma_T$, $G$, $m_p$, $c$, are cross section of Tompson scattering, gravitational constant, proton mass, and
    the speed of light, respectively.
    Their model also underestimates the number density of less luminous AGNs at $z < 1.5$.
    They solve this problem by introducing a disc instability as a triggering mechanism of AGNs.
    Other SA models also try to reproduce AGN LFs by introducing additional triggering mechanisms of SMBH growth
    without reconsidering the accretion timescale.

    In this paper, we test a new phenomenological and physically-motivated model of the accretion timescale
    and investigate whether the effect of the accretion timescale
    can explain the cosmic evolution of the AGN number density.
    We employ a revised version of an SA model, \textit{ ``New Numerical Galaxy Catalogue''}
   (hereafter \nugc; \citealt[][hereafter M16]{nu2gc}).
    The model more accurately explains statistical properties of galaxies and AGNs at various redshifts
    than the model of \citetalias{nu2gc}.
    We show the statistical properties of AGNs and SMBHs
    obtained by the model.
    This paper is organized as follows.
    In Sec.~\ref{sec:model} we present the details of the
    modellings which is relevant to the growth of SMBHs and their host bulges.
    In Sec.~\ref{sec:AGN}, we present the statistical properties of model SMBHs and AGNs.
    We mainly focus on the effect of the accretion timescale on AGN properties.
    Finally, in Sec \ref{sec:discussion}, we discuss and summarise the results.

  \section{Model Descriptions}
  \label{sec:model}
    We create merging histories of DM haloes from
    large cosmological $N$-body simulations \citep{Ishiyama15}
    \footnote{Cosmological simulation data are available from the following link:
              \protect\url{http://hpc.imit.chiba-u.jp/~ishiymtm/db.html}},
    which have higher mass resolution and larger volume compared with previous simulations
    (e.g. 4 times better mass resolution
    compared with Millennium simulations, \citealt{Springel05June}).
    Table~\ref{tab:N-body} summarises basic properties of the simulations.
    The \nugc~-M, and -SS simulations have the same mass resolution with different box sizes
    ($L = 560$ and $70~h^{-1}\mathrm{Mpc}$, respectively).
    The \nugc~-H2 simulation is one of the high-resolution simulations for our SA model
    which has $\sim 64$ times higher mass resolution than the \nugc~-SS simulation with the same box size.
    Throughout this paper, we assume a $\Lambda$CDM universe with
    the following parameters:
    $\Omega_{0}=0.31$,  $\lambda_{0}=0.69$,
    $\Omega_\mathrm{b} = 0.048$,
    $\sigma_8 = 0.83$, $n_\mathrm{s} = 0.96$, and a Hubble constant of
    $H_0 = 100~h~\mathrm{km}~\mathrm{s}^{-1}$~Mpc$^{-1}$, where $h=0.68$
    \citep{Planck,Planck16}.

    \begin{table*}
    \centering
      \begin{tabular}{lllccc}
      \hline
        Name & $N$ & $L$~[$h^{-1}$ Mpc] & $m~[h^{-1} M_\odot]$ & $M_\mathrm{min}~[h^{-1} M_\odot]$ & $M_\mathrm{max}~[h^{-1} M_\odot]$ \\
          \hline
        \nugc {\scriptsize -M} & $4096^3$ & 560.0 & $2.20\times 10^8 $ & $8.79\times 10^9$ & $2.67\times 10^{15}$ \\
        \nugc {\scriptsize -SS} & $512^3$ & 70.0 & $2.20\times 10^8 $ & $8.79\times 10^9$ & $6.58\times 10^{14}$ \\
        \nugc {\scriptsize -H2} & $2048^3$ & 70.0 & $3.44\times 10^6 $ & $1.37\times 10^8$ & $4.00\times 10^{14}$ \\
        \hline
      \end{tabular}
      \caption{Properties of the \nugc~simulations we have employed in this paper. $N$ is the number of simulated particles, $L$ is the comoving box size,
      $m$ is the individual mass of a dark matter particle, $M_\mathrm{min}$ is the mass of the smallest haloes ($= 40\times m$) which corresponds to
      the mass resolution, and $M_\mathrm{max}$ is the mass of the largest halo in each simulation.}
    \label{tab:N-body}
    \end{table*}

\begin{figure}
      \begin{center}
        \includegraphics[width=\hsize]{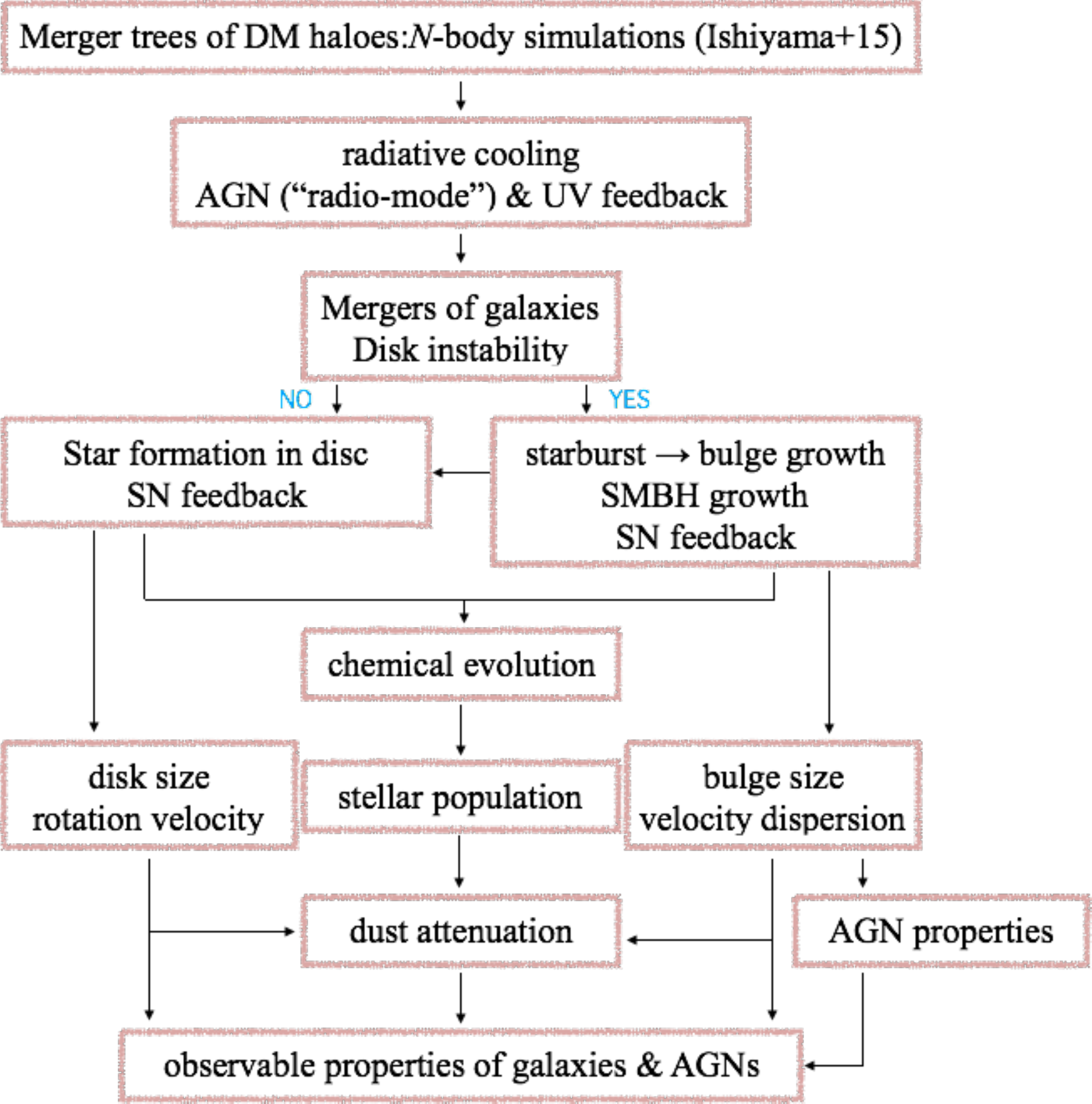}
      \end{center}
      \caption{Schematics of the model showing the determination of observable properties of galaxies and AGNs.}
      \label{fig:flows}
    \end{figure}

    The model used in this work is based upon the
    original $\nu^2 GC$ model of \citetalias{nu2gc},
    although it has undergone numerous improvements.
    This model originates from \cite{NY04} and \citet{nugc} ($\nu GC$ model).
    Both $\nu GC$ and $\nu^2 GC$ models have been used for a variety of astrophysical studies including
    gravitational waves,
    Ly$\alpha$ emitters, star formation, and AGN clustering
    \citep{Enoki03,EN07,MARK07,Makiya14,Oogi16,Oogi17}.
    The SMBH growth and AGN properties in \citetalias{nu2gc}
    are based on \cite{Enoki03}, \cite{Enoki14}, and \cite{Shirakata15}.
    In this section, we describe the processes which relate to
    the growth of SMBHs (and their host bulges).
    Other modellings are shown in Appendix \ref{App:Model}.
    Schematics of the model are shown in Fig. \ref{fig:flows}.
    In our model, the values of adjustive parameters are
    determined by a Markov Chain Monte Carlo (MCMC) method.
    The details of fitting procedures, its result, and the resulting statistical properties of galaxies
    with the fiducial model are shown in Appendix \ref{App:ModelResults}.

    \subsection{Bulge growth by mergers and disc instability}
    \label{sec:MergerandDI}
      We assume that the bulge (spheroid) component within a galaxy grows via starbursts
      and the migration of disc stars, both of which are triggered by
      mergers of galaxies and disc instabilities.
      Our model for these processes is based on \cite{Shirakata16}.

      \subsubsection{Mergers of galaxies}
      \label{Merger}
        When DM haloes merge with each other, the newly formed halo should contain several galaxies
        which are classified as satellite galaxies and a single central galaxy.
        All members of this galaxy group would eventually merge
        under the gravitational attraction of the resultant halo.
        Mergers of galaxies occur via dynamical friction (central-satellite merger)
        and random collision (satellite-satellite merger).
        We estimate the timescales of dynamical friction and random collision
        in the same manner as \citetalias{nu2gc}.
        For the dynamical friction, we set the merger timescale, $\tau_\mathrm{mrg}$,
        as $\tau_\mathrm{mrg}~=~f_\mathrm{mrg}~\tau_\mathrm{fric}$,
        where $f_\mathrm{mrg}$ is an adjustable parameter (in this paper, $f_\mathrm{mrg} = 0.81$)
        and $\tau_\mathrm{fric}$ is
        the timescale of dynamical friction, for which we adopt the formula by \cite{Jiang08} and  \cite{Jiang10}.
        \footnote{\citetalias{nu2gc} set the orbital circularity as 0.5 for determining
          $\tau_\mathrm{fric}$, which is the average value obtained from \cite{Wetzel11}.
          In this paper, we consider the halo mass dependence on the circularity obtained from
          the same previous work \citep{Wetzel11}.}

        These types of mergers induce bulge formation and growth within a galaxy.
        We introduce the model of the merger-driven bulge growth proposed by \cite{Hopkins09F}
        based on hydrodynamic simulations.
        When galaxies merge, stars and gas lose their angular momentum
        through bar instabilities induced by the merger.

        We define a primary galaxy as the galaxy with a larger baryon mass, $M_\mathrm{1}$
        (cold gas $+$ stars $+$ a central BH), between the merging pair,
        and secondary galaxy as the one with smaller baryon mass, $M_\mathrm{2}$.
        We assume that the secondary is absorbed in the bulge of the primary.
        The bulge also obtains the cold gas and stars from the primary's disc.
        The migrated stellar mass, $\Delta M_\mathrm{1ds}$, is determined as
        $MIN(f_\mathrm{*}M_\mathrm{2}, M_\mathrm{1,ds})$, where
        $f_\mathrm{*}~=~G (\mu)~=~2\mu~/~(1+\mu)$ is the mass fraction of the disc that is destroyed
        as a function of $\mu = M_\mathrm{2}/M_\mathrm{1}$ \citep{Hopkins09F}.
        This results in the bulge of the primary gaining the stellar mass of
        $M_\mathrm{2} + \Delta M_\mathrm{1ds} \lesssim 2M_\mathrm{2}$ per a merger.

        The gas mass which migrates in from the primary's disc is assumed to depend on
        the disc fraction of the primary, $f_\mathrm{1d} = (M_\mathrm{1,ds} + M_\mathrm{1,dg})/M_\mathrm{1}$
        ($M_\mathrm{1dg}$ is the cold gas mass in a primary's disc before the merger), the gas mass fraction in the primary's disc,
        $f_\mathrm{1g}$, and a pair of orbital parameters, $b$ and $\theta$.
        The parameter, $b$, is the peri-galacticon distance before coalescence
        and $\theta$ is the inclination of the orbit
        of the secondary relative to the primary's disc.
        Assuming the disc has an exponential surface density profile, we obtain the radius in which the gas migrates
        to the bulge, $R_\mathrm{gas}$ following the Eq. 7 of \cite{Hopkins09F}:
        \begin{equation}
          \label{eq:Rgas}
          \frac{R_\mathrm{gas}}{r_\mathrm{ds}} = (1 - f_\mathrm{1g})f_\mathrm{1d} F(\theta, b) G(\mu),
        \end{equation}
        where $r_\mathrm{ds}$ is the scale radius of the disc and $F(\theta, b)$ is a function of $b$ and $\theta$.
        \footnote{We assume that gas and stars in the disc have the same scale radius
        \citep[see, however,][]{Mitchell18}.}
        Since we cannot obtain $b$ and $\theta$ from merger trees of the DM haloes,
        we employ the average value of $F(\theta, b)$ suggested by \cite{Hopkins09F},
        $\langle F(\theta, b) \rangle = 1.2$.
        The mass of the cold gas inside $R_\mathrm{gas}$, $\Delta M_\mathrm{1dg}(<R_\mathrm{gas})$, migrates to the bulge
        and is exhausted by a starburst. The mass is described as follows:
        \begin{equation}
          \label{eq:Dgas}
          \Delta M_\mathrm{1dg} = M_\mathrm{1dg} \times \left\{1 - \left(1 + \frac{R_\mathrm{gas}}{r_\mathrm{ds}}\right) \exp(-R_\mathrm{gas}/r_\mathrm{ds})\right\}.
        \end{equation}
        As seen in Eq.~\ref{eq:Rgas}, $R_\mathrm{gas}$ is larger
        for smaller $f_\mathrm{1g}$
        because gas can lose its angular momentum by the torques induced by stars
        \citep{Hopkins09F}.

        As shown in Eqs.~\ref{eq:Rgas} and~\ref{eq:Dgas}, $\Delta M_\mathrm{1dg}$
        is smaller than $M_\mathrm{1dg}$
        even when $\mu = 1$ (i.e., an equal-mass merger).
        In this case, we cannot form pure bulge galaxies.
        We thus assume that the disc of the primary galaxy is completely destroyed when $\mu > f_\mathrm{major}$,
        where $f_\mathrm{major}$ is a free parameter ($f_\mathrm{major} = 0.89$).
        We then set $\Delta M_\mathrm{1ds} = M_\mathrm{1ds}$ and
        $\Delta M_\mathrm{1dg} = M_\mathrm{1dg}$.

        The cold gas in the bulge is consumed by a starburst
        even when only a minor merger occurs.
        The time evolution of the mass of stars, gas, metals (hot and cold phases),
        and BHs are calculated by Eqs.~\ref{eq:dotstar}, \ref{eq:dotBH},
        \ref{eq:dothot}, \ref{eq:dotcold}, \ref{eq:dotcoldZ},
        and \ref{eq:dothotZ}
        with $\tau_\mathrm{star} \rightarrow 0$.
        The mass of newly formed stars by a starburst, $\Delta M_\mathrm{star, burst}$ is described as:
        \begin{equation}
          \Delta M_\mathrm{star, burst} = \frac{\alpha}{\alpha + \beta + f_\mathrm{BH}} M_\mathrm{cold}^0,
          \label{eq:SB}
        \end{equation}
        where $M_\mathrm{cold}^0$ is the cold gas mass in the bulge immediately after a merger,
        $\alpha$ is the locked-up mass fraction, $f_\mathrm{BH}$ is the fraction of the gas
        which gets accreted onto the SMBH, and $\beta$ is defined in Eq \ref{eq:SNFB} in Appendix \ref{SF}.
        Most of the cold gas in the bulge is turned into stars by the starburst
        and the remaining small fraction of the gas is accreted onto the central BH
        as described in Sec.~\ref{SMBH}.

      \subsubsection{Disc instability}
      \label{DI}
         We also consider bulge growths via disc instabilities.
         When a galactic disc becomes gravitationally unstable,
         a small fraction, $f_\mathrm{bar}$, of the galactic disc
         is assumed to migrate to the bulge.

         Following \cite{ELN82}, a galactic disc becomes bar unstable when
         \begin{equation}
           \frac{V_\mathrm{max}}{(GM_\mathrm{disc}/r_\mathrm{ds})^{1/2}} < \epsilon_\mathrm{DI,crit},
           \label{eq:DI}
         \end{equation}
         where $V_\mathrm{max}$ is
         the maximum rotation velocity.
         The scale length, $r_\mathrm{ds}$, is estimated as $r_\mathrm{ds} = (1/\sqrt{2})\langle \lambda_\mathrm{H} \rangle R_\mathrm{init}$,
         where $R_\mathrm{init}$ is the initial radius of the hot gas sphere and $\langle \lambda_\mathrm{H} \rangle$ is
         the mean value of the dimensionless spin parameter.
         We employ $\langle \lambda_\mathrm{H}\rangle = 0.042$ \citep{Bett07}, for simplicity,
         because the time evolution of the spin parameter is unclear.
         Note that to calculate the statistical properties of galaxies, such as the size distribution
         of the discs at $z \sim 0$, we take the distribution of the spin parameter into account (Sec. \ref{Size}).

         Galactic discs are more stable when bulges are present.
         We consider this effect by calculating $V_\mathrm{max}$ as follows:
         \begin{align}
           &V_\mathrm{max} = \sqrt{V_\mathrm{max,NFW}^2 + V_\mathrm{max,bulge}^2}, \\[5mm]
           &V_\mathrm{max,NFW} \sim 0.465\sqrt\frac{c}{\ln(1+c) - c/(1+c)} V_\mathrm{circ}, \\[5mm]
           &V_\mathrm{max,bulge} = \left \{ \begin{array}{ll}
                                    \sigma_\mathrm{1D} & (r_\mathrm{ds} \lesssim r_\mathrm{b}) \\[5mm]
                                    \sqrt{\frac{M_\mathrm{bulge}G}{r_\mathrm{ds}}} & (r_\mathrm{ds} > r_\mathrm{b}), \\
                                   \end{array} \right.
         \end{align}
         where $c$ is the concentration parameter of a DM halo, $\sigma_\mathrm{1D},$ and $r_\mathrm{b}$
         are the 1D velocity dispersion and the size of the bulge, respectively.
         We assume that a bulge has the isothermal density profile (see Sec.~\ref{sec:bulgeprop}).

         The critical value for disc stabilities, $\epsilon_\mathrm{DI,crit}$ (Eq. \ref{eq:DI}),
         depends on the gas fraction and density profile of a galactic disc
         \citep[e.g.][]{ELN82, CST95}.
         If the velocity dispersion of galactic discs is neglected,
         the value of $\epsilon_\mathrm{DI,crit}$ is $\sim 1.1$ for the exponential stellar disc
         \citep{ELN82} and $\sim 0.9$ for the gaseous disc \citep{CST95}.
         We, however, treat $\epsilon_\mathrm{DI,crit}$ as an adjustable parameter,
         whose value should be $\leq 1.1$ since the disc actually has the velocity dispersion
         and becomes more stable.
         We set $\epsilon_\mathrm{DI,crit}~=~0.75$ to explain the observed cosmic SFR density.
         If we set $\epsilon_\mathrm{DI,crit} = 1.1$, the cosmic SFR density
         becomes constant at $4 < z < 6$, which is inconsistent with the previous suggestions
         and such model cannot explain the observed stellar mass -- SFR relation.

         We note that some other SA models \cite[e.g.][]{Cole00, Lacey16} use
         the circular velocity and the half-mass radius of the disc instead of $V_\mathrm{max}$ and $r_\mathrm{ds}$.
         The circular velocity would change by the effect of the supernovae (SNe) explosions.
         We thus use $V_\mathrm{max}$ following original prescription by \cite{ELN82}.
         If we assume an exponential disc, the effective radius is only $\sim 1.67$ times larger than the scale length.

         When a galactic disc becomes gravitationally unstable, a fraction of the cold gas and stars
         in the disc is added to the bulge component.
         The migrated stellar mass from the disc to bulge, $\Delta M_\mathrm{ds,DI}$,
         is determined as:
         \begin{equation}
           \Delta M_\mathrm{ds,DI} = f_\mathrm{bar} M_\mathrm{ds},
           \label{eq:starDI}
         \end{equation}
         where $f_\mathrm{bar}$ is a free parameter and $M_\mathrm{ds}$ is the stellar mass of the disc.
         The gas mass which migrates in from the disc, $\Delta M_\mathrm{dg,DI}$,
         is determined as:
         \begin{align}
           &\Delta M_\mathrm{dg,DI} = M_\mathrm{1dg} \times \left\{1 - \left(1 + \frac{R_\mathrm{gas}}{r_\mathrm{ds}}\right) \exp(-R_\mathrm{gas}/r_\mathrm{ds})\right\}, \label{eq:gasDI}\\
           &\frac{R_\mathrm{gas}}{r_\mathrm{ds}} = (1 - f_\mathrm{1g})f_\mathrm{1d} f_\mathrm{bar}, \label{eq:RgasDI} \\
         \end{align}
         where $M_\mathrm{dg}$ is the gas mass of the disc.
         Eqs. \ref{eq:gasDI}, and \ref{eq:RgasDI} are analogous to our galaxy merger case
         with $G (\mu) = f_\mathrm{bar}$ and $F(\theta, b) = 1.0$.
         The value of the free parameter, $f_\mathrm{bar}$, is set to $0.63$.

         The spheroids formed through this process might be
         so-called `pseudo-bulges', although we do not differentiate between
         bulges formed by these instabilities and those formed by mergers.
         Starbursts triggered by these instabilities are also treated in the same way
         as those by mergers.

    \subsection{Growth of SMBHs and properties of AGNs}
    \label{SMBH}
      \subsubsection{BH seeding}
      \label{seed}
        A seed BH is immediately placed within a newly formed galaxy.
        We use a mass of the seed BHs,  $M_\mathrm{BH,seed} = 10^{3} M_\odot$, for all galaxies
        independent from the redshift.
        The minimum mass of the halo in which the gas cools and possibly forms a galaxy
        depends on redshift and the mass resolution of $N$-body simulations (see Fig. 2 in \citetalias{nu2gc}).
        A seed BH is, therefore, placed a halo with different mass with different mass resolution and/or at different redshift.
        The seed BH mass, however, does not affect the main results of this paper,
        focusing mainly on AGNs at $z \lesssim 6$,
        since the seed mass is negligible compared with the total amount of the accreted gas onto a BH
        \citep[see ][]{Shirakata16}.
        \cite{Shirakata16} suggest that the mass of the seed BHs
        should be dominated by $\sim~10^3~M_\odot$ to reproduce
        the $M_\mathrm{BH} - M_\mathrm{bulge}$ relation at $z \sim 0$,
        including galaxies with $M_\mathrm{bulge}~<~10^{10} M_\odot$.

      \subsubsection{Mass accreted by SMBHs}
        When a starburst is triggered by a galaxy merger or disc instability (Sec.~\ref{sec:MergerandDI}),
        a small fraction of the gas is supplied to the central SMBH.
        The accreted gas mass per starburst, $\Delta M_\mathrm{acc}$, is given by:
        \begin{equation}
          \label{eq:DetM}
          \Delta M_\mathrm{acc} = f_\mathrm{BH} \Delta M_\mathrm{star, burst},
        \end{equation}
        where $f_\mathrm{BH} = 0.02$, in this paper.
        We calculate the time evolution of the mass accretion rate, $\dot{M}_\mathrm{BH}$,
        from $\Delta M_\mathrm{acc}$ and the accretion timescale, $t_\mathrm{acc}$, as
        \begin{equation}
          \label{eq:Mdot}
          \dot{M}_\mathrm{BH} = \frac{\Delta M_\mathrm{acc}}{t_\mathrm{acc}} \exp\left(\frac{t-t_\mathrm{start}}{t_\mathrm{acc}}\right),
        \end{equation}
        where $t_\mathrm{start}$ is the starting time of accretion,
        which is the same as that of the starburst.
        The prescription for $t_\mathrm{acc}$ is the main topic of this paper and
        will be described in Sec.~\ref{sec:lifetime} in detail.
        The starting time of the starburst, $t_\mathrm{start}$, is assigned randomly within the time step.
        \cite{Shirakata15} suggests that $t_\mathrm{start}$ must be delayed from the starting time of the starburst
        so that the dust extinction of a galaxy becomes negligible for AGNs.
        In this paper, we do not include this delay to show clearly the effect of varying the modelling of
        the accretion timescale.

        We note that Eqs. \ref{eq:DetM} and \ref{eq:Mdot} are valid
        for SMBH growth via both galaxy mergers and disc instabilities.
        Practically, the value of $f_\mathrm{BH}$ is not necessarily the same for both galaxy mergers
        and disc instabilities. There are, however, almost no suggestions about the difference of
        the fraction of the cold gas mass which gets accreted onto an SMBH with different triggering mechanisms.
        We, thus, employ the common $f_\mathrm{BH}$, for diminishing the degree of freedom.

        SMBHs also increase their mass via SMBH-SMBH coalescence following mergers of galaxies.
        As in \citetalias{nu2gc}, we simply assume that SMBHs merge instantaneously
        after the merger of their host galaxies.

      \subsubsection{The accretion timescale for SMBHs}
      \label{sec:lifetime}
        \begin{table}
          \begin{center}
            \begin{tabular}{rlc}\hline
              Model Name  & $t_\mathrm{acc}$ & free parameters\\ \hline
              \texttt{KH00model}  & $3\times10^7 (1+z)^{-1.5}$ yr & None\\
              \texttt{Galmodel}  & $\alpha_\mathrm{bulge} t_\mathrm{dyn,bulge}$ & $\alpha_\mathrm{bulge}$\\
              \texttt{GalADmodel} & $\alpha_\mathrm{bulge} t_\mathrm{dyn,bulge} + t_\mathrm{loss}$ & $\alpha_\mathrm{bulge}, t_\mathrm{loss,0}, \gamma_\mathrm{BH}, \gamma_\mathrm{gas}$\\ \hline
            \end{tabular}
            \caption{Summary of the accretion timescale model (Sec. \protect\ref{sec:lifetime}).}
            \label{tab:acctime}
          \end{center}
        \end{table}

        In this paper, we test three types of the accretion timescale
        summarised in Table \ref{tab:acctime}.
        The \texttt{KH00model}, $t_\mathrm{acc}~=~3\times10^7~(1 + z)^{-1.5}$~yr,
        means that the accretion timescale is proportional to the dynamical time of the host halo
        (originally introduced by \citetalias{KH00}).

        Some SA models \citep[e.g.][]{Fanidakis12, Pezzulli17}
        instead use the \texttt{GalModel}, $t_\mathrm{acc}~=~\alpha_\mathrm{bulge}~t_\mathrm{dyn,bulge}$
        by assuming the accretion continues until the gas supply from the host galaxy continues.
        The accretion timescale is proportional to the dynamical time of the host bulge,
        $t_\mathrm{dyn,bulge}~=~r_\mathrm{b}~/~V_\mathrm{b}$
        (where $r_\mathrm{b}$ and $V_\mathrm{b}$ are the size and 3D velocity dispersion
        of the bulge, respectively), and the coefficient, $\alpha_\mathrm{bulge}$,
        is a free parameter.
        We choose the value of $\alpha_\mathrm{bulge}$
        so that the bright-end of the model AGN LFs are consistent with observed AGN LFs.
        In this paper, we set $\alpha_\mathrm{bulge} = 0.58$.

        We newly introduce the \texttt{GalADmodel} considering that
        the accretion would continue when gas is left
        in the circumnuclear torus or the accretion disc
        even when there is no gas supply from the host galaxy.
        We assume that $t_\mathrm{acc}$ is the sum of the gas supply timescale
        from its host galaxy, which is assumed to relate with the dynamical time of the bulge,
         \footnote{This also corresponds to the star formation timescale for a starburst \citep{nugc}.}
        and the timescale for the angular momentum loss
        of the accreted gas at $\lesssim 100$ pc, $t_\mathrm{loss}$:
        \begin{equation}
          t_\mathrm{acc} = \alpha_\mathrm{bulge} t_\mathrm{dyn,bulge} + t_\mathrm{loss},
          \label{eq:tacc}
        \end{equation}

        The second term of Eq.~\ref{eq:tacc} includes the angular momentum loss timescale
        in a circumnuclear torus and/or in the accretion disc.
        We construct a simplified and phenomenological model for the angular momentum loss
        in the central region.
        The gas accretion should continue beyond the starburst phase of the host galaxies
        if the accreted gas requires a longer timescale
        to lose its angular momentum in the circumnuclear torus and the accretion disc .
        In this region, the gravitational potential is dominated by the SMBH.
        The timescale thus should depend on the mass of the SMBH.
        Considering a circumnuclear torus in which
        the mass accretion rate depends on the gravitational stability \citep[e.g.][]{KW08},
        the accretion timescale would become longer for the more massive SMBH.
        This timescale would also depend on the mass ratio between the accreted gas
        and the SMBH.
        When this ratio becomes higher,
        the self-gravity of the accreted gas works more effectively and thus
        the outer edge of the accretion disc becomes smaller.
        The dynamical timescale then becomes shorter.
        We hence describe $t_\mathrm{loss}$ as a function of $M_\mathrm{BH}$
        and $\Delta M_\mathrm{acc}$:
        \begin{equation}
          t_\mathrm{loss} = \frac{t_\mathrm{loss, 0}}{\mathrm{Gyr}} \left(\frac{M_\mathrm{BH}}{M_\odot}\right)^{\gamma_\mathrm{BH}} \left(\frac{\Delta M_\mathrm{acc}}{M_\odot}\right)^{\gamma_\mathrm{gas}},
          \label{eq:tad}
        \end{equation}
        where $t_\mathrm{loss, 0}$, $\gamma_\mathrm{BH}$, and $\gamma_\mathrm{gas}$ are free parameters
        which are tailored to match the observed AGN LFs from $z \sim 0$ to $5$.
        We set values of $t_\mathrm{loss, 0}$, $\gamma_\mathrm{BH}$, and $\gamma_\mathrm{gas}$
        to be $1$ Gyr, $3.5$, and $-4.0$, respectively.
        We show that $\gamma_\mathrm{BH}$ would be $> 0$ and $\gamma_\mathrm{gas}$ would be $\lesssim 0$,
        considering the $\alpha-$viscosity in the accretion disc,
        and these signs would be the same by considering CNDs (Appendix~\ref{App:Tloss}).

        When we use this model, we find that
        there are SMBHs whose accretion timescale exceeds the age of the universe.
        In this case, we set $\dot{M}_\mathrm{BH} = 0$ implicitly assuming that accreted gas becomes gravitationally stable
        in a circumnuclear torus and/or a accretion disc, which cannot be accreted onto an SMBH.
        This treatment does not affect the shape of the AGN LFs since the accretion rates of such SMBHs are negligibly small.

        There are some analytical estimates for the timescale of the angular momentum loss
        in a circumnuclear torus
        \citep[e.g.][]{KU02, KW08}, which have been employed by
        some SA models \citep[e.g.,][]{Antonini15,Bromley14,Granato04}.
        We note that there are large uncertainties as to whether a circumnuclear torus
        with some common properties exists for all types of AGNs.

        We do not consider an obscured phase \citep[e.g.][]{Hopkins05S625},
        in which SMBHs do not appear as luminous AGNs at optical bands
        despite sufficiently large accretion rates onto SMBHs.
        To avoid this uncertainty, we compare the model results with observations
        by using AGN LFs in hard $X$-ray~(2-10 keV) (see also Sec. \ref{sec:ObsFrac}).

      \subsubsection{AGN luminosity}
      \label{AGNlum}
        We calculate the AGN bolometric luminosity, $L_\mathrm{bol}$, from the accretion rate (Eq.~\ref{eq:Mdot}).
        Hereafter we define the bolometric luminosity normalised
        by the Eddington luminosity ($L_\mathrm{Edd}$) as $\lambda_\mathrm{Edd} \equiv L_\mathrm{bol}/L_\mathrm{Edd}$
        and the accretion rate normalised by Eddington rate
        ($\dot{M}_\mathrm{Edd} = L_\mathrm{Edd} / c^2$) as $\dot{m}$.
        We employ the following relation
        between $\lambda_\mathrm{Edd}$ and $\dot{m}$ (based on \citealt{Kawaguchi03}):
        \begin{equation}
          \lambda_\mathrm{Edd} = \left[\frac{1}{1+3.5\{1+\tanh(\log(\dot{m}/\dot{m}_\mathrm{crit}))\}} + \frac{\dot{m}_\mathrm{crit}}{\dot{m}}\right]^{-1},
          \label{eq:M2L}
        \end{equation}
        where $\dot{m}_\mathrm{crit}$ is an adjustable parameter, whose value should be
        $2.5 \lesssim \dot{m}_\mathrm{crit} \lesssim 16.0$.
        We set $\dot{m}_\mathrm{crit} = 10.0$ and in this case,
        $\lambda_\mathrm{Edd}$
        has similar dependence on $\dot{m}$ to that obtained by \cite{Watarai00} and \cite{MK00}.

\begin{figure}
          \begin{center}
            \includegraphics[width=\hsize]{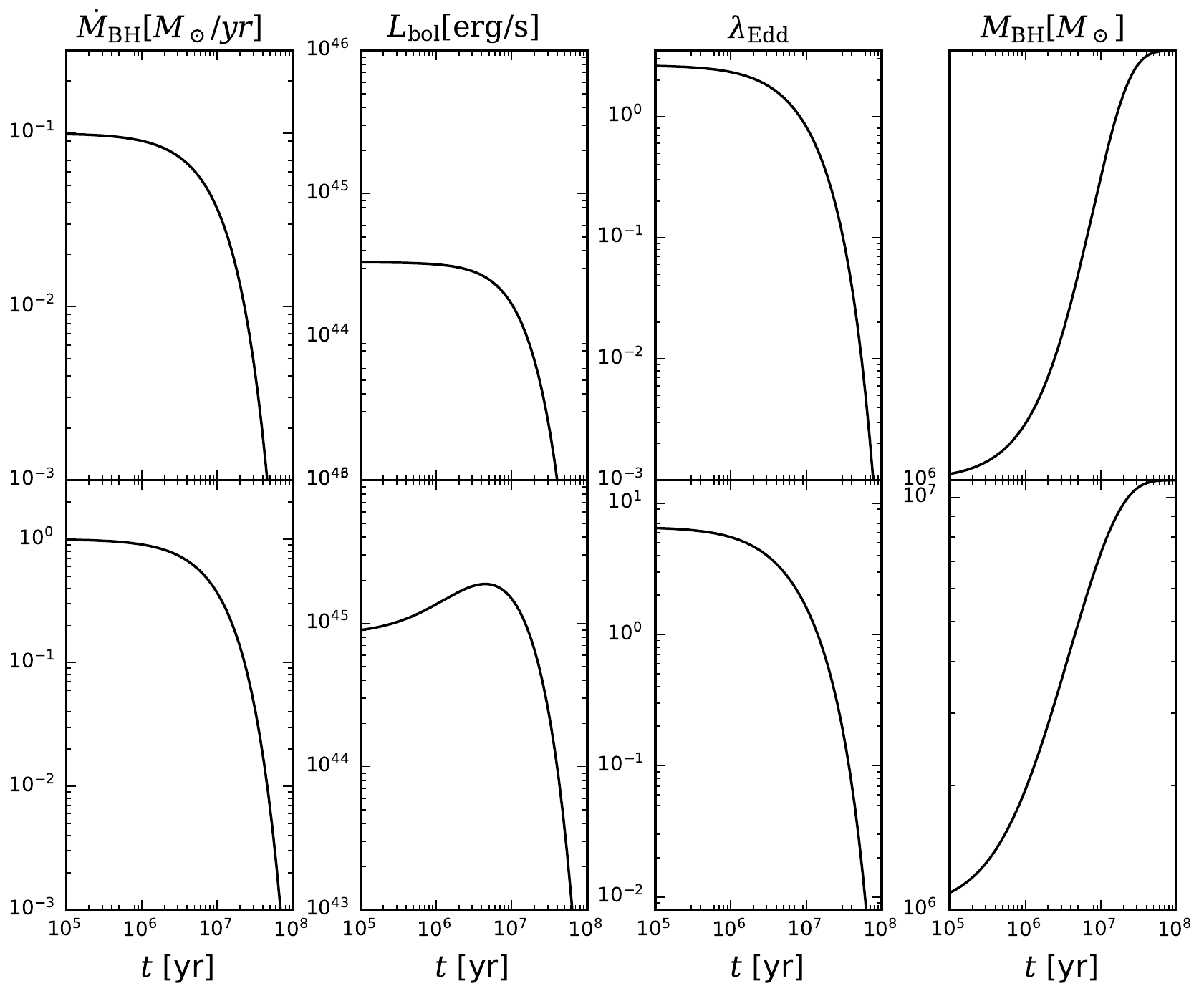}
          \end{center}
          \caption{Examples of the growth history of model SMBHs with the initial SMBH mass of $10^6 M_\odot$.
          We assume $t_\mathrm{acc} = 10^7$ yr and $\Delta M_\mathrm{acc} = 10^6$ and $10^7 M_\odot$ in top and bottom panels, respectively.
          In this figure, we show the evolution of $\dot{M}_\mathrm{BH}, L_\mathrm{bol}, \lambda_\mathrm{Edd},$ and $M_\mathrm{BH}$ from left to right panels.}
          \label{fig:growthhist}
        \end{figure}

        Although the gas accretion rate (Eq. \ref{eq:Mdot}) decreases monotonically
        with time,  $L_\mathrm{bol}$ does not necessarily decrease with time
        due to the difference of the change rate between $\lambda_\mathrm{Edd}$ and $L_\mathrm{Edd}$.
        When the following condition is satisfied, $L_\mathrm{bol} (t)$
        becomes larger than $L_\mathrm{bol} (t_\mathrm{start})$:
        \begin{equation}
          \frac{\lambda_\mathrm{Edd} (t)}{\lambda_\mathrm{Edd} (t_\mathrm{start})} > \frac{L_\mathrm{Edd} (t_\mathrm{start})}{L_\mathrm{Edd} (t)}.
        \end{equation}
        A part of AGNs with $\lambda_\mathrm{Edd} > 1.0$ satisfies this condition.
        We show the evolution of two SMBHs with $M_\mathrm{BH} = 10^6 M_\odot$ in Fig. \ref{fig:growthhist}.
        We assume $t_\mathrm{acc} = 10^7$ yr and $\Delta M_\mathrm{acc} = 10^6$ and $10^7 M_\odot$ (top and bottom panels, respectively).

        In order to obtain AGN luminosity in the optical or $X$-ray range, we employ the bolometric correction
        estimated by \cite{Marconi04}:
        \begin{equation}
        \label{eq:bolcor}
          \log [L/L_\mathrm{Y}] = a + b\mathcal{L} + c\mathcal{L}^{2} + d\mathcal{L}^{3},
        \end{equation}
        where $\mathcal{L} = (\log L - 12)$, $L$ is the intrinsic
        bolometric luminosity in units of $L_{\odot}$ ($= 3.826\times 10^{33}$ erg/s),
        and $L_\mathrm{Y}$ is the luminosity in hard $X$-ray (2-10 keV), $L_X$, or
        B-band luminosity, $\nu_B L_\mathrm{B}$ ($\nu_B$ is a central frequency of the B-band
        corresponding to $4400$ \AA).
        Parameters ($a,b,c,d$) are $(1.54, 0.24, 0.012, -0.0015)$ for hard $X$-ray,
        and $(0.80, -0.067, 0.017, -0.0023)$ for B-band.
        To obtain UV ($1450$ \AA) luminosity, $L_\mathrm{UV}$, we use
        \begin{equation}
          M_\mathrm{UV}~=~M_B + 0.85,
          \label{eq:BUV}
        \end{equation}
        where $M_\mathrm{UV}$ and $M_B$ are $UV-$ and $B-$ band magnitudes,
        respectively. The $B$-band magnitude, $M_B$, is calculated with Eq. \ref{eq:bolcor}.
        The Eq. \ref{eq:BUV} is obtained by assuming the template SED presented in \cite{KSM01}.
        By using this template SED, we also obtain
        \begin{equation}
          L_\mathrm{UV} = 0.26 L_\mathrm{bol}.
        \end{equation}

        We note that we do not consider the change of the radiation efficiency in
          the low-Eddington accreting regime (namely, $\dot{m} < 0.01 \dot{m}_\mathrm{crit}$)
          since the bolometric correction for AGNs with $\dot{m} < 0.01 \dot{m}_\mathrm{crit}$ is unclear.
          The bolometric correction obtained by \cite{Marconi04} consider the dependency on the bolometric luminosity.
          It would actually depend not only on the bolometric luminosity but also on the Eddington ratio \citep[e.g.][]{Lusso12}.
          It means that although the radiation efficiency should decrease in the low-Eddington accreting regime,
          the bolometric correction should become smaller (i.e. the fraction of $X$-ray radiation becomes larger).
          This effect is not considered in, e.g. \cite{Fanidakis12}.
          They introduce the change of the radiative efficiency without considering the shift of the bolometric correction.
          In this paper, we do not introduce the change of the radiative efficiency to keep the consistency
          and to diminish the degree of freedom of the model.

      \subsubsection{``Observable fraction'' of AGNs}
        \label{sec:ObsFrac}
        To compare the calculated AGN LFs with observed UV AGN LFs, we need to define
        ``observable fraction'' in $UV$-band, $f_{obs,UV}$, because we can only obtain
        the intrinsic luminosity of AGNs from our model.
        Since AGN obscuration and absorption processes are very complicated,
        we derive an empirical formula by the following procedures.
        Recent work \citep[e.g.][]{Aird15A, Ueda14May} has estimated the hydrogen column density distribution
        around AGNs by a compilation of available samples obtained by
        \textit{Swift}/BAT, MAXI, \textsl{ASCA, XMM-}\textit{Newton}, \textsl{Chandra} and \textsl{ROSAT}.
        Therefore, one can estimate the ``intrinsic'' luminosity in hard $X$-ray of observed AGNs
        by utilizing the hydrogen column density distribution.
        We thus use the observed hard $X$-ray LFs \citep[][Table 9]{Aird15A} to obtain
        the ``observable fraction''. The procedures are as follows.

        First, we convert hard $X$-ray luminosities to UV luminosities with Eqs~\ref{eq:bolcor} and~\ref{eq:BUV}
        and we obtain ``intrinsic'' UV LFs.
        Second, we assume the shape of the observable fraction as
        \begin{equation}
          f_{obs, UV} = A(z) \left(\frac{L_\mathrm{bol}}{10^{46} \mathrm{erg/s}}\right)^{\beta(z)},
          \label{eq:ObsFrac}
        \end{equation}
        where $L_\mathrm{bol}$ is the bolometric luminosity.
        We assume that $A$ and $\beta$ are a function of redshift,
        $A (z)~=~A_0~(1 + z)^{A_1}$ and $\beta (z)~=~\beta_0~(1~+~z)^{\beta_1}$,
        considering that the dust-to-gas ratio evolves with redshift.
        The value of $\beta_0$ should be positive, considering
        the luminosity dependence of AGN obscuration \citep[e.g.][]{Lawrence91}.
        Third, we fit parameters, $A_0, A_1, \beta_0,$ and $\beta_1$ by a Markov Chain Monte Carlo (MCMC)
        method to fit observed UV LFs (see the caption of Fig.~\ref{fig:AGNLFUV_ev}).
        After $10^5$ iterations of the MCMC fitting, we obtain the best fit values
        $(A_0,~A_1,~\beta_0,~\beta_1)~=~(0.16,~-0.05,~0.07,~0.00)$
        with which the observable fraction does not exceed $1$.

        \cite{Hopkins07J} propose an alternative formula for the ``observable fraction''.
        They employ an observed distribution of hydrogen column density
        and assume a dust attenuation curve, then they derive intrinsic AGN LFs
        in hard $X$-ray (2-10 keV), soft $X$-ray (0.5-2 keV), optical $B$,
        and mid-IR (15 $\mu$ m).
        We show the difference between observable fractions obtained from
        \cite{Hopkins07J} and this paper in Appendix~\ref{App:ObsFrac}.

        \cite{Ricci17} suggest that observed UV LFs of AGNs are well explained
        by their hard $X$-ray LFs, whose hydrogen column densities are
        less than $10^{21-22} \mathrm{cm^{-2}}$.
        Since the modelling of the distribution of gas around an SMBH is difficult
        for SA models, we estimate the observable fraction by an empirical formulation.

    \subsection{``Radio mode'' AGN feedback}
      \label{AGNFB}
      We introduce the so-called radio-mode AGN feedback process
      to prevent gas in massive haloes from cooling and forming stars.
      Following \cite{Bower06}, gas cooling in a halo is quenched
      when the following two conditions are satisfied:
      \begin{equation}
        t_\mathrm{dyn}(r_\mathrm{cool}) < \alpha_\mathrm{cool} t_\mathrm{cool},
        \label{eq:AGNFB1}
      \end{equation}
      and
      \begin{equation}
        \epsilon_\mathrm{SMBH} L_\mathrm{Edd} > L_\mathrm{cool},
        \label{eq:AGNFB2}
      \end{equation}
      where $L_\mathrm{cool}$ is the cooling luminosity of the gas,
      $t_\mathrm{dyn}$ is the dynamical time of the halo,
      $\alpha_\mathrm{cool}$ and $\epsilon_\mathrm{SMBH}$ are free parameters
      which are determined to reproduce the bright-end
      of the LFs of galaxies at $z~\sim~0$.
      We set $(\alpha_\mathrm{cool}, \epsilon_\mathrm{SMBH}) = (1.14, 2.19 \times 10^{-3})$.

  \section{Statistical Properties of AGNs and SMBHs}
  \label{sec:AGN}
\begin{figure}
      \begin{center}
        \includegraphics[width=\hsize]{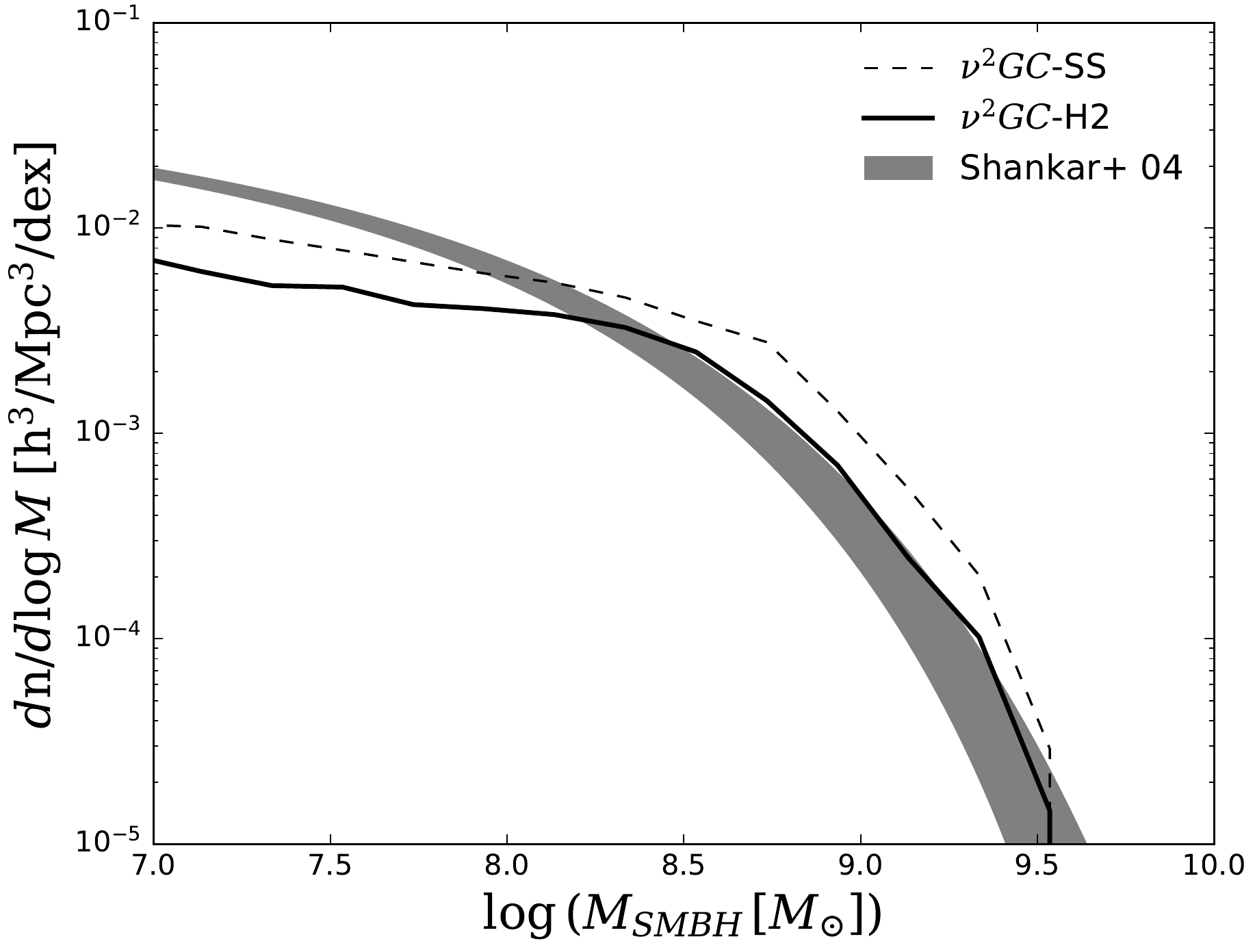}
      \end{center}
      \caption{SMBH MF at $z~\sim~0$. The model result obtained with the
      \nugc-SS and \nugc-H2 simulations
      appear in grey dashed and black solid lines with analytical fit
      to the observational data obtained from \protect\cite{Shankar04}
      in grey shaded region.}
      \label{fig:SMBHMF_z0}
    \end{figure}

\begin{figure}
      \begin{center}
        \includegraphics[width=\hsize]{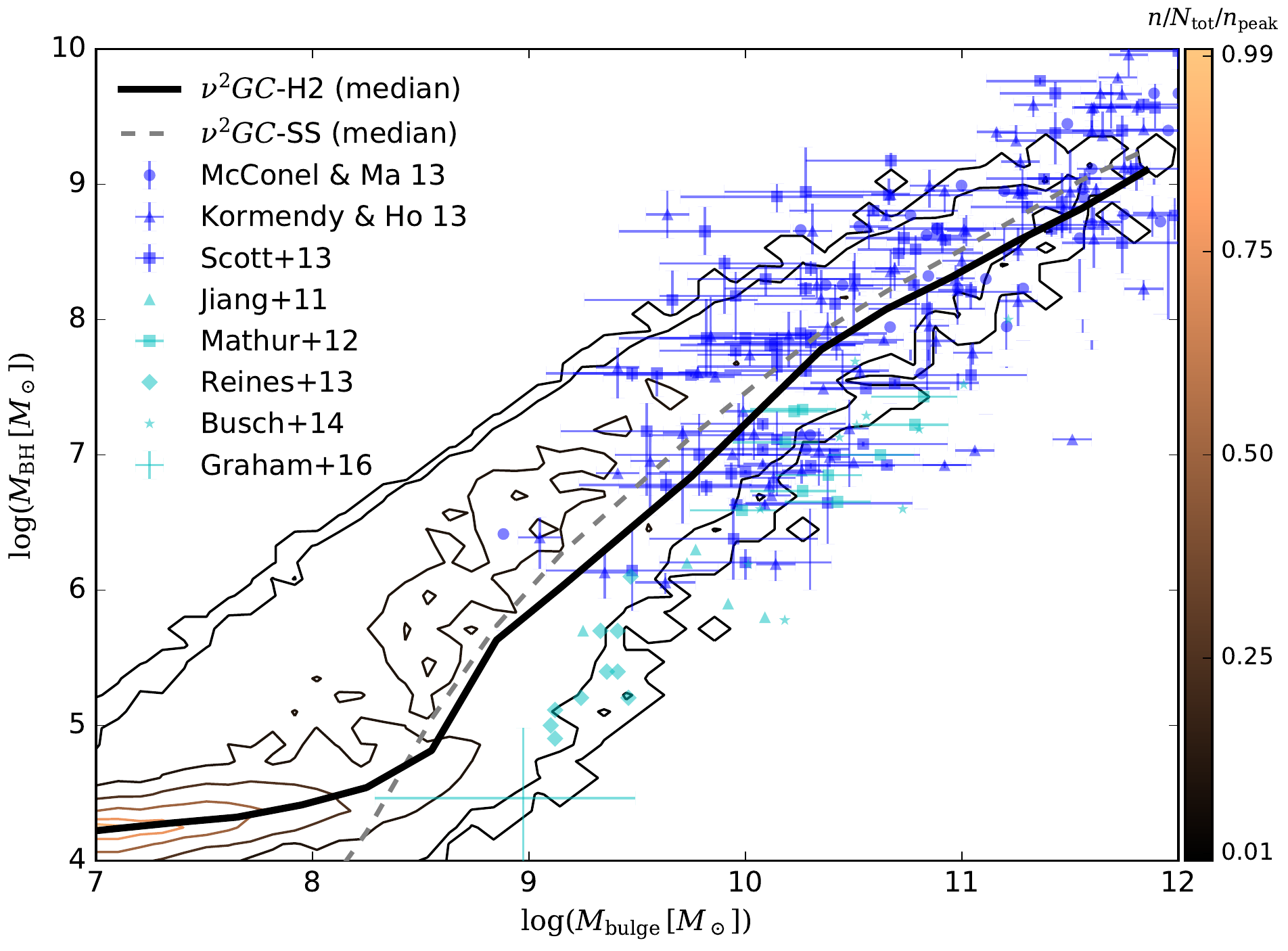}
      \end{center}
      \caption{The relation between bulge mass and SMBH mass at $z \sim 0$.
               The colour contour and black solid line show the distribution and the median value of mock galaxies
               obtained from the fiducial model with the \nugc-H2 simulation, respectively.
               We overplot the result with the \nugc-SS simulation,
               for checking the effect of the mass resolution.
               Blue filled circles, triangles, and squares are observational results for
               quiescent BH systems \protect\citep[][respectively]{MM13,KH13_review,SGS13}.
               Cyan filled triangles, squares, diamonds, stars, and pluses are observational results for AGNs
               \protect\cite{Jiang11}, \protect\cite{Mathur12}, \protect\cite{RGG13}, \protect\cite{Busch14},
               and \protect\cite{GCS16}, respectively.}
      \label{fig:Maggorian}
    \end{figure}

    We present statistical properties of model AGNs and SMBHs,
    and show their dependence on the models of the accretion timescale
    onto SMBHs.
    We first present the local SMBH MF in Fig.~\ref{fig:SMBHMF_z0} and
    the $M_\mathrm{BH}$ -- $M_\mathrm{bulge}$ relation  (including both AGNs and quiescent BHs) in Fig.~\ref{fig:Maggorian}.
    We show the results with the \nugc-SS and \nugc-H2 simulations in both figures for checking the effect of the mass resolution.
    The model SMBH MF at $z \sim 0$
    are shown as the grey dashed and black solid lines in Fig.~\ref{fig:SMBHMF_z0}.
    The SMBH MF is roughly consistent with the observational estimate \citep{Shankar04} (grey shaded region).
    The $M_\mathrm{BH}$--$M_\mathrm{bulge}$ relation at $z \sim 0$ is consistent with
    observations at $M_\mathrm{BH} > 10^{9.5} M_\odot$ (Fig. \ref{fig:Maggorian})
    since we adjust the parameter, $f_\mathrm{BH}$, to reproduce this relation.
    We, however, find that the median value of the $M_\mathrm{BH}$ -- $M_\mathrm{bulge}$ relation
    obtained by the fiducial model deviates from
    the observational estimates for $M_\mathrm{bulge} < 10^{9.5} M_\odot$.
    We do not use such low mass galaxies for the model calibration
    since the observed sample is too small.
    Most observational data for less massive galaxies with $M_\mathrm{bulge} < 10^{9.5} M_\odot$
    are AGN data. It is unclear whether the quiescent BHs with $M_\mathrm{bulge} < 10^{9.5} M_\odot$
    have the same relation as the AGNs.
    In addition, the bulge mass of less massive galaxies is difficult to estimate by observations
    since the bulge is more rotational-support.

    \subsection{The effect of the accretion timescale on AGN LFs}
    \label{Main}
\begin{figure*}
        \begin{center}
          \includegraphics[width=\hsize]{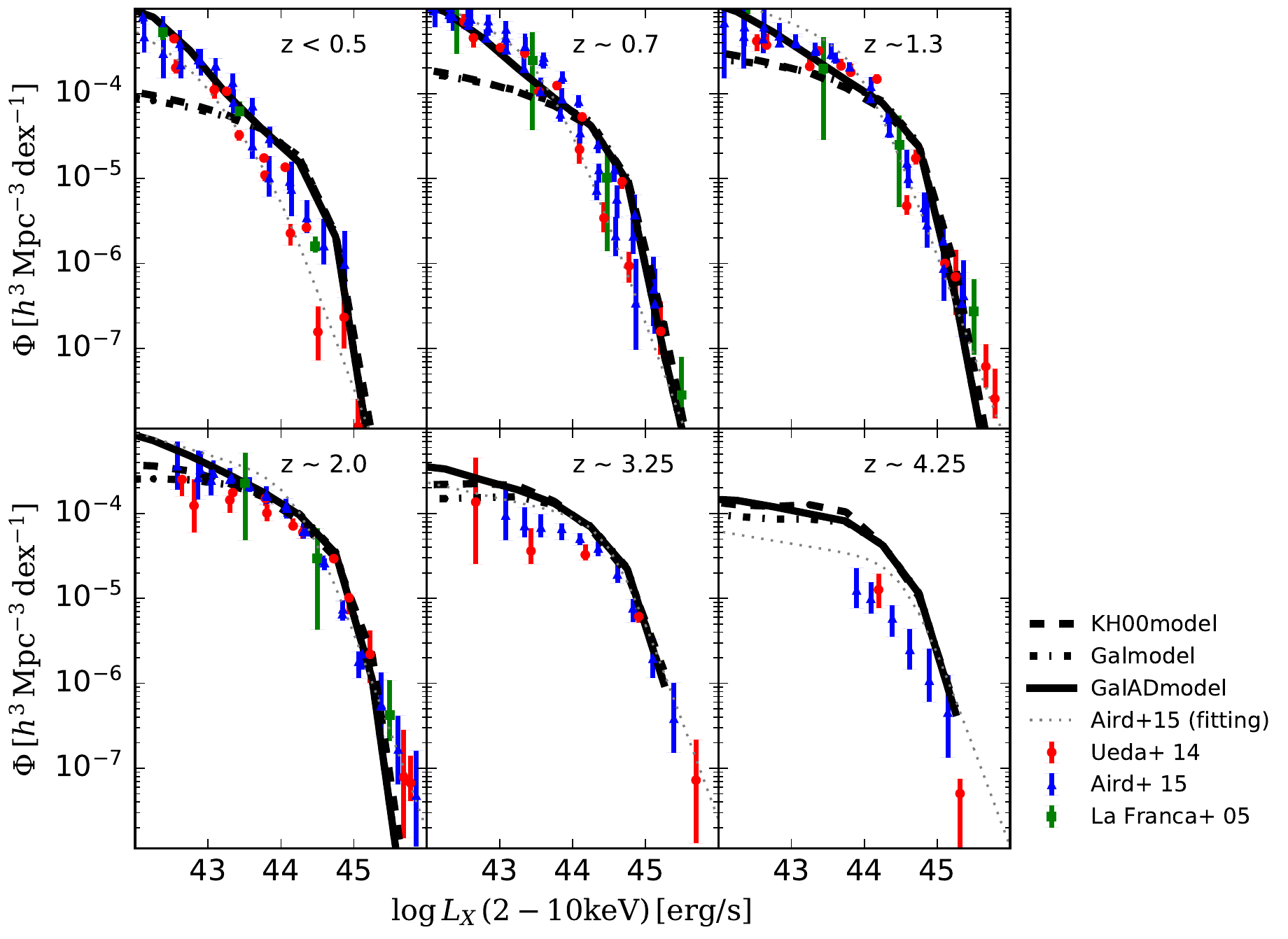}
        \end{center}
        \caption{AGN LFs in hard $X-$ ray (2-10 keV) at $z < 0.5$, $z \sim 0.7$, $z \sim 1.3$,
          $z \sim 2.0$, $z \sim 3.25$, and $z \sim 4.25$.
          The model LFs are obtained with the \nugc-M simulation.
          Black dashed, dot-dashed, and solid lines are the model LFs with different models of
          accretion timescale; the \texttt{KH00model}, \texttt{Galmodel}, and \texttt{GalADmodel}, respectively.
          Observational results are obtained from
          Red circles, blue triangles, and green squares are the data taken from
          \protect\cite{Ueda14May}, \protect\cite{Aird15A}, and \protect\cite{Franca05}, respectively.
          Grey dotted lines show the fitting LFs of observed data \protect\citep{Aird15A}.}
        \label{fig:AGNLFX}
      \end{figure*}

    We show the AGN properties obtained with \nugc.
      We present the luminosity of AGNs in the hard $X$-ray (2-10 keV) band
      because the effect of obscuration and absorption is small.
      We show how AGN LFs change when we use three different models of the accretion timescale in Fig. \ref{fig:AGNLFX}.
      Black lines show the model hard $X$-ray LFs with different accretion timescales.
      We also show the fitting function of the LFs from \cite{Aird15A} with grey dotted lines
      and observed data from \cite{Aird15A}, \cite{Ueda14May}, and \cite{Franca05}.
      We have confirmed that the results have no statistical differences
      when we employ the high resolution $N$-body simulations.

      Black dashed lines show the hard $X$-ray~(2-10~keV) AGN LFs with the \texttt{KH00model},
      which is the timescale proportional to the dynamical time of the host halo.
      The model is consistent with
      observational results at \LX~$> 43.5$
      within the dispersion of the observed data.
      We, however, find that the model underestimates the number density of AGNs at $z~<~1.0$
      with \LX~$< 43.5$ (i.e., nuclei of Seyfert galaxies),
      whose UV (1450\AA) magnitude, $M_\mathrm{UV}$, corresponds to $\sim -20.6$.
      Such less luminous AGNs are not considered in the estimation of the AGN lifetimes
      in \citetalias{KH00} and their lifetimes could significantly differ for luminous AGNs.

\begin{figure}
        \begin{center}
          \includegraphics[width=\hsize]{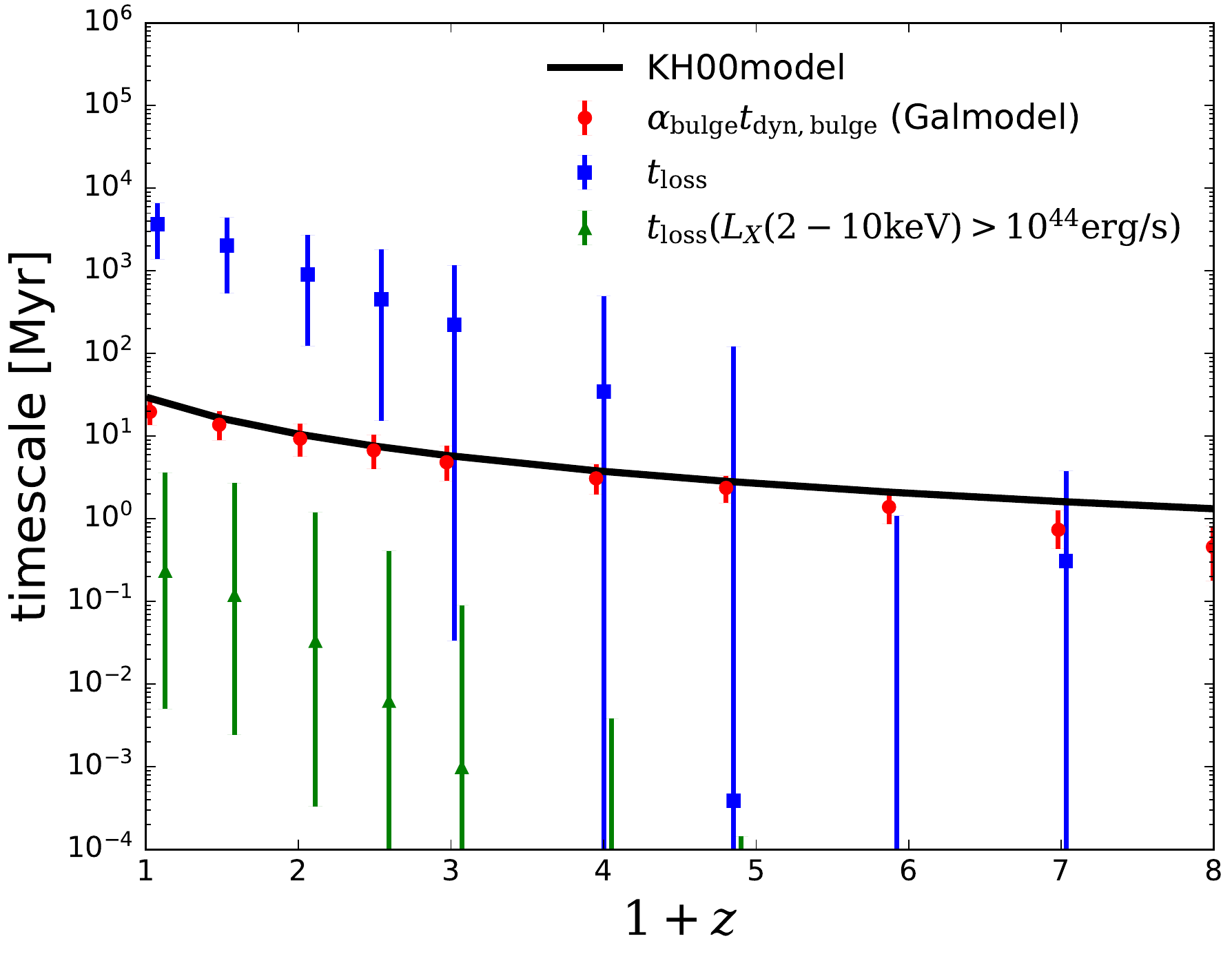}
        \end{center}
        \caption{The redshift evolution of the accretion timescale with \texttt{KH00model}, \texttt{Galmodel},
        and $t_\mathrm{loss}$.
        The black solid line shows the \texttt{KH00model}, which corresponds to the dynamical time of haloes.
        The red circles and blue squares with error bars show the median value of $t_\mathrm{dyn,bulge}$ and $t_\mathrm{loss}$
        of AGNs with \LX~$> 41.0$
        obtained by the \texttt{GalADmodel}.
        The errorbars are 25th and 75th percentiles.
        We also show the value of $t_\mathrm{loss}$ of AGNs with \LX~$> 44.0$ by green triangles.
        }
        \label{fig:tgal-tconst}
      \end{figure}

      Black dot-dashed lines show hard $X$-ray AGN LFs by the model in which the \texttt{Galmodel}.
      This modelling is similar to previous SA models \citep[e.g.][]{Fanidakis12,Shirakata16,Pezzulli17}.
      The accretion timescale does not cause a big difference in the faint-end slope of AGN LFs compared with that with the \texttt{KH00model},
      since the \texttt{Galmodel} has the accretion timescale with the same order as the \texttt{KH00model}
      as shown later in Fig.~\ref{fig:tgal-tconst}.

      Black solid lines show the hard $X$-ray AGN LFs with \texttt{GalADmodel},
      implicitly considering the timescale of angular momentum loss
      in the circumnuclear torus and the accretion disc.
      The model enables us to reproduce not only bright-ends of the LFs
      but also the faint-ends, especially at $z < 1.5$.
      When this model of the accretion timescale is employed,
      a significant fraction of low-luminosity AGNs sustain their activity for a long time as we will show later.
      The model thus reproduces the both the bright and faint-ends of AGN LFs
      much better than the other models.

      Next, Fig. \ref{fig:tgal-tconst} shows the redshift evolution of the accretion timescale of
      \texttt{KH00model} and \texttt{Galmodel},
      and $t_\mathrm{loss}$.
      We select AGNs with \LX~$> 41.0$.
      The red circles and blue squares with error bars show the median value of $\alpha_\mathrm{bulge} t_\mathrm{dyn,bulge}$
      and $t_\mathrm{loss}$ with 25th and 75th percentiles.
      The redshift evolution of the dynamical time of the bulge (red circles) and the halo (black solid line)
      are similar although the difference becomes larger at higher redshift.
      This explains why the AGN LFs with the \texttt{KH00model} and \texttt{Galmodel} are similar.
      While $t_\mathrm{loss}$ distributes broadly,
      it is longer especially at lower redshift.
      This results in the increase of the number density of AGNs at \LX~$< 43.5$ and $z < 1.5$.
      We also plot $t_\mathrm{loss}$ only for luminous AGNs with \LX~$> 43.5$ as green triangles.
      The timescale is more than 1 order of magnitude shorter than that of AGNs with \LX~$> 41.0$ at all redshifts.

\begin{figure}
       \begin{center}
         \includegraphics[width=\hsize]{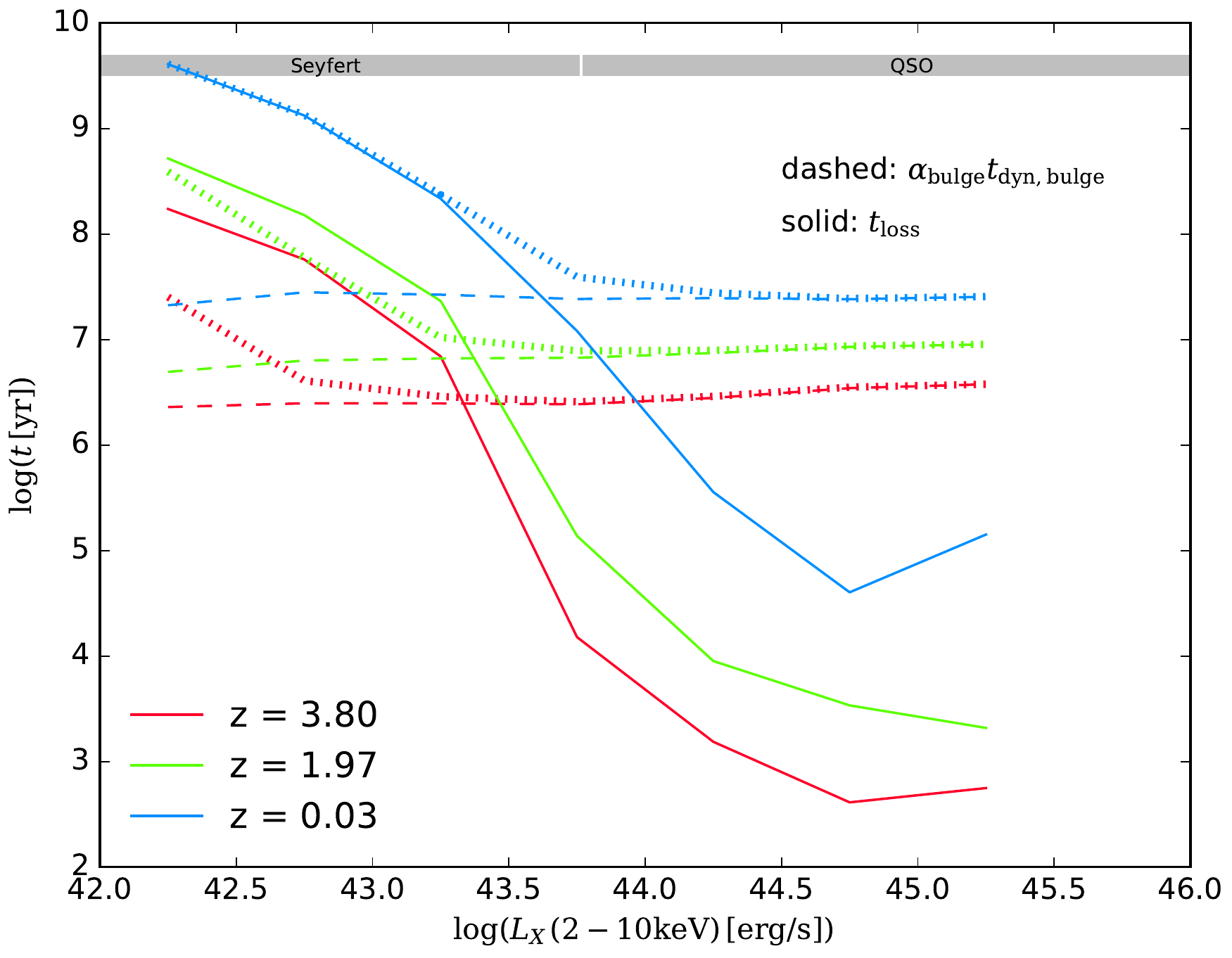}
       \end{center}
       \caption{The relation between hard $X$-ray luminosity and two different timescales
           at $z \sim 0, 2,$ and $4$ (blue, green, and red lines, respectively) obtained with the \texttt{GalADmodel}.
           Solid and dashed lines describe $t_\mathrm{loss}$ and $\alpha_\mathrm{bulge} t_\mathrm{dyn,bulge}$, respectively.}
       \label{fig:L-tacc}
     \end{figure}

      The \texttt{GalADmodel} predicts the longer accretion timescales for the less luminous AGNs
      due to the effect of $t_\mathrm{loss}$ as shown in Fig. \ref{fig:L-tacc}.
      This figure shows the relation between hard $X$-ray luminosity and
      timescales ($t_\mathrm{loss}$ and $\alpha_\mathrm{bulge} t_\mathrm{dyn,bulge}$)
      at $z \sim 0, 2,$ and $4$.
      We find that the timescale is almost constant ($\sim 2\times 10^7$ yr) for AGNs with \LX~$>44.0$
      (corresponds to $M_{UV} < -22.3$),
      which is consistent with the constraints obtained by previous studies \citep{YT02,KH00,Hopkins05S625}.
      Less luminous AGNs, in contrast, have negative correlations between the timescale and $L_X$.
      We also find that the total accretion timescale becomes longer at lower redshift for all AGNs.

\begin{figure}
        \begin{center}
          \includegraphics[width=\hsize]{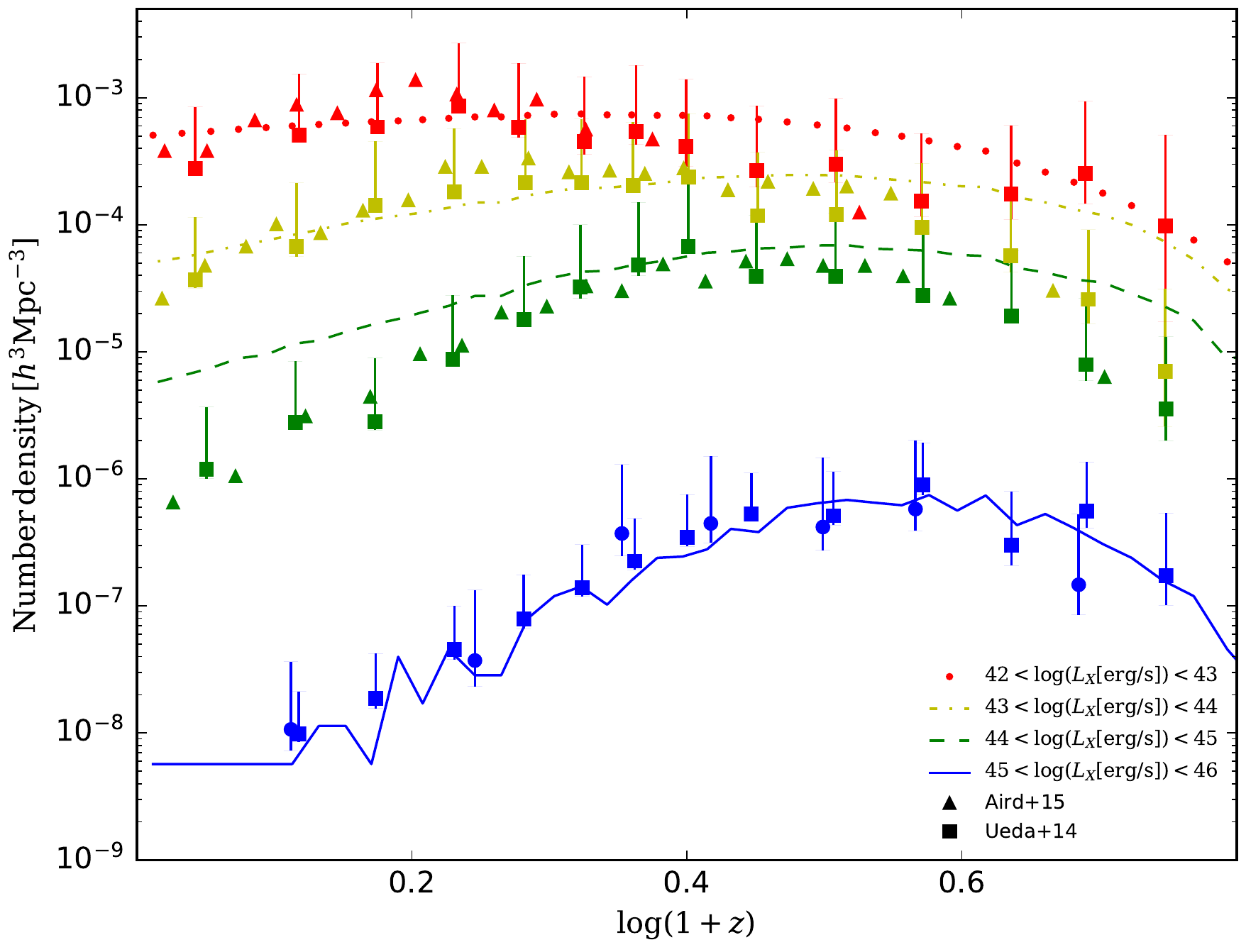}
        \end{center}
        \caption{The redshift evolution of the AGN number density.
          Colour describes the luminosity bins
          ($\log(L_X / \mathrm{erg s^{-1}}) = [42,43]$; red,
          ($\log(L_X / \mathrm{erg s^{-1}}) = [43,44]$; yellow,
          ($\log(L_X / \mathrm{erg s^{-1}}) = [44,45]$; green, and
          ($\log(L_X / \mathrm{erg s^{-1}}) = [45,46]$; blue).
          The results obtained with \texttt{GalADmodel} are shown with lines.
          Filled squares with error bars and triangles are observational results
      \protect\citep[respectively]{Ueda14May, Aird15A}.}
        \label{fig:DownSizing}
        \end{figure}

      The results obtained with the \texttt{GalADmodel} naturally explains the
      evolution of the AGN number density, which is sometimes called as
      ``anti-hierarchical trend'' of SMBH growth.
      Fig. \ref{fig:DownSizing} shows the number density of AGNs obtained with the \texttt{GalADmodel}, and
      those obtained from observations \citep{Ueda14May,Aird15A}.
      The reason why the model result shows mild anti-hierarchical trends would be
      partially because we consider the obscured fraction in hard $X$-ray (2-10 keV) is 0
      at all redshift.
      We will show a more detailed analysis in future.

      \subsection{The effect of the timescale on other properties of AGNs}
      \label{sec:others}
\begin{figure}
        \begin{center}
          \includegraphics[width=\hsize]{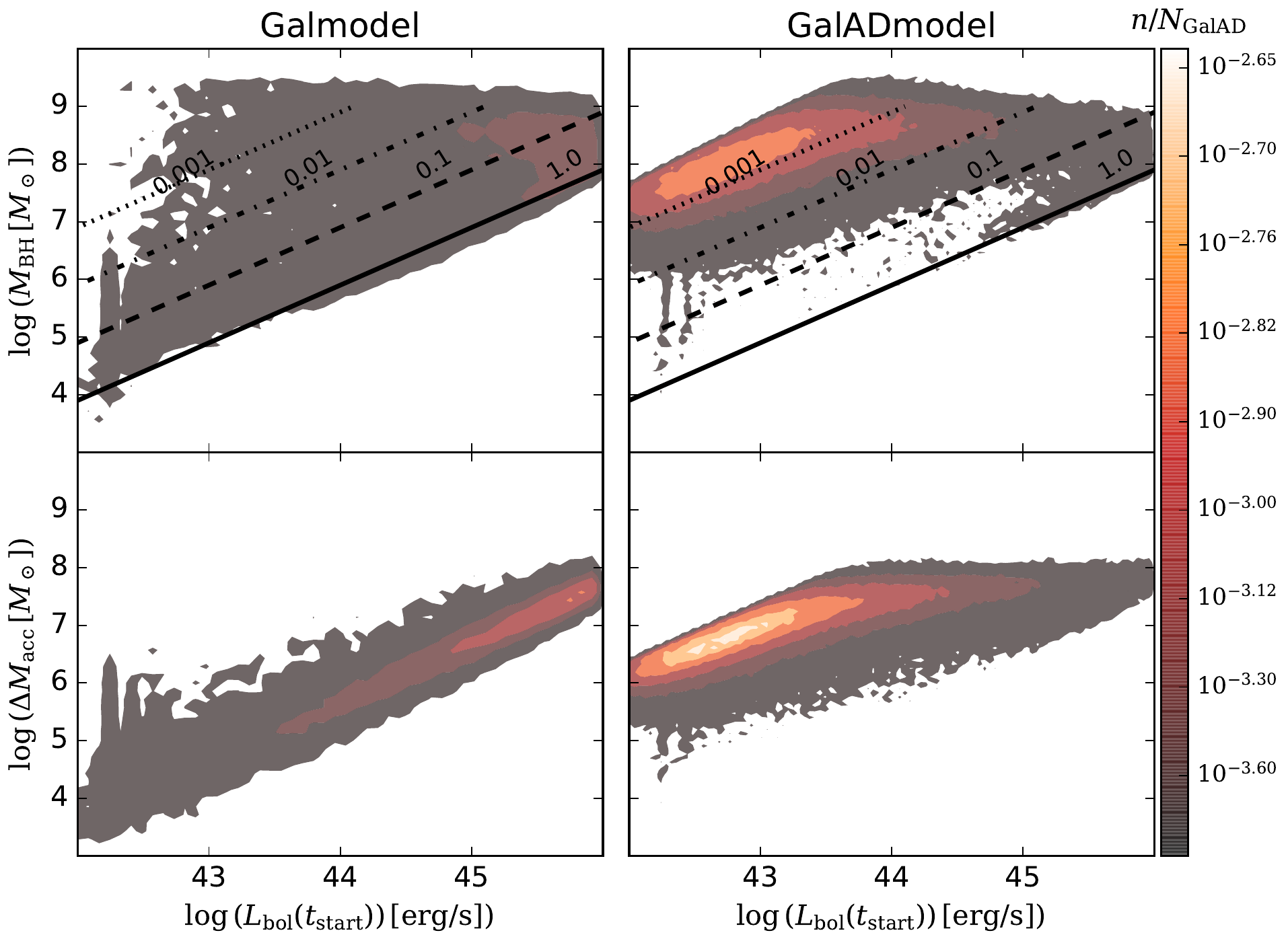}
        \end{center}
        \caption{The relation between the AGN bolometric luminosity at $t = t_\mathrm{start}$, $L_\mathrm{bol} (t_\mathrm{start})$,
                 and BH mass, $M_\mathrm{BH}$ (top) and
                 accreted gas mass onto an SMBH, $\Delta M_\mathrm{acc}$ (bottom) at $z \sim 0$.
                 The distributions are described as density contours,
                 whose value is normalized by the number of total AGNs with \texttt{GalADmodel}.
                 Left and right panels show the results obtained with
                 the \texttt{Galmodel} and \texttt{GalADmodel}, respectively.
                 Black solid, dashed, dot-dashed, and dashed lines show $\lambda_\mathrm{Edd} = 1, 0.1, 0.01,$ and $0.001$, respectively.}
        \label{fig:tloss_prop_peak}
      \end{figure}

\begin{figure}
        \begin{center}
          \includegraphics[width=\hsize]{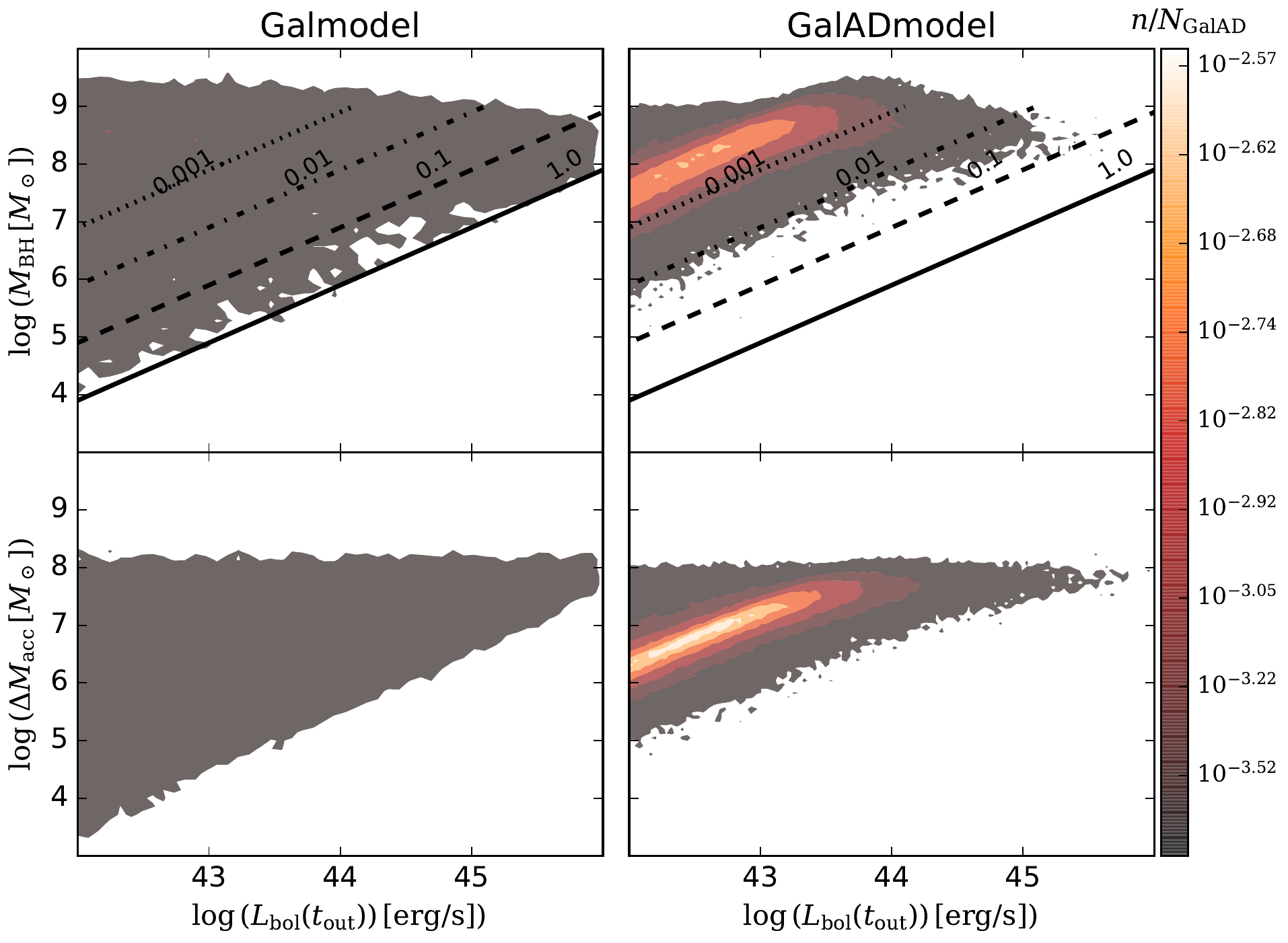}
        \end{center}
        \caption{The same figure as \protect\ref{fig:tloss_prop_peak} although
                 the $x$-axis show the AGN bolometric luminosity at output time, $L_\mathrm{bol}$
                 instead of $L_\mathrm{bol,peak}$.}
        \label{fig:tloss_prop_out}
      \end{figure}

      To see dependencies of the accretion timescale on $M_\mathrm{BH}$ and $\Delta M_\mathrm{acc}$,
      we show  the relation between AGN bolometric luminosity and BH mass, $M_\mathrm{BH}$ (top panels), and
      accreted gas mass onto an SMBH, $\Delta M_\mathrm{acc}$ (bottom panels) at $z \sim 0$,
      in Figs.~\ref{fig:tloss_prop_peak} and \ref{fig:tloss_prop_out}.
      In Fig. \ref{fig:tloss_prop_peak}, $x$-axes are the AGN bolometric luminosity at $t = t_\mathrm{start}$, $L_\mathrm{bol} (t_\mathrm{start})$,
      while these are AGN bolometric luminosity at the output time, $L_\mathrm{bol} (t_\mathrm{out})$, in Fig. \ref{fig:tloss_prop_out}.
      The left panels show the result with the \texttt{Galmodel}
      and the right panels show that obtained by the \texttt{GalADmodel}.
      We note that the model AGNs have a weak correlation between $M_\mathrm{BH}$ and $\Delta M_\mathrm{acc}$,
      of the form $M_\mathrm{BH} \propto \Delta M_\mathrm{acc}^{1.1}$, with a large dispersion.
      This positive correlation comes from the fact that the host galaxy of the heavier SMBH
      is more massive and has large amount of the cold gas.

      Fig. \ref{fig:tloss_prop_peak} shows the clear correlation between $L_\mathrm{bol} (t_\mathrm{start})$ and $\Delta M_\mathrm{acc}$
      with the \texttt{Galmodel} (bottom left panel).
      Since $t_\mathrm{dyn,bulge}$ is similar for galaxies at the same redshift (see Fig. \ref{fig:tgal-tconst}),
      the peak accretion rate, $\dot{M}_\mathrm{peak} \equiv \Delta M_\mathrm{acc} / t_\mathrm{acc}$,
      is mainly determined by $\Delta M_\mathrm{acc}$.
      The higher peak bolometric luminosity therefore implies a larger amount of the accreted gas.
      The relation between $L_\mathrm{bol} (t_\mathrm{start})$ and $M_\mathrm{BH}$ with the same model (top left panel) comes from
      the correlation, $M_\mathrm{BH} \propto \Delta M_\mathrm{acc}^{1.1}$.

      The correlations obtained by the \texttt{GalADmodel} (right panels) show bimodal distributions,
      which are quite different from the model with the \texttt{Galmodel}.
      The peak accretion rate is proportional to $M_\mathrm{BH}^{-\gamma_\mathrm{BH}} \Delta M_\mathrm{acc}^{1 - \gamma_\mathrm{gas}}$
      if $\alpha_\mathrm{bulge} t_\mathrm{dyn,bulge}$ is smaller than $t_\mathrm{loss}$.
      Since $\gamma_\mathrm{BH} = 3.5$ and $\gamma_\mathrm{gas} = -4.0$,
      $\dot{M}_\mathrm{peak} \propto M_\mathrm{BH}^{-3.5} \Delta M_\mathrm{acc}^{5.0}$.
      The peak accretion rate, thus, can be written as
      $\dot{M}_\mathrm{peak} \propto \Delta M_\mathrm{acc}^{1.15}$ (or $\propto M_\mathrm{BH}^{1.05}$).
      These positive correlations appear as contour peaks at $\log(L_\mathrm{bol} (t_\mathrm{start})/\mathrm{erg~s^{-1}}) < 44.0$.

      Fig \ref{fig:tloss_prop_out} shows the same relations as shown in Fig. \ref{fig:tloss_prop_peak},
      but instead plotting bolometric luminosity estimated at an output time.
      Since $L_\mathrm{bol} (t_\mathrm{start})$ has positive correlations with $M_\mathrm{BH}$ and $\Delta M_\mathrm{acc}$
      when the \texttt{Galmodel} is employed,
      the dispersions of the correlation between $L_\mathrm{bol}$ and $M_\mathrm{BH}$ and $\Delta M_\mathrm{acc}$
      (left panels) reflect the elapsed time from their AGN activity.

      The relation between AGN luminosity and SMBH mass allows us to compare
      theoretical models with observations and to potentially place a stronger constraint on the accretion timescale.
      There are numerous previous studies which present the relation between AGN luminosities and
      the SMBH mass at various redshifts \citep[e.g.,][]{SW10,Nobuta12,Ikeda17}.
      \cite{SW10} and \cite{SE10} show the relation between the bolometric luminosity and the SMBH mass
      for broad line AGNs at $z < 0.3$ and $0.2 < z < 2.0$, respectively.
      Since their sample are limited at $\lambda_\mathrm{Edd} > 0.01$,
      we cannot distinguish the two models of the accretion timescale.
      If complete AGN sample with $\lambda_\mathrm{Edd} > 0.001$ are obtained,
      we could put a stronger constraint on the accretion timescale.

\begin{figure*}
        \begin{center}
          \includegraphics[width=\hsize]{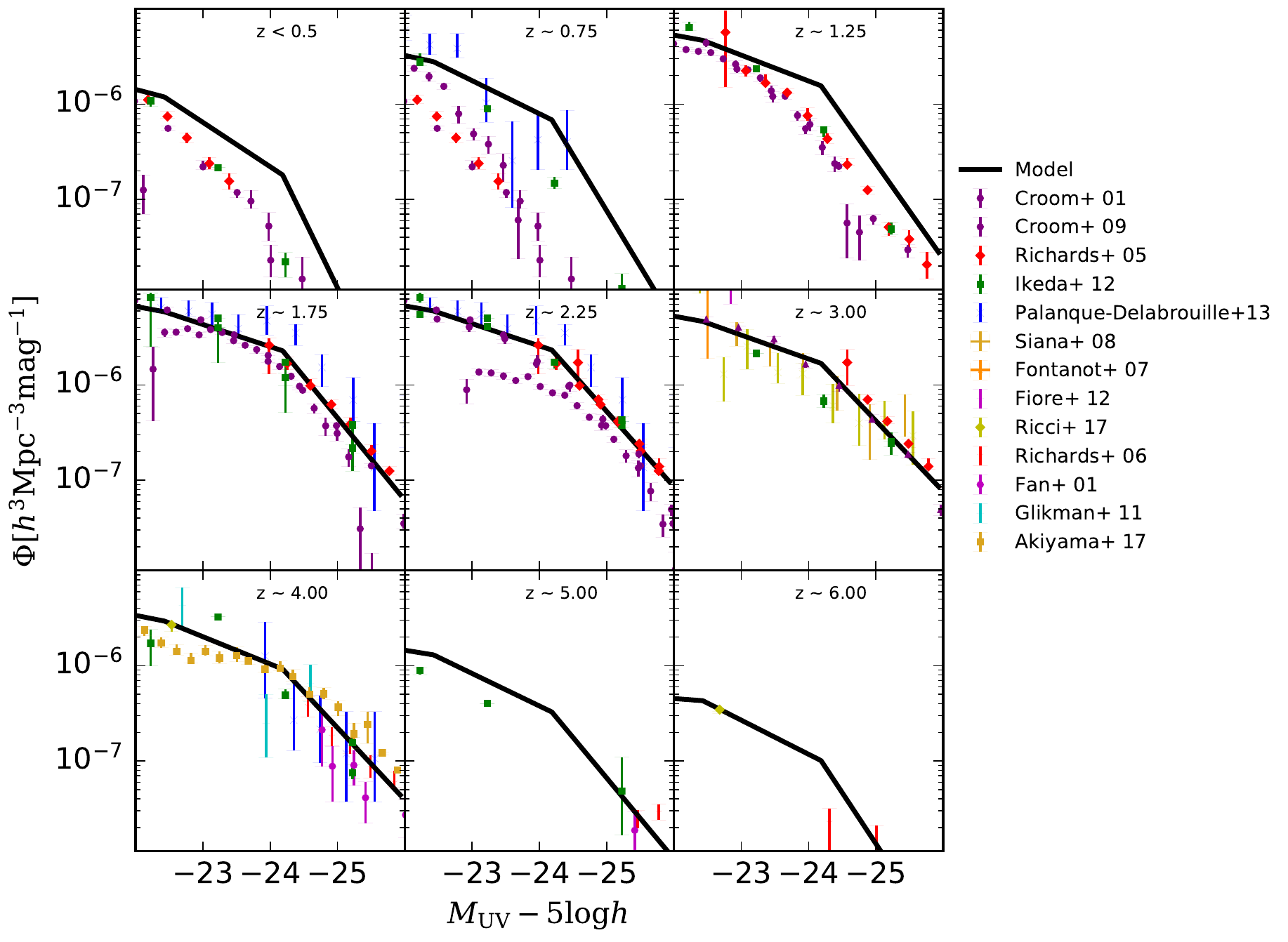}
        \end{center}
        \caption{AGN LFs in UV- band(1450 \AA) at $z < 0.5$, $z \sim 0.75$, $z \sim 1.25$, $z \sim 1.75$,
          $z \sim 2.25$, $z \sim 3.00$, $z \sim 4.00$, $z \sim 5.00$, and $z \sim 6.00$.
              The model LFs (volume-weighted) obtained with the \nugc-M simulation
              appear in black solid lines.
              Observational results are obtained from
              \protect\cite{Croom01}, \protect\cite{Croom09Nov}, \protect\cite{Fan01},
              \protect\cite{Richards2005July}, \protect\cite{Richards06June}, \protect\cite{Fontanot07},
              \protect\cite{Siana08}, \protect\cite{Glikman11},\protect\cite{Fiore12},
              \protect\cite{Ikeda12}, \protect\cite{Palanque-Delabrouille13}, \protect\cite{Ricci17},
              and \protect\cite{Akiyama18}.}
        \label{fig:AGNLFUV_ev}
      \end{figure*}

      In Fig.~\ref{fig:AGNLFUV_ev}, we present AGN LFs in $UV-$ band (1450 \AA) from $z \sim 6.0$ to $0.0$
      obtained by the \texttt{GalADmodel}.
      The observable fraction is defined by Eq.~\ref{eq:ObsFrac}.
      The results are roughly consistent with observed UV AGN LFs
      \citep{Croom01,Croom09Nov,Fan01,Richards2005July,Richards06June,Fontanot07,
      Siana08,Glikman11,Fiore12,Ikeda12,Palanque-Delabrouille13,Ricci17,Akiyama18},
      especially at $z > 1.5$.
      We, however, overproduce UV LFs at lower redshift.
      In such redshift range, we also overproduce hard $X$-ray LFs (see Fig. \ref{fig:AGNLFX})
      compared with the fitting LFs of \cite{Aird15A}
      although the model LFs are consistent with observed data points within the range of a dispersion.
      We need to take the dispersion of observed hard $X$-ray LFs into account
      for estimating the observable fraction although we leave it for future studies.
      The UV LFs do not place a strong constraint on the accretion timescale
      since the observed UV LFs are well determined only at $M_\mathrm{UV} < -20.8$
      (corresponds to \LX~$> 44.6$)
      because of the contamination of galaxies' emission \citep{Parsa16}.
      The hard $X$-ray LFs obtained from models
      with the different assumption of the accretion timescale show little difference
      at \LX~$> 44.6$.

\begin{figure}
       \begin{center}
         \includegraphics[width=\hsize]{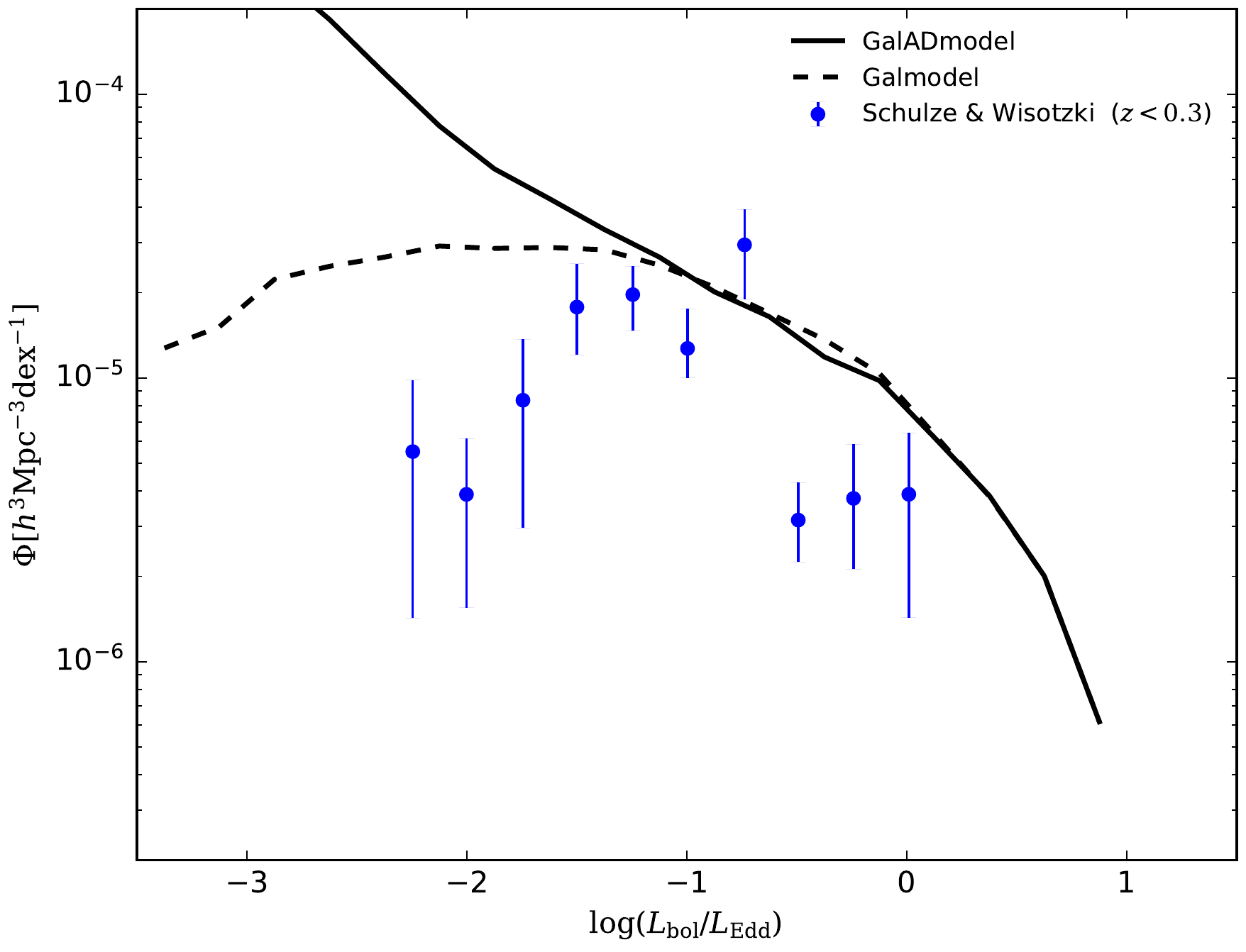}
       \end{center}
       \caption{The Eddington ratio distribution functions at $z \sim 0$ obtained with \texttt{GalADmodel}
       and \texttt{Galmodel} (black solid and dashed lines, respectively).
       In both models, AGNs with
       $M_\mathrm{BH} > 10^6 M_\odot$ and $L_X > 10^{43}$ erg/s are selected.
       Also, we compare the results with that obtained by \protect\cite{SW10} at $z \sim 0.3$
       (blue filled circles with error bars).}
       \label{fig:ERDF_z03}
     \end{figure}

      We show Fig. \ref{fig:ERDF_z03}
      to show the effect of the timescale on the Eddington ratio distribution function.
      The black solid and dashed lines are results obtained with \texttt{GalADmodel} and \texttt{Galmodel}, respectively.
      We select all AGNs with $M_\mathrm{BH} > 10^6 M_\odot$ and $L_\mathrm{bol} > 10^{43.5}$ erg/s at $z \sim 0$.
      The results at $\log(\lambda_\mathrm{Edd}) > -1.5$
      are roughly consistent with that obtained by the observation \citep{SW10} at $z \sim 0.3$.
      We, however, note that it is difficult to compare model Eddington ratio distribution functions  with observations
      since (1) optical observational sample is limited in type-1 AGNs with well-estimated SMBH mass,
      (2) SMBH masses of $X$-ray AGNs are simply estimated from e.g., the BH mass -- stellar mass relation,
      (3) observational sample seems to be incomplete for less massive SMBHs, and
      (4) the obscured fraction of AGNs would depend on both their luminosity and Eddington ratio \citep[e.g.][]{Oh15,Khim17}.
      Also, if there is an obscured growing phase before visible AGN phase suggested by, e.g. \cite{Hopkins05S625},
      then the super-Eddington accreting phase should be preferentially missed.

      Fig. \ref{fig:ERDF_z03} clearly show the difference
      caused by the implementation of the accretion timescale.
      The \texttt{GalADmodel} increases the number of AGNs with $\log(\lambda_\mathrm{Edd}) < -1.5$
      and the difference between the two models becomes larger at smaller Eddington ratio.
      We find that the \texttt{GalADmodel} and \texttt{Galmodel} have no difference for active BHMF
      with AGNs $M_\mathrm{BH} > 10^6 M_\odot$, $L_\mathrm{bol} > 10^{43.5}$ erg/s, and
      $\lambda_\mathrm{Edd} > 0.03$ (roughly similar selection as that of \citealt{SW10}).
      As we can expected from AGN LFs (Fig. \ref{fig:AGNLFX}), the Eddington ratio distribution functions at $z > 1.0$ also have little difference
      between the \texttt{GalADmodel} and \texttt{Galmodel}.
      The evolution of the Eddington ratio will appear in a future paper.

    \subsection{Triggers of the gas supply from host galaxies}
\begin{figure}
        \begin{center}
          \includegraphics[width=\hsize]{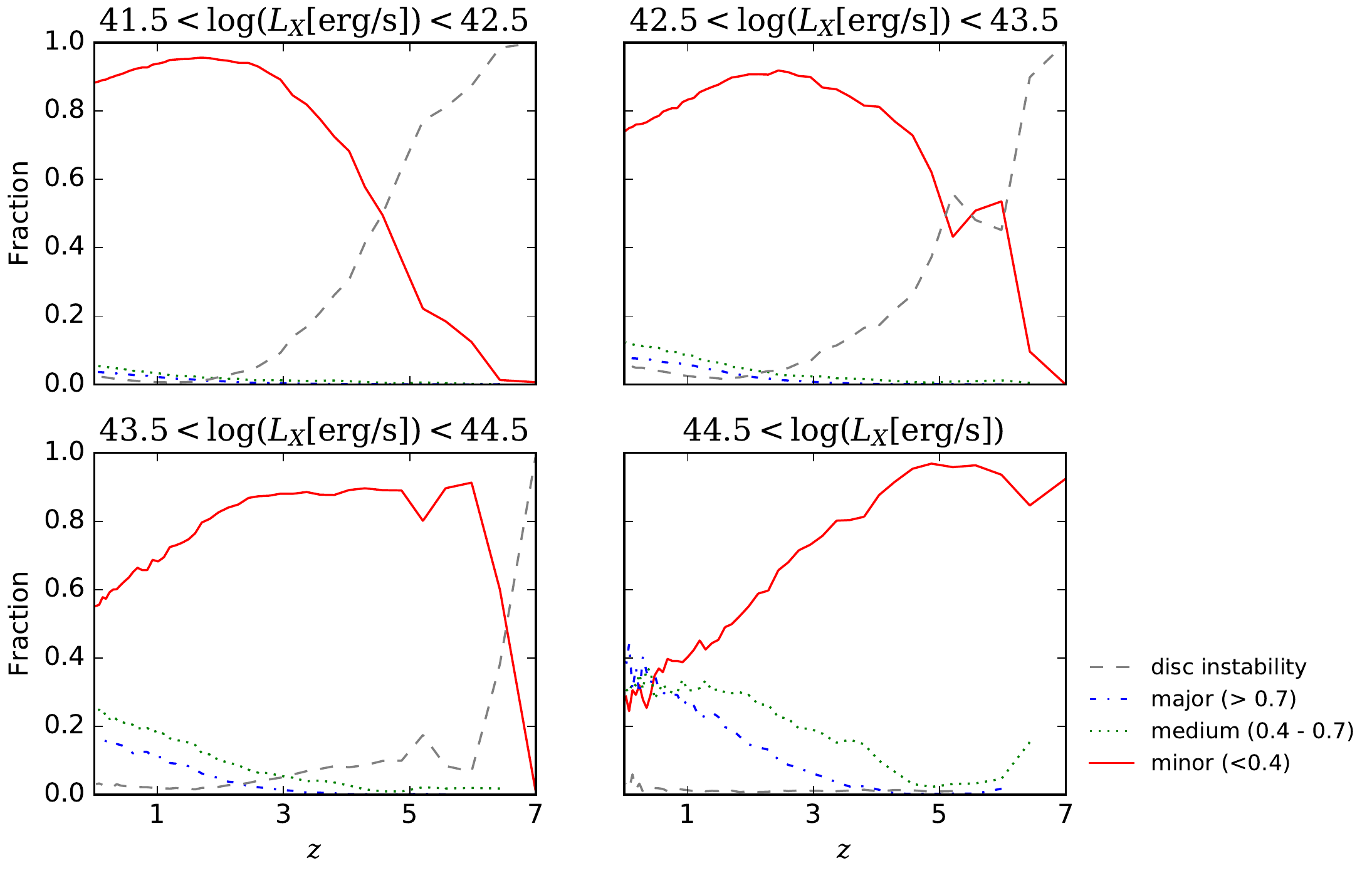}
        \end{center}
        \caption{Fraction of the AGN host galaxies whose AGN activity is triggered by mergers of galaxies or disc instabilities.
            We pick out AGNs (in \nugc-M box) with $\log(L_X / \mathrm{erg s^{-1}}) = $ [41.5, 42.5], [42.5, 43.5], [43.5, 44.5], and $> 44.5$.
            Mergers are classified according to the mass ratio of merging galaxies: $> 0.70$ (major, blue dash dotted), between 0.4 and 0.7 (middle, grean dotted)
            and $< 0.4$ (minor,red solid). We also show the fraction of AGNs triggered only by the disc instability (grey dashed).}
        \label{fig:MergerFrac}
      \end{figure}

      Fig.~\ref{fig:MergerFrac} shows the fraction of AGN host galaxies at $0.0 < z < 7.0$ in each luminosity bin,
      divided by triggering situations.
      We classify the galaxies by the mass ratio of the merging galaxies; major (mass ratio $> 0.7 = f_\mathrm{major}$; blue dash dotted line),
      intermediate ($0.4-0.7$; green dotted line), and minor (< 0.4; red solid line).
      The grey dashed line shows the fraction of AGNs triggered only by a disc instability.
      For merger-driven AGN activities, the typical merging mass ratio becomes larger for more luminous AGNs.
      Interestingly, we find that the primary trigger of AGNs at $z < 4.0$ is mergers of galaxies,
      although, at higher redshift, disc instabilities become essential for less luminous AGNs.
      This result is inconsistent with \cite{Fanidakis12} and \cite{Griffin18}, who suggest that
      disc instabilities and ``hot halo mode accretion'' are dominant triggering mechanisms of AGNs even at $z < 4.0$.
      As we described in Sec. \ref{sec:MergerandDI}, we employ the smaller $\epsilon_\mathrm{DI,crit}$
      for reproducing the properties of star formation galaxies at $z > 4$.
      Also, we consider the effect of the bulge potential on
      the stability of galactic discs. With this effect, the number of disc-unstable galaxies becomes
      60 \% smaller at $z \sim 1$ with $\epsilon_\mathrm{DI,crit} = 0.75$.
      These are why our model suggests such low efficiency of disc instabilities as a triggering mechanism.
      The critical point is that \textit{the observed number density of AGNs can be sufficiently reproduced
      at $z < 4$ only by mergers of galaxies, and the importance of disc instabilities and other processes should be
      investigated in more detail.}  Our model predicts disc instabilities drive only
      less than 20\% of AGNs at $z \sim 0$. We will come back this topic in Sec. \ref{sec:discussion}.

  \section{Discussion and conclusions}
  \label{sec:discussion}
      We have presented the latest results of an updated version of an SA model, $\nu^2 GC$.
      The most important changes are related to the bulge and SMBH growth model.
      We assume that the gas accretion onto the SMBH and the bulge growth are triggered by mergers of galaxies and disc instabilities.
      For bulge and SMBH growths by mergers of galaxies, we employ a phenomenological model proposed by \cite{Hopkins09F},
      whose model is based on results of hydrodynamic simulations.
      Along with this revision, we have also updated the way of calculating the velocity dispersion and size of bulges
      when bulges grow via minor mergers.
      For bulge and SMBH growths by disc instabilities, we employ a classical model originally proposed by \cite{ELN82}.
      We consider the effect of the bulge potential on the gravitational stability of the disc.

      We have investigated the effect of the accretion timescale on statistical properties of AGNs,
      such as their luminosity functions.
      We stress that the impact of the accretion timescale
      especially for low luminosity ($L_X < 10^{44}$ erg/s) AGNs
      has been almost neglected in previous SA models.
      When we assume that the accretion timescale is proportional to the dynamical time of the
      host halo or the host bulge, as in the previous SA models,
      the number density of the low luminosity AGNs is one order of magnitude smaller than
      observational estimates.
      We have found that the number density of such less luminous AGNs becomes consistent with the observational data when we take
      a phenomenological and physically-motivated model for the timescale of the angular momentum loss in the circumnuclear torus and/or the accretion disc
      into account.
      The \texttt{GalADmodel} predicts that low luminosity AGNs at $z < 1.0$, such as local Seyfert-like AGNs,
      are mainly triggered by minor mergers. The contribution of disc instabilities is only less than 20 \%.

      Previous studies with SA models solve the inconsistent number density of less luminous AGNs
      by considering other AGN triggering mechanisms
      such as ``efficient'' disc instabilities \citep[e.g.][]{Hirschmann12},
      fly-by interactions of galaxies \citep[e.g.][]{Menci14}, and
      the direct gas accretion from the hot halo \citep[e.g.][``hot halo mode'']{Fanidakis12}.
      \cite{Hirschmann12} suggest the importance of disc instabilities as a triggering mechanism of less luminous AGNs.
      We, however, have to note that the phenomenological modelling of disc instabilities in SA models
      would be too simple and is not supported by numerical simulations \citep[see][]{Athanassoula08}.
      We have tried to make more physically reasonable modelling of disc instabilities in this paper.
      For the first step, we include the stabilising effect by the bulge component and take smaller $\epsilon_\mathrm{DI}$ (Sec. \ref{DI}).
      We then find that disc instabilities are not the main contributor to AGN triggering mechanisms.
      As another point, some SA models \citep[e.g.][]{Fanidakis12,Griffin18} assume that
      a disc instability destroys a galactic disc entirely and
      all the gas is exhausted by a starburst forming a spheroidal galaxy just as major mergers.
      By these two effects (ignoring bulge potential and the complete destruction of a disc), some SA models
      are likely to overproduce the number density of AGNs induced by disc instabilities.
      Further updates are necessary, and we leave it for future studies.
      \cite{Menci14} suggest that fly-by interactions are important instead of disc instabilities.
      Although we do not introduce fly-by interactions, the random collision of galaxies may have similar effects.
      The ``hot halo mode'' \citep{Fanidakis12, Griffin18} is the same as our ``radio mode'' AGN feedback model,
      both of which are based on \cite{Bower06}. In our fiducial models, we do not calculate the AGN luminosity with this mode
      because the bolometric correction and the radiative efficiency are unclear.
      When we assume the same bolometric correction as that of QSOs, and the radiative efficiency is 0.1,
      the contribution of the radio mode AGN to the AGN LFs becomes the same order as that of AGNs induced by mergers of galaxies and disc instabilities
      at $L_X \sim 10^{41}$ erg/s at $z \sim 0$.
      The contribution becomes smaller at more luminous regime and at higher redshift.
      Our results based on the timescales show that observed AGN LFs
      can be reproduced without ``radio mode'' or ``hot halo mode'' accretions.
      Even without the ``radio mode'' AGN feedback, \texttt{GalADmodel} produces
      a large number of AGNs with low Eddington ratios,
      which would be AGN jet and outflow sources.
      Considering the injected energy and momentum from the
      low Eddington ratio AGNs, they may have non-negligible impact on the star formation quenching
      of massive galaxies.
      We will examine which explanation is more plausible in a future study.

      \cite{Marulli08} suggest the importance of AGN light curve
      for determining the shape of AGN LFs.
      They assume three types of the Eddington ratio evolution models
      based on observations and hydrodynamical simulations.
      The faint end slope of AGN LFs at $z < 1.0$ are well fitted when they assume the constant Eddington ratio, namely,
      $= 0.3[(1+z)/4]^{1.4}$ at $z < 3$, and $= 1$ at $z > 3$.
      By using this Eddington ratio, the accretion timescale should be $\sim 0.17$ Gyr at $z \sim 0$,
      which is larger than the dynamical time of bulges (Fig. \ref{fig:tgal-tconst}) and is qualitatively consistent with our suggestion.
      However, the model with this assumption of the constant Eddington ratio underestimates the number density of luminous AGNs at $z > 1$.
      They also introduce introduce AGN light curve with two stages; rapid, Eddington-limited growth phase, and
      longer quiescent phase with lower Eddington ratios.
      By using this light curve, the accretion timescale should be longer when the SMBH mass is smaller or the accreted gas mass is larger,
      which is the opposite to that suggested in the \texttt{GalADmodel}.
      The resulting faint end slope of AGN LFs at $z < 1$ is shallower than observations.
      They cannot explain the shape of the AGN LFs by changing just the Eddington ratio distribution.
      Finally they introduce SMBH mass dependency to the $f_\mathrm{BH}$ and successfully reproduce AGN LFs at $z < 5$.

      Hydrodynamic simulations \citep[e.g.][]{Sijacki15,Khandai15,Hirschmann14} do explain AGN LFs well,
        assuming Bondi-Hoyle-Littleton (BHL) accretion for all SMBH growths.
      Generally, hydrodynamic simulations assume that the ``effective'' accretion rate
      onto SMBHs is roughly 200 times larger than the BHL accretion rate,
      which is too small compared to that of observed AGNs
      \citep[e.g.][]{Ho09}. The assumption of the accretion rate with $\sim 200$ times larger than
      the BHL accretion, independent of any properties of galaxies and SMBHs, might be a too simplified assumption.
      Besides, we must care about another uncertainty; different AGN feedback models
      are employed in different cosmological simulations, which reproduce AGN LFs at the same extent.

      As we have shown, there are several prescriptions to explain the faint end slopes of AGNLFs at $z < 1$.
      For discriminating the models, comparisons of model results with observed properties of AGNs and their host galaxies
      are necessary.
      We have shown the relation between $M_\mathrm{BH}$ and $L_X$ (Figs. \ref{fig:tloss_prop_peak} and \ref{fig:tloss_prop_out},)
      the Eddington ratio distribution function at $z \sim 0.3$ (Fig. \ref{fig:ERDF_z03}),
      and the fraction of AGNs with different triggering mechanisms (Fig. \ref{fig:MergerFrac}).
      Since the difference between the \texttt{Galmodel} and \texttt{GalADmodel}
      is clear for low luminosity AGNs with the smaller SMBH masses,
      the comparisons with observations are challenging.
      The other possible way would be comparing the clustering properties with observations.
      \cite{Fanidakis13} suggest that the host halo mass of luminous AGNs like QSOs and low luminosity ones
      is different. In their model, luminous AGNs are triggered by starbursts induced by mainly disc instabilities (and mergers of galaxies)
      and their typical host halo mass is $\sim 10^{12} M_\odot$. Low luminosity AGNs, on the other hand, are triggered mainly
      ``hot halo mode'' and their halo mass is larger than those of luminous AGNs, namely $\sim 10^{13} M_\odot$.
      The ``hot halo mode'' is efficient for cluster galaxies whose host halo is cooling inefficient.
      On the other hand, \cite{Oogi16} suggest that when they assume AGNs are mainly triggered by mergers of galaxies,
      the host halo mass weakly depends on the AGN luminosities at $1 < z < 4$.
      The \texttt{GalADmodel} also shows the same trend as \cite{Oogi16} at $1 < z < 4$.
      We, thus, can discriminate effects of the accretion timescale and
      AGN triggering mechanisms by detailed comparisons with observational results.

\begin{figure}
        \begin{center}
          \includegraphics[width=\hsize]{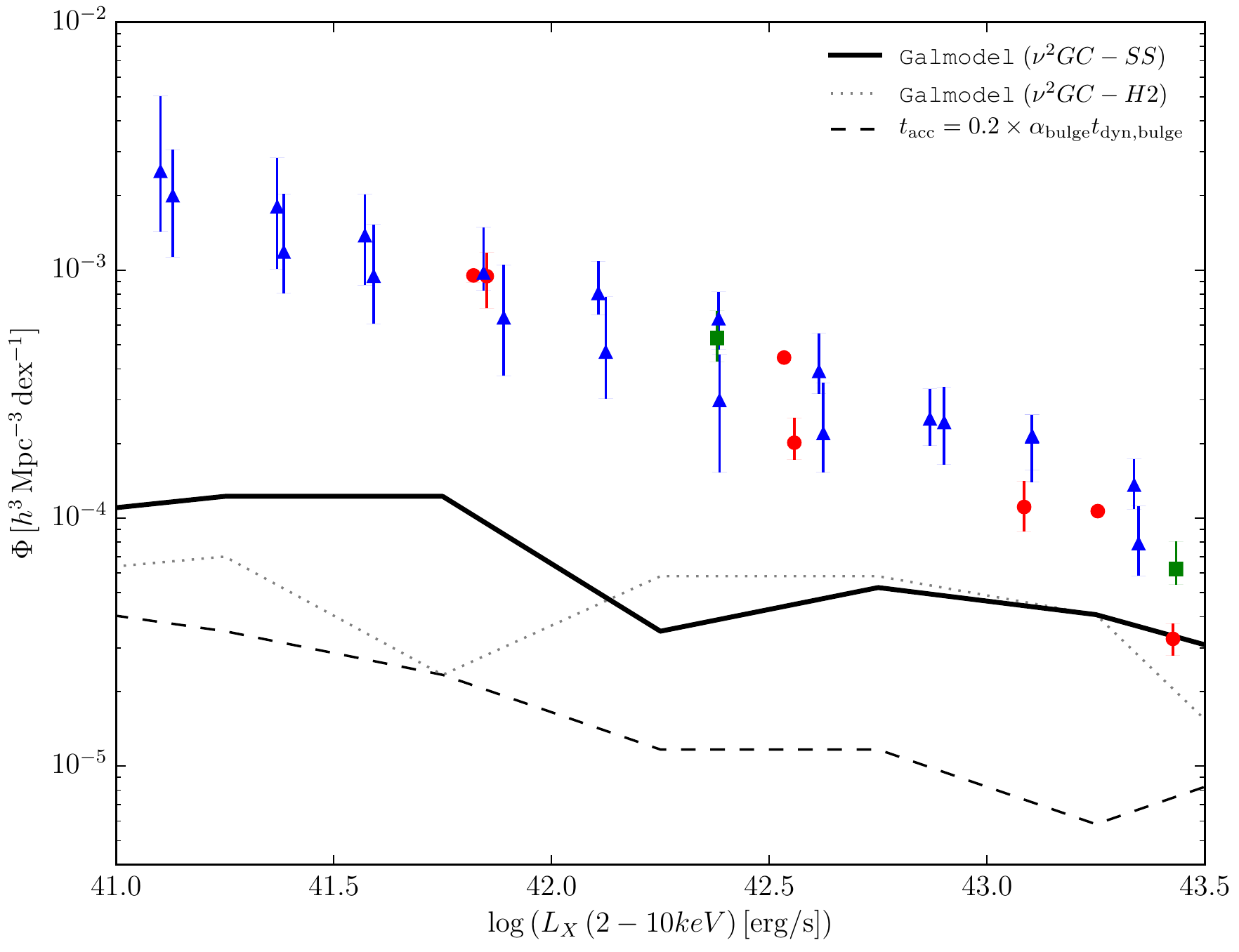}
        \end{center}
        \caption{AGNLFs at $z~\sim~0$. To check the effect of the determination of $t_\mathrm{dyn,bulge}$ and the accreted gas mass,
          we compare three models: (1) the \texttt{Galmodel} with the \nugc-SS simulation(black solid line),
          (2) the \texttt{Galmodel} with the \nugc-H2 simulation (black dotted line), and
          (3) the model in which $t_\mathrm{acc} = 0.2\times \alpha_\mathrm{bulge} t_\mathrm{dyn,bulge}$ (black dashed line).
          Observational results is the same as the top left panel of Fig. ~\protect\ref{fig:AGNLFX}.}
        \label{fig:test}
      \end{figure}

      One might think that the underproduction of less luminous AGNs results from the underestimation of
      the velocity dispersion of the bulge and/or the underestimation of the cold gas mass in galaxies.
      As shown in Fig.~\ref{fig:FJ}, the velocity dispersion of the bulge tends to be smaller than those obtained from observations,
      although the bulge size is broadly consistent with the observational data at $z \sim 0$ (Fig. \ref{fig:bulgesize}).
      The dynamical time of the bulge evaluated in the fiducial model
      is statistically longer than the value estimated from the observed velocity dispersion and bulge size.
      We thus underestimate the gas accretion rate onto SMBHs since the peak accretion rate is proportional to $t_\mathrm{dyn,bulge}^{-1}$.
      In addition, low mass galaxies in the model seem to have smaller gas masses than observed galaxies (Fig.~\ref{fig:HIMF})
      due to the insufficient resolution,
      which could also cause the underestimation of the gas accretion rate.
      In Fig.~\ref{fig:test}, we check these effects and find that
      both are insufficient to compensate the underproduction of the less luminous AGNs.
      We compare hard $X$-ray LFs at $z~\sim~0$ obtained by the following three models:
      (1) the \texttt{Galmodel} with the \nugc-SS simulation (black solid line),
      (2) the \texttt{Galmodel} with the \nugc-H2 simulation (black dotted line), and
      (3) the model with $t_\mathrm{acc} = 0.2 \times \alpha_\mathrm{bulge} t_\mathrm{dyn,bulge}$
      (black dashed line).
      The number density of AGNs obtained by the model (3) becomes smaller than that obtained by the model (1)
      since $t_\mathrm{dyn,bulge}$ is set to be smaller, and the AGN activity shut off sooner.
      Also, we find no effect of the gas deficiency by comparing (1) and (2),
      while the number of galaxies with $M_{HI}< 10^8 M_\odot$ increases
      when we employ the \nugc-H2 simulation.
      The comparison (1) and (2), therefore, suggests the gas deficiency is not the cause of
      the underestimation of the abundance of the less luminous AGNs.
      Even at $z \sim 1$, the model (3) does not solve the inconsistency of the faint-end slope
      since the shorter accretion timescale causes the shallower slope.
      We have confirmed that the faint-end slope of the AGN LF at $z \sim 1$ also does not change with model (3).
      We conclude the underestimation of the gas mass of galaxies is not a primary cause of the  underestimation
      of the number density of faint AGNs.

\begin{figure}
        \begin{center}
          \includegraphics[width=\hsize]{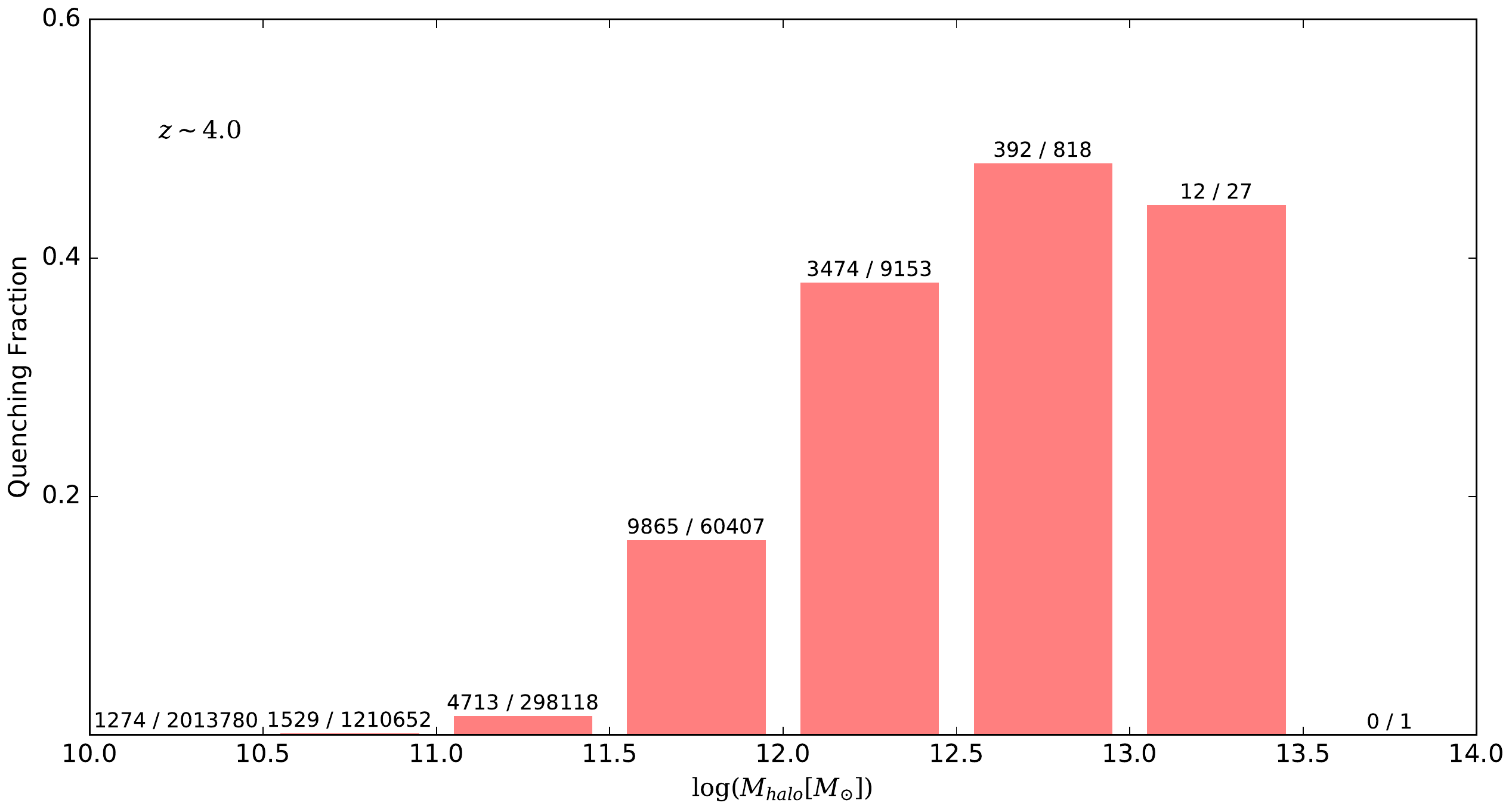}
        \end{center}
        \caption{The fraction of central galaxies whose gas cooling is shut off the by radio-mode AGN feedback at $z \sim 4$.
        The $x$ axis is the host halo mass of the galaxies. The number means $[Quenched halo] / [Total halo]$.}
        \label{fig:Qfrac}
      \end{figure}

      Another problem of the AGN LFs obtained with the $\nu^2$GC is that there are no AGNs with \LX~$> 45.3$
      at $z > 2.6$. Such luminous AGNs do not appear even when we employ $N$-body simulations with larger volumes.
      The modelling of the radio-mode AGN feedback is likely to be responsible for this,
      which was originally proposed by \cite{Bower06} and is similar to other SA models.
      Host halo masses of AGNs with \LX~$\sim 45.0$
      at $z \sim 4$ in the fiducial AGN model are $10^{12-13} M_\odot$.
      Such massive haloes could satisfy conditions of Eqs.~\ref{eq:AGNFB1} and
      ~\ref{eq:AGNFB2} and the gas cooling is quenched even at high redshifts.
      This is shown in Fig.~\ref{fig:Qfrac}, which shows the fraction of galaxies
      whose gas cooling is quenched by the radio-mode AGN feedback.
      We find that about the half of galaxies are quenched when $M_\mathrm{halo} > 10^{12.5} M_\odot$
      at $z \sim 4$.
      We will address this problem in future studies.

  \section*{Acknowledgements}
    We appreciate the detailed review and useful suggestions
    by the anonymous referee, which have improved our paper.
    We would like to express the deepest gratitude to A. R. Pettitt
    for thorough English proofreading that drastically improves the paper.
    We thank J. Aird to give the fitting function of hard $X$-ray
    luminosity functions of AGNs.
    We appreciate the fruitful comments from the observational side
    by T. Izumi, M. Onoue, Y. Ueda, D. Zhao, T. Nagao, Y. Matsuoka, Y. Kimura and M. Akiyama.
    We also thank K. Wada for theoretical comments.
    H.~Shirakata has been supported by the Sasakawa Scientific Research Grant from
    The Japan Science Society (29-214) and JSPS KAKENHI (18J12081).
    T.~Okamoto has been financially supported by MEXT KAKENHI (16H01085).
    T.~Kawaguchi was supported in part by an University Research
    Support Grant from the NAOJ and JSPS KAKENHI (17K05389).
    M.~Nagashima has been supported by the Grant-in-Aid (25287041 and 17H02867)
    from the MEXT of Japan.
    T.~Ishiyama and T.~Oogi has been supported by MEXT as ``Priority Issue on Post-K computer''
    (Elucidation of the Fundamental Laws and Evolution of the Universe)
    JICFuS.
    T.~Ishiyama has been supported JSPS KAKENHI Grant Number 15K12031.
    R.~Makiya was supported in part by MEXT KAKENHI (15H05896).
    T.~Oogi was supported by World Premier International Research Center Initiative (WPI).
    K.~Okoshi has been supported by JSPS KAKENHI (16K05299).




  \bibliographystyle{mnras}
  \bibliography{Astrophysics} 




  \appendix

  \section{Galaxy modellings}
  \label{App:Model}
    \subsection{Gas cooling}
    \label{GasCooling}
      Here we describe the calculation of the amount of the cold gas,
      which is accreted onto a central galaxy.
      In the model, we define a central galaxy of a new common halo as the
      central galaxy of the most massive progenitor halo.

      The mass fraction of the baryonic matter in a DM halo has been calculated
      with the following procedures, identical to that of \citetalias{nu2gc}.
      Before reionization of the universe, the mass fraction is given as
      $\langle f_b \rangle \equiv \Omega_\mathrm{b} / \Omega_0$.
      The mass fraction, however, deviates from $\langle f_b \rangle$
      after cosmic reionization because of the photoionization heating
      due to the UV radiation from galaxies and quasars.
      Small haloes with shallow gravitational potential wells
      cannot hold the gas heated by photoionization.
      We treat this effect following \cite{Okamoto08N}
      who performed high-resolution cosmological hydrodynamical simulations
      with a time-dependent UV background radiation field.
      They proposed the fitting formulae of the mass fraction of the baryonic matter
      as a function of the halo mass, $M_h$, and redshift, $z$,
      which was originally proposed by \cite{Gnedin00}:
      \begin{equation}
        f_\mathrm{b} (M_h, z) = \langle f_\mathrm{b} \rangle \times \left\{ 1+(2^{\alpha_\mathrm{UV}/3}-1)\left[\frac{M_h}{M_c (z)}\right]^{-\alpha_\mathrm{UV}}\right\}^{-3/\alpha_\mathrm{UV}},
      \end{equation}
      where $\alpha_\mathrm{UV}~=~2$ controls the rate of decrease of $f_\mathrm{b}$
      in low mass haloes.
      The characteristic mass as a function of $z$, $M_c (z)$, is described by using the fitting formula
      to the simulation results of \cite{Okamoto08N}:
      \begin{equation}
        M_c (z) = 6.5\times10^9 \exp(-0.604z) \exp[-(z/8.37)^{17.6}] h^{-1} M_\odot.
      \end{equation}
      We assume reionization occurs at $z~=~9.0$.
      See Sec. 2.3 of \citetalias{nu2gc} for a more in-depth description.

      All baryonic matter in a halo is diffuse hot gas soon after halo formation.
      To calculate the cold gas mass,
      we firstly calculate cooling radius, $r_\mathrm{cool} (t)$.
      We assume the Navarro-Frenk-White (NFW) density profile
      \citep{NFW97} for DM haloes and
      the isothermal density profile with a finite core radius, $r_c$,
      for hot gas haloes;
      \begin{align}
        &\rho_\mathrm{NFW} (r) = \frac{\rho_\mathrm{DM,0}}{(r/r_\mathrm{s})(1 + r/r_\mathrm{s})^2}, \label{eq:NFW}\\[5mm]
        &\rho_\mathrm{hot} (r) = \frac{\rho_\mathrm{hot,0}}{1 + (r/r_\mathrm{c})^2}, \label{eq:hotdens}
      \end{align}
      where $r_\mathrm{s}$ is the scale radius of the DM halo,
      which is described by using the concentration parameter, $c$,
      and virial radius, $R_\mathrm{vir}$,
      as $R_\mathrm{vir}/r_s~\equiv~c$.
      We assume $r_\mathrm{c}~=~0.22~r_\mathrm{s}$ \citep{MSS98},
      and use an analytical formulation of $c$ obtained by
      fitting to the results of cosmological $N$-body simulations \citep{Prada12}.
      After the collapse of a DM halo, the hot gas gradually
      cools via radiative cooling.
      The cooling time at a radius, $r$, is defined as
      \begin{equation}
      \label{eq:tcool}
      t_\mathrm{cool}(r) = \frac{3}{2} \frac{\rho_\mathrm{hot} (r)}{\mu m_\mathrm{p}} \frac{k_\mathrm{B} T_\mathrm{vir}}{n^2_\mathrm{e} (r) \Lambda(T_\mathrm{vir}, Z_\mathrm{hot})},
      \end{equation}
      where $\mu, m_\mathrm{p}, k_\mathrm{B},$ and $ n_\mathrm{e}$ are the mean molecular weight, proton mass,
      Boltzmann constant, and electron number density, respectively.
      We employ a cooling function, $\Lambda$,
      provided by \cite{SD93}, which is a function of
      hot gas metallicity, $Z_\mathrm{hot}$, and virial temperature, $T_\mathrm{vir}$.
      Virial temperature is calculated from the circular velocity of the host DM halo,
      $V_\mathrm{circ}$, as
      \begin{equation}
        T_\mathrm{vir} = \frac{1}{2} \frac{\mu m_\mathrm{p}}{k_\mathrm{B}} V^2_\mathrm{circ}.
      \end{equation}
      The cooling radius, $r_\mathrm{cool} (t)$, is defined as the radius at which $t_\mathrm{cool}$ (Eq.~\ref{eq:tcool})
      is equal to the time elapsed since the halo formation epoch.
      We can calculate the mass which cools in a given time step from Eqs.~\ref{eq:hotdens} and~\ref{eq:tcool}.

      We evaluate the accretion radius, $r_\mathrm{acc}(t)$,
      in which gas can cool and be accreted onto the central galaxy.
      We set $r_\mathrm{acc}$ as $MIN\{r_\mathrm{cool}, r_\mathrm{ff}(t_\mathrm{ff} = t_\mathrm{cool}),
      R_\mathrm{vir}\}$, similar to \cite{Lacey16}.
      Free-fall time, $t_\mathrm{ff}$, and free-fall radius, $r_\mathrm{ff}$, have the following relationship:
      \begin{equation}
        t_\mathrm{ff}(r_\mathrm{ff}) = \frac{\pi}{2}\sqrt{\frac{r_\mathrm{ff}^3}{2GM(r < r_\mathrm{ff})}},
      \end{equation}
      where $G$ is the gravitational constant and
      $M(r < r_\mathrm{ff})$ is obtained by the volume integration of Eq.~\ref{eq:NFW} from $r = 0$ to $r~=~r_\mathrm{ff}$.

      We note that we assume the existence of a ``cooling hole''
      in the same way as \citetalias{nu2gc}.
      Since we assume that the radial profile of the remaining hot gas is unchanged
      until the DM halo mass doubles,
      there is no hot gas at $r < r_\mathrm{cool}$ once the gas cools and is accreted onto the central galaxy.

    \subsection{Star formation}
    \label{SF}
      Our model includes star formation in cold gas discs and
      reheating of the gas by SNe.
      The implementation is similar to that of \citetalias{nu2gc}.

      When the diffuse hot gas cools, it forms a cold gas disc and triggers star formation.
      The SFR, $\Psi$, is given by $\Psi=M_\mathrm{cold}/\tau_\mathrm{star}$,
      where $M_\mathrm{cold}$ is the cold gas mass in a disc and $\tau_\mathrm{star}$  is the
      star formation timescale.
      We assume that $\tau_\mathrm{star}$ can be described with the dynamical timescale of the disc,
      $\tau_\mathrm{d}~=~r_\mathrm{d}/V_\mathrm{d}$ (where $r_\mathrm{d}$ and $V_\mathrm{d}$ are
      the half-mass radius and the circular velocity of the disc, respectively):
      \begin{equation}
        \label{eq:SF}
        \tau_\mathrm{star} = \epsilon_\mathrm{star}^{-1}\tau_\mathrm{d} \left[1 + \left(\frac{V_\mathrm{d}}{V_\mathrm{star}}\right)^{\alpha_\mathrm{star}}\right],
      \end{equation}
      where $\epsilon_\mathrm{star}, V_\mathrm{star}$, and $\alpha_\mathrm{star}$ are free parameters,
      \footnote{In \citetalias{nu2gc}, $V_\mathrm{star}$ is assumed to be identical to $V_\mathrm{hot}$,
      defined in Eq.~\ref{eq:SNFB}.}
      whose values are $0.46,197~\mathrm{km/s}$, and $-2.14$, respectively.
      The cold gas is reheated by SNe explosions
      at a rate of $M_\mathrm{cold} / \tau_\mathrm{reheat}$.
      The timescale for the reheating is given as follows:
      \begin{equation}
        \tau_\mathrm{reheat} = \frac{\tau_\mathrm{star}}{\beta (V_\mathrm{d})}, \label{eq:reheat}
      \end{equation}
      and
      \begin{equation}
        \beta (V_\mathrm{d}) = \left(\frac{V_\mathrm{d}}{V_\mathrm{hot}}\right)^{-\alpha_\mathrm{hot}}. \label{eq:SNFB}
      \end{equation}
      We calculate the chemical enrichment associated with the star formation and SNe explosions
      following \cite{Maeder92}.
      We assume instantaneous recycling for SNe \ii~and
      neglect any effects by SNe Ia.

      The gas reheated by SNe would not be available for gas cooling immediately.
      We do not severely differentiate the ejected and reheated gas by SNe.
      In our model, the gas with mass $\beta M_\mathrm{reheat}$ cannot cool immediately
      and is stored in a reservoir due to the reheating and ejection by SNe.
      A fraction of this gas might return to the hot gas halo and cool with some timescale.
      \cite{Lacey16} assume the returned gas mass as $\alpha_\mathrm{return}~M_\mathrm{ej}$
      \footnote{$M_\mathrm{ej}$ in \cite{Lacey16} is the same as $M_\mathrm{reheat}$
      in \nugc~although both are calculated with the same procedure.},
      where $\alpha_\mathrm{return}$ is a free parameter.
      We, however, simply assume that $\alpha_\mathrm{return} = 0$ and
      that all of the reheated gas falls back to the halo as hot gas
      when the halo mass doubles without escaping from the halo.
      If we set $\alpha_\mathrm{return} = 1.0$, the cosmic star formation density at $z < 1.0$
      becomes only $\sim 1.3$ times larger.

      We obtain the time evolution of the masses of stars, hot gas, BHs,
      cold gas, and metals in cold and hot gas for a given SFR, $\Psi (t)$, as follows:
      \begin{align}
        &\dot{M}_\mathrm{star} &=& \alpha         \Psi (t), \label{eq:dotstar}\\
        &\dot{M}_\mathrm{BH}   &=& f_\mathrm{BH} \Psi (t), \label{eq:dotBH}\\
        &\dot{M}_\mathrm{reheat}  &=& \beta          \Psi (t), \label{eq:dothot}\\
        &\dot{M}_\mathrm{cold} &=& - (\alpha + \beta + f_\mathrm{BH}) \Psi (t), \label{eq:dotcold}\\
        &\dot{(M_\mathrm{cold} Z_\mathrm{cold})} &=& [p - (\alpha + \beta + f_\mathrm{BH}) Z_\mathrm{cold}] \Psi (t), \label{eq:dotcoldZ}\\
        &\dot{(M_\mathrm{reheat} Z_\mathrm{hot})} &=& \beta Z_\mathrm{cold} \Psi (t), \label{eq:dothotZ}
      \end{align}
      where $M_\mathrm{star}, M_\mathrm{BH}$, and $M_\mathrm{reheat}$
      \footnote{$M_\mathrm{reheat}$ is given as $M_\mathrm{hot}$ in \citetalias{nu2gc}.}
      are the masses of stars, central BHs, and reheated gas mass by SNe in a galaxy, respectively, and
      $f_\mathrm{BH}$ is a free parameter tuned to match observational estimates of
      the relation between masses of bulges and SMBHs at $z \sim 0$.
      The metallicities of the cold and hot gas are denoted by $Z_\mathrm{cold}$ and $Z_\mathrm{hot}$, respectively.
      The value of the locked-up mass fraction, $\alpha$, and chemical yield, $p$, depend on
      the initial mass function (IMF). We adopt the Chabrier IMF \citep{Chabrier03}
      with which the corresponding values are $(\alpha, p)~=~(0.52, 1.68 Z_\odot)$.
      In this paper, we assume $Z_\odot = 0.019$.
      From Eq.~\ref{eq:dotstar} to~\ref{eq:dothotZ}, we analytically derive increments/decrements
      of the mass and metallicity of each component during a time step (see Eq.15 - 19 of \citetalias{nu2gc}).

    \subsection{Size of galaxies}
    \label{Size}
      Here we describe how to estimate galaxy size, the circular velocity of galactic discs,
      and the velocity dispersion of bulges.

      \subsubsection{Disc size and circular velocity}
        We assume that DM and hot gas haloes have the same specific angular momentum
        and that the angular momentum is conserved during the formation of a cold gas disc.
        We adopt the log-normal distribution for the dimensionless spin parameter,
        $\lambda_\mathrm{H}~\equiv~L\left|E\right|^{1/2}/GM^{5/2}$,
        where $L, E$, and $M$ are the angular momentum, binding energy, and DM halo mass, respectively,
        the same prescription as \citetalias{nu2gc}.
        The mean value of $\lambda_\mathrm{H}$ is $0.042$ and the logarithmic variance
        is $0.26$, which are obtained from $N$-body simulations of \cite{Bett07}.

        The effective radius of a cold gas disc, $R_\mathrm{d}$, is given by the following relation:
        \begin{equation}
          R_\mathrm{d} = (1.68/\sqrt{2})\lambda_\mathrm{H} R_\mathrm{init},
          \label{eq:discsize}
        \end{equation}
        where the initial radius of the hot gas sphere, $R_\mathrm{init}$, is set to
        the accretion radius, $r_\mathrm{acc}$, introduced in Sec. \ref{GasCooling}.
        Disc rotation velocity, $V_\mathrm{d}$, is given as the circular velocity of its host halo.
        In the model, $R_\mathrm{init}$ and $V_\mathrm{d}$ are renewed when the disc mass increases from the previous time step
        and when the new $R_\mathrm{init}$ is larger than the previous time step.

        We note that $R_\mathrm{d}$ becomes smaller than that at the previous time step when a merger of galaxies
        or disc instability occurs,
        by which the disc mass of the primary galaxy decreases.
        We then consider the conservation of the angular momentum and
        set the new effective radius, $R_{d,new}$, as $R_{d,new} = (M_{0d}/M_{1d}) \times R_\mathrm{d}$,
        where $M_{0d}$ and $M_{1d}$ are the disc mass (stellar $+$ cold gas) of the primary galaxy after and before
        the merger or disc instability, respectively.

      \subsubsection{Bulge size and velocity dispersion}
      \label{sec:bulgeprop}
        We describe how to estimate bulge size and velocity dispersion
        when a merger of galaxies or a disc instability occurs.
        There have been several previous studies \citep[e.g.][]{Hopkins09Feb, Covington11, Shankar13}
        which investigate how to calculate the size and velocity dispersion of the bulge
        from the Virial theorem and energy conservation.
        They, however, only study the major merger case.
        Applying their result to galaxies experiencing minor mergers or a disc instability,
        by which a galactic disc is not completely destroyed, is not straightforward.
        In this paper, we apply the similar formula to \citetalias{nu2gc}
        \footnote{\citetalias{nu2gc} assume that only major mergers are induced starbursts in bulges
        and a galactic disc is completely destroyed by a major merger
        while it does not change by a minor merger.}
        to obtain size and velocity dispersion of bulges formed  not only by major mergers
        but also by minor mergers and disc instability.

        We first consider merging galaxies.
        The total energy of each galaxy which contributes to the bulge formation is given by the Virial theorem:
        \begin{equation}
          E_i = -\frac{1}{2}[(M_{\mathrm{b},i} + M_{\mathrm{BH},i})V_{\mathrm{b},i}^2 + (M_{d,i} + M_{\mathrm{cold},i}) V_{\mathrm{d},i}^2],
          \label{eq:Etot}
        \end{equation}
        where $M_\mathrm{b}, M_\mathrm{d}$, and $M_\mathrm{cold}$ are the masses of the bulge stars, disc stars, and cold gas, respectively,
        and $V_\mathrm{b}$ and $V_\mathrm{d}$ denote the velocity dispersion of the bulge and the rotation velocity of the disc, respectively.
        The subscripts, $i~=~\{0,1,2\}$, indicate the merger remnant, the primary progenitor, and the secondary
        progenitor, respectively.

        We consider the effect of the gravitational potential of the DM halo
        which hosts the primary galaxy on the bulge dynamics.
        The method is similar but slightly different from \cite{Lacey16}.
        Assuming that a fraction of the DM halo mass, $M_{\mathrm{DM,1}}$, affects the bulge dynamics,
        we simply replace $M_\mathrm{b,1}$ to $M_\mathrm{b,1}~+~M_\mathrm{DM,1}$ in Eq.~\ref{eq:Etot}.
        The mass, $M_\mathrm{DM,1}$ is given by:
        \begin{equation}
          M_\mathrm{DM,1} = \frac{\Omega_\mathrm{0}}{\Omega_\mathrm{b}} \left(\frac{M_\mathrm{h}}{M_\mathrm{h0}}\right)^{\alpha_\mathrm{h}},
          \label{eq:EDM}
        \end{equation}
        where $M_\mathrm{h0}$ and $\alpha_\mathrm{h}$ are free parameters and the values are determined
        to reproduce the observed relation between the bulge size and $K$-band magnitude
        of galaxies at $z~\sim~0$.
        In this paper, the values of $M_\mathrm{h0}$ and $\alpha_\mathrm{h}$ are $10^{14} M_\odot$ and $1.82$, respectively.
        Since we do not utilize sub-halo merger trees, we ignore the effect of the DM potential
        for the secondary galaxies. We will update the model in the near future
        by including this effect.

        As described in Sec.~\ref{Merger}, a fraction of the disc mass in the primary galaxy,
        $\Delta M_\mathrm{1ds} + \Delta M_\mathrm{1dg}$,
        migrates to the bulge.
        The remaining energy in the disc, $E_\mathrm{0,d}$, is then:
        \begin{equation}
          E_\mathrm{0,d}~=~-\frac{1}{2}~\{M_\mathrm{d,1}~+~M_\mathrm{cold,1}~-(~\Delta~M_\mathrm{1ds}~+~\Delta M_\mathrm{1dg})\}V_\mathrm{d,1}^2.
        \label{eq:Edisc}
        \end{equation}
        The total energy of the bulge of the merger remnant, $E_\mathrm{0,b}$, can be described as follows:
        \begin{equation}
          E_\mathrm{0,b} = E_\mathrm{0} - E_\mathrm{0,d}.
          \label{eq:erem}
        \end{equation}

        Considering the energy dissipation, we obtain the energy conservation relation as follows:
        \begin{equation}
          f_\mathrm{diss}(E_\mathrm{1} + E_\mathrm{2} + E_\mathrm{orb}) = E_\mathrm{0,b},
        \end{equation}
        where $f_\mathrm{diss}$ is the fraction of energy dissipated from the merging system.
        We simply parameterize $f_\mathrm{diss}$ by following \citetalias{nu2gc}:
        \begin{equation}
          f_\mathrm{diss} = 1 + \kappa_\mathrm{diss} f_\mathrm{gas},
          \label{eq:dissipation}
        \end{equation}
        where
        \begin{equation}
          f_\mathrm{gas} = \frac{\Delta M_\mathrm{1g} + M_\mathrm{2g}}{M_\mathrm{1} + M_\mathrm{2}}.
          \label{eq:gasfrac}
        \end{equation}
        The orbital energy, $E_\mathrm{orb}$, is given as follows:
        \begin{equation}
          E_\mathrm{orb} = - \frac{E_\mathrm{1} E_\mathrm{2}}{(M_\mathrm{2}/(M_\mathrm{1}+M_\mathrm{DM,1}))E_\mathrm{1} + ((M_\mathrm{1}+M_\mathrm{DM,1})/M_\mathrm{2})E_\mathrm{2}},
        \end{equation}
        where $M_1$ and $M_2$ are the total mass of each galaxy (cold gas $+$ stars $+$ a BH).

        We calculate the velocity dispersion and the size of a bulge, $r_\mathrm{b}$, as
        \begin{align}
          &V_\mathrm{b,0}^2 = -\frac{2E_\mathrm{0,b}}{M_\mathrm{tot,0}}, \label{eq:sigma}\\
          &r_\mathrm{b,0} = \frac{GM_\mathrm{tot,0}}{2V_\mathrm{b,0}^2}, \label{eq:rb}
        \end{align}
        where $M_\mathrm{tot,0}$ is the total mass of the merger remnant
        (including $M_{\mathrm{DM},1}$).
        To obtain the 1D velocity dispersions, $\sigma_\mathrm{1D}$, we assume the bulge structure can
        be described by an isothermal sphere. The 1D velocity dispersion is simply given by
        $\sigma_\mathrm{1D}~=~V_\mathrm{b,0}~/~\sqrt{3}$.

        For the disc instability, we employ the same formulae as those for
        the merger of galaxies while subscripts, $i~=~\{1,2\}$, indicate the bulge and disc, respectively
        and the orbital energy, $E_\mathrm{orb}$, is set to be $0$.

      \subsubsection{Dynamical response caused by SNe feedback}
        We consider the change of the size and velocity
        caused by SN feedback.
        The SN feedback continuously expels gas from a galaxy.
        As a result, the gravitational potential well becomes shallower
        and the gravitationally bound system expands and its rotation speed slows down
        \citep{YA87}.
        We refer to this effect as \textit{dynamical response}, which
        is taken into account the same way as \citetalias{nu2gc}.
        This affects the size of galactic discs and bulges,
        the rotation velocity of galactic discs and their host haloes,
        and the velocity dispersion of galactic bulges.
        See Sec. 2.8 of \citetalias{nu2gc} for farther details.

      \subsection{Photometric properties and morphological identification}
      \label{LumMor}
        In order to compare our results with observations, we have to convert
        the mass of galaxies to observed luminosities.
        We employ a stellar population synthesis model of \cite{BC03} and obtain
        the spectral energy distribution (SED) of model galaxies.
        To estimate the extinction effect for galaxies,
        we make the same assumptions as \citetalias{nu2gc}; first, the dust-to-cold gas mass ratio is proportional to
        the metallicity of the cold gas; second, the dust optical depth is proportional to
        the dust column density.
        The dust optical depth, $\tau_\mathrm{dust}$, is then calculated from the following relation:
        \begin{equation}
          \tau_\mathrm{dust} = \tau_0 \left(\frac{M_\mathrm{cold}}{M_\odot}\right)\left(\frac{Z_\mathrm{cold}}{Z_\odot}\right)\left(\frac{R_\mathrm{e}}{\mathrm{kpc}}\right)^{-2},
        \end{equation}
        where $R_\mathrm{e}$ is the effective radius of the galaxy, and $\tau_0$ is a tunable parameter determined to reproduce the local galactic properties,
        such as LFs. We set $\tau_\mathrm{V0}~=~2.5~\times~10^{-9}$ following \cite{nugc}, which is the dust attenuation coefficient
        in $V$-band.
        We calculate the optical depth of the disc and bulge separately.
        The effective radius, $R_\mathrm{e}$ is $R_\mathrm{d}$ for the disc,
        and $R_\mathrm{b} = 0.744 r_\mathrm{b}$ for the bulge \citep{NY03}.
        We employ the Calzetti extinction law \citep{Calzetti00},
        and assume a slab model for the dust distribution in the disc and the bulge.

        The morphological types of model galaxies are determined in the same manner as \citetalias{nu2gc};
        using bulge-to-total ($B/T$) luminosity ratio in $B$-band, galaxies with $B/T~>~0.6$, $0.4~< B/T~< 0.6$, and $B/T~<~0.4$
        are classified as elliptical, lenticular, and spiral galaxies, respectively \citep{SV86}.

\section{General Results of Galaxies}
  \label{App:ModelResults}
    In this section, we present properties of galaxies obtained from the fiducial model
    and compare them with those obtained from observations.
    Firstly, we run the MCMC fitting with the \nugc-SS simulation to tune parameters.
    For the model calibration, we use observed $K-$ and $r-$ band LFs at $z \sim 0$ obtained from
    the Galaxy and Mass Assembly (GAMA) survey, $\mathrm{H_I}$ mass function at $z \sim 0$ extracted from
    the data of the Arecibo Legacy Fast ALFA (ALFALFA) survey,
    $M_\mathrm{BH}$ -- $M_\mathrm{bulge}$ relation at $z \sim 0$ \citep[Eq. 11][]{KH13_review},
    scaling relations of galactic discs and bulges at $z \sim 0$ \citep[][respectively]{Courteau07,Forbes08}
    cosmic SFR density obtained from observations (UV- and IR-bands, and radio 1.4 GHz),
    $K-$ band LFs at $z = 1, 2, 3$ obtained with the UKIDSS Deep Survey \citep{Cirasuolo10},
    and AGN hard $X-$ray LFs at $z = 0.4, 1, 2$ \citep{Ueda14May}.

    We summarised the fiducial values of our free parameters and related equations in Table~\ref{tab:params}.
    We run the calculation with 50000 realisations, excluding the initial 10000 steps of the
    ``burn-in'' phase (for more details, see Sec. 3.2 in \citealt{nu2gc}).
    The reduced $\chi^2$ decreases at 3.4 \% of the initial value after the first 10000 iterations,
    and at 1.5 \% after 20000 iterations.
    After 20000 iterations, $\chi^2$ becomes a little larger
    (2.2 \%/2.3 \% of the initial value after 40000/50000 iterations).
    The dispersion of values of MCMC-fitted parameters after 50000 iterations is
    1.69 / 1.29 times larger than that after 20000 / 40000 iterations.
    The averaged values of parameters, on the other hand, seems to be converged.
    The change of the averaged values of parameters is 4.7 \% from 20000 to 50000 iterations
    and 1.4 \% from 400000 to 50000 iterations.
    The increase of the iterations would thus cause the increase of the dispersion values.

    We have checked the correlations between values of two different parameters by using the Pearson's $r$
    (Table ~\ref{tab:Pearson}). The correlation is weak for most combinations of two parameters
    although some ($\alpha_\mathrm{star}$ -- $V_\mathrm{star}$, $\kappa_\mathrm{diss}$ -- $\epsilon_\mathrm{SMBH}$,
    $M_\mathrm{h0}$ -- $\alpha_\mathrm{bulge}$, $M_\mathrm{h0}$ -- $t_\mathrm{loss,0}$,
    $\alpha_\mathrm{bulge}$--$t_\mathrm{loss,0}$, and $\gamma_\mathrm{gas}$ -- $\gamma_\mathrm{BH}$)
    have strong correlations, $|r| \gtrsim 0.8$.

    The MCMC fitting has two crucial problems.
    First, since the \nugc-SS simulation has only $70^3 h^{-3} \mathrm{Mpc}^3$,
    we cannot fit the bright end slope of AGN LFs.
    The larger box simulations are not realistic considering the computational cost.
    Second, we have to fit parameter values so that all observational results are
    equally well reproduced. In other words, we cannot prioritise observational properties
    to fit.
    We, therefore, use the \nugc-SS simulation and refit some ill-fitted parameters by hand
    so that they are in $1 \sigma$ in the MCMC-fitted values.
    The parameters which are refitted by hand are shown in Table \ref{tab:params}.
    We cannot determine the values of $f_\mathrm{mrg}$, $\epsilon_\mathrm{DI,crit}$,
    $f_\mathrm{BH}$, $\gamma_\mathrm{gas}$, and $\gamma_\mathrm{BH}$ because of the degeneracy
    and the small box size.

    The main results of this paper on the statistical properties of SMBHs and AGNs
    appear in Sec~\ref{sec:AGN}.
    Additional properties of galaxies such as size/velocity -- magnitude relations of galactic discs,
    stellar mass -- SFR relations appear in Appendix. \ref{App:Galev}.

    \begin{table*}
      \begin{center}
        \begin{tabular}{llcccc}
          Galaxies:\\
          \hline
          parameter & related equation & value range & MCMC best & MCMC dispersion & adopted value \\
          \hline
          $\alpha_\mathrm{star}$          & Eq.~\ref{eq:SF}          & [-3.0,0.0]    & -2.14  & 0.10  & -2.14  \\
          $V_\mathrm{star}$ [km/s]        & Eq.~\ref{eq:SF}          & [100.0,400.0] & 211.30 & 14.37 & 197.00 \\
          $\epsilon_\mathrm{star}$        & Eq.~\ref{eq:SF}          & [0.05,0.50]   & 0.48   & 0.02  & 0.46   \\
          $V_\mathrm{hot}$ [km/s]         & Eq.~\ref{eq:SNFB}        & [50.0,400.0]  & 121.64 & 2.74  & 121.64 \\
          $\alpha_\mathrm{hot}$           & Eq.~\ref{eq:SNFB}        & [0.0,4.0]     & 3.92   & 0.07  & 3.92   \\
          $\alpha_\mathrm{return}$        & Sec.~\ref{SF}            &               &        &       & 0.00   \\
          $f_\mathrm{mrg}$                & Sec.~\ref{Merger}        & [0.8,1.0]     & 0.98   & 0.01  & 0.81   \\
          $f_\mathrm{major}$              & Sec.~\ref{Merger}        & [0.3,1.0]     & 0.89   & 0.08  & 0.89   \\
          $\kappa_\mathrm{diss}$          & Eq.~\ref{eq:dissipation} & [1.0,3.0]     & 2.70   & 0.20  & 2.75   \\
          $M_\mathrm{h0} [10^{14} M_\odot]$ & Eq.~\ref{eq:EDM}         & [0.1,10.0]    & 2.10   & 1.43  & 1.00   \\
          $\alpha_\mathrm{h}$             & Eq.~\ref{eq:EDM}         & [0.5,2.0]     & 1.82   & 0.13  & 1.82   \\
          $\epsilon_\mathrm{DI,crit}$     & Eq.~\ref{eq:DI}          & [0.7,1.1]     & 1.05   & 0.01  & 0.75   \\
          $f_\mathrm{bar}$                & Sec.~\ref{DI}            & [1e-3,1.0]    & 0.63   & 0.10  & 0.63   \\
          $\tau_\mathrm{V0}$              & Sec.~\ref{LumMor}        &               &        &       & $2.5 \times 10^{-9}$ \\
          \hline
          \\
          SMBHs and AGNs: \\
          \hline
          $\alpha_\mathrm{cool}$         & Eq.\ref{eq:AGNFB1} & [0.8,1.2]   & 1.14   & 0.04 &  1.14  \\
          $\log(\epsilon_\mathrm{SMBH})$ & Eq.\ref{eq:AGNFB2} & [-3.0,0.0]  & -2.66  & 0.53 & -2.66  \\
          $f_\mathrm{BH}$                & Eq.\ref{eq:dotBH}  & [1e-3,8e-2] & 0.06   & 0.01 & 0.02   \\
          $M_\mathrm{seed} [M_\odot]$    & Sec.\ref{seed}     &             &        &      & $10^3$ \\
          $\alpha_\mathrm{bulge}$        & Eq.\ref{eq:tacc}   & [0.1,1.2]   & 0.77   & 0.24 & 0.58   \\
          $\tau_\mathrm{loss,0}$ [Gyr]   & Eq.\ref{eq:tad}    & [0.1,5.0]   & 1.56   & 0.71 & 1.00   \\
          $\gamma_\mathrm{gas}$          & Eq.\ref{eq:tad}    & [-5.0,0.0]  & -3.28  & 0.41 & -4.0   \\
          $\gamma_\mathrm{BH}$           & Eq.\ref{eq:tad}    & [0.0,5.0]   & 4.40   & 0.42 &  3.5   \\
          $\dot{m}_\mathrm{crit}$        & Eq.\ref{eq:M2L}    &             &        &      &  10.0  \\
          \hline
        \end{tabular}
        \caption{Summary of free parameters in the fiducial model. Almost all parameters are fitted
                 with the MCMC method (iteration = 50000).
                 We show the (1) parameter name, (2) related equation or section,
                 (3-5) parameter range, best fit value, and dispersion (if MCMC fitted parameter),
                 and (6) adopted value.}
        \label{tab:params}
      \end{center}
    \end{table*}

    \begin{table*}
      \begin{center}
        \scalebox{0.8}[0.8]{
          \begin{tabular}{lrrrrrrrrrrrrrrrrrrr}
            \hline
            & $\gamma_\mathrm{BH}$ & $\gamma_\mathrm{gas}$ & $\alpha_\mathrm{bulge}$ & $\tau_\mathrm{loss,0}$ & $f_\mathrm{BH}$ & $f_\mathrm{bar}$ & $\epsilon_\mathrm{DI,crit}$ & $\log(\epsilon_\mathrm{SMBH})$ & $\alpha_\mathrm{cool}$ & $\alpha_\mathrm{h}$ & $M_\mathrm{h0}$ & $\kappa_\mathrm{diss}$ & $f_\mathrm{major}$ & $f_\mathrm{mrg}$ & $\alpha_\mathrm{return}$ & $\alpha_\mathrm{hot}$ & $V_\mathrm{hot}$ & $\epsilon_\mathrm{star}$ & $V_\mathrm{star}$ \\
            \hline
            $\alpha_\mathrm{star}$         & -0.25 &  0.19 &  0.39 &  0.34 &  0.08 & -0.16 &  0.06 &  0.31 & -0.30 &  0.19 & -0.34 & -0.08 &  0.21 &  0.48 & -0.17 &  0.05 & -0.30 &  0.05&   0.80 \\
            $V_\mathrm{star}$              & -0.27 &  0.23 &  0.54 &  0.42 &  0.20 & -0.13 & -0.25 &  0.39 & -0.37 &  0.12 & -0.46 & -0.21 &  0.28 &  0.51 & -0.12 &  0.20 & -0.02 &  0.44 \\
            $\epsilon_\mathrm{star}$       & -0.02 &  0.10 &  0.31 &  0.27 &  0.38 & -0.22 & -0.09 &  0.16 & -0.18 &  0.11 & -0.32 & -0.20 &  0.26 &  0.17 &  0.13 &  0.20 & -0.05 \\
            $V_\mathrm{hot}$               &  0.22 & -0.25 & -0.20 & -0.29 & -0.36 &  0.33 & -0.74 & -0.21 &  0.16 & -0.47 &  0.23 &  0.12 & -0.35 & -0.08 &  0.08 & -0.62 \\
            $\alpha_\mathrm{hot}$          & -0.24 &  0.25 &  0.38 &  0.36 &  0.48 & -0.31 &  0.29 &  0.26 & -0.24 &  0.39 & -0.33 & -0.24 &  0.42 &  0.15 & -0.07 \\
            $\alpha_\mathrm{return}$       & -0.06 &  0.07 & -0.17 & -0.26 & -0.25 &  0.07 & -0.09 & -0.03 &  0.39 & -0.03 &  0.35 &  0.14 & -0.45 & -0.19 \\
            $f_\mathrm{mrg}$               &  0.23 & -0.24 &  0.27 &  0.22 &  0.18 & -0.29 & -0.12 &  0.04 & -0.29 & -0.11 & -0.38 &  0.08 &  0.24 \\
            $f_\mathrm{major}$             & -0.38 &  0.41 &  0.62 &  0.69 &  0.75 & -0.47 &  0.18 &  0.23 & -0.33 &  0.57 & -0.71 & -0.20 \\
            $\kappa_\mathrm{diss}$         &  0.52 & -0.59 & -0.70 & -0.71 & -0.52 & -0.28 & -0.24 & -0.80 &  0.69 & -0.36 &  0.49 \\
            $M_\mathrm{h0}$                &  0.29 & -0.36 & -0.82 & -0.89 & -0.79 &  0.16 & -0.17 & -0.45 &  0.73 & -0.42 \\
            $\alpha_\mathrm{h}$            & -0.57 &  0.61 &  0.54 &  0.59 &  0.64 & -0.50 &  0.34 &  0.46 & -0.32 \\
            $\alpha_\mathrm{cool}$         &  0.13 & -0.19 & -0.74 & -0.70 & -0.67 &  0.02 & -0.19 & -0.60 \\
            $\log(\epsilon_\mathrm{SMBH})$ & -0.61 &  0.67 &  0.72 &  0.68 &  0.42 &  0.10 &  0.23 \\
            $\epsilon_\mathrm{DI,crit}$    & -0.18 &  0.21 &  0.12 &  0.26 &  0.27 &  0.07 \\
            $f_\mathrm{bar}$               & -0.02 &  0.02 & -0.14 & -0.11 & -0.36 \\
            $f_\mathrm{BH}$                & -0.26 &  0.38 &  0.74 &  0.79 \\
            $\alpha_\mathrm{bulge}$        & -0.59 &  0.66 &  0.91 \\
            $\tau_\mathrm{loss,0}$         & -0.57 &  0.61 \\
            $\gamma_\mathrm{gas}$          & -0.94 \\
            \hline
          \end{tabular}
        }
        \caption{List of the Pearson's $r$.}
        \label{tab:Pearson}
      \end{center}
    \end{table*}

    \subsection{Properties of galaxies at $z \sim 0$}
    \label{local}
\begin{figure*}
        \begin{center}
          \includegraphics[width=\hsize]{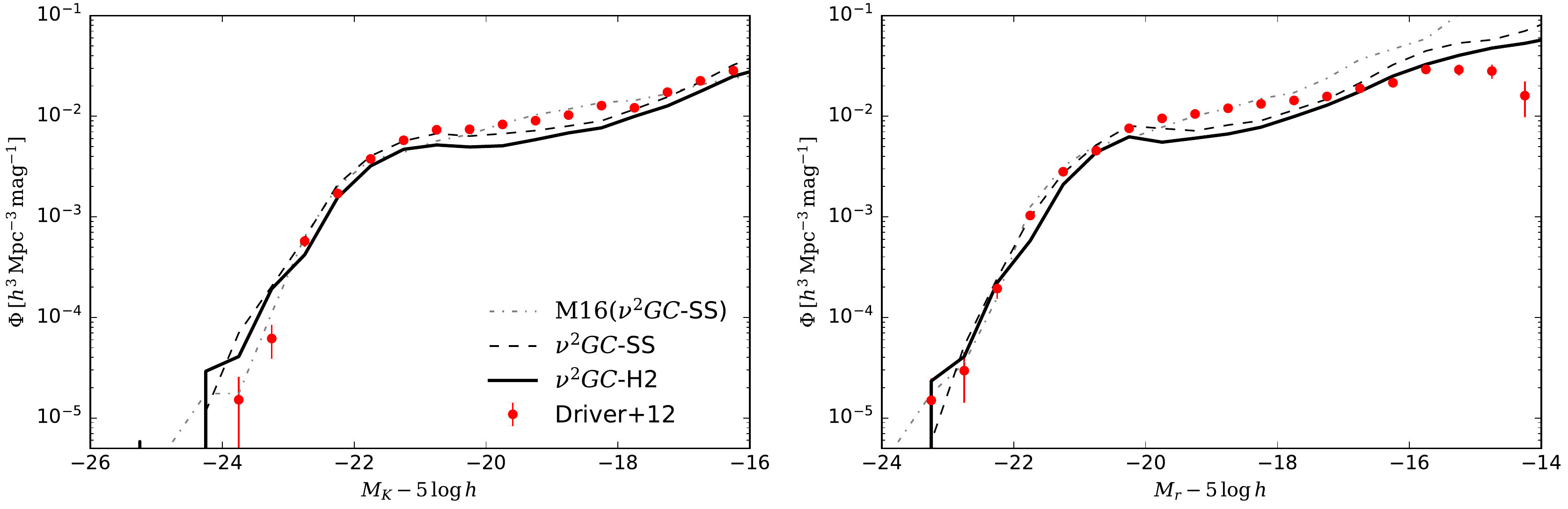}
        \end{center}
        \caption{$K-$ and $r-$ band LFs of galaxies.
                 Black dashed and solid lines show the results by the fiducial model
                 with \nugc~-SS and \nugc~-H2 simulations, respectively.
                 We show the result of \protect\citetalias{nu2gc} as grey dot-dashed lines.
                 Red filled circles with error bars are observational estimates by the GAMA survey \citep{Driver12}.}
        \label{fig:GLF-Kandr}
      \end{figure*}

\begin{figure}
        \begin{center}
          \includegraphics[width=\hsize]{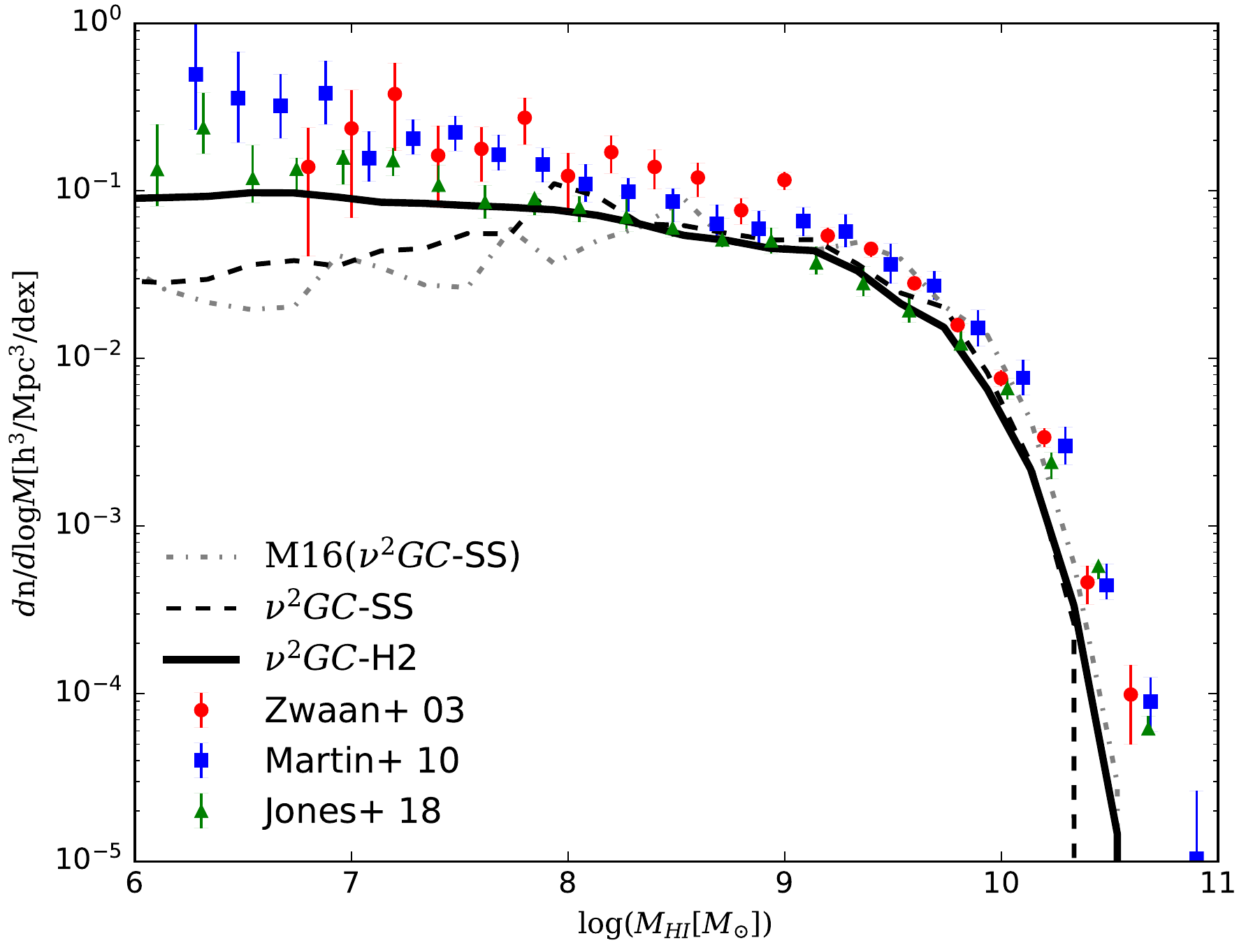}
        \end{center}
        \caption{$\mathrm{H_I}$ MF at $z \sim 0$. Black dashed and solid lines show the results obtained from the fiducial model
          with \nugc~-SS and \nugc~-H2 simulations, respectively.
                 We show the result of \protect\citetalias{nu2gc} as grey dot-dashed lines.
                 Red filled circles, blue filled squares, and green filled triangles with error bars are observational data obtained from the
               HIPASS \citep{Zwaan03} and ALFALFA surveys (\citealt{Martin10} and \citealt{Jones18}), respectively.}
        \label{fig:HIMF}
      \end{figure}

      Fig.~\ref{fig:GLF-Kandr} shows the $K$- and $r$- band LFs at $z \sim 0$.
      The results of the fiducial model with the \nugc-SS and -H2 simulations shown
      to test the resolution effect.
      We overplot the results obtained by \citetalias{nu2gc} in grey dash-doted lines.
      Red points with errorbars are the observational estimates by
      the GAMA survey \citep{Driver12}.
      Fig.~\ref{fig:HIMF} shows the $\mathrm{H_I}$ mass function (MF) at $z \sim 0$.
      We assume the relation between the cold gas mass and
      the atomic hydrogen gas mass, $M_\mathrm{HI}$,
      as $M_\mathrm{HI}~=~0.54 M_\mathrm{cold}$, which is the same relation
      used in \citetalias{nu2gc}.

      The bright-end slopes of the LFs and the massive-end slope of the $\mathrm{H_I}$ MF are sensitive to the values
      of $\alpha_\mathrm{cool}$ and $\epsilon_\mathrm{SMBH}$, which
      are both related to the radio-mode AGN feedback.
      The faint-end slopes are determined by
      the energy of the SN feedback determined by $\alpha_\mathrm{hot}$ and $V_\mathrm{hot}$.
      The low mass end slope of the $H_I$ MF is also sensitive to the values of
      $\alpha_\mathrm{star}$ and $V_\mathrm{star}$, which determine
      the gas consumption timescale by star formation.
      Although the model explains the wide range of the observed LFs and $\mathrm{H_I}$ MF at $z \sim 0$,
      the number of galaxies with smaller $\mathrm{H_I}$ gas mass ($M_\mathrm{HI} < 10^8 M_\odot$) is under-predicted,
      which is the same trend as \citetalias{nu2gc} and other SA models
      \citep[e.g.][]{Gonzalez-Perez14, Lagos14Sep, Lacey16}.
      This is partly due to the insufficient resolution of the employed $N-$ body simulation.
      As shown in Fig.~\ref{fig:HIMF}, the result with the \nugc-H2 simulation
      ($\sim 4^3$ times higher mass resolution than the \nugc-SS simulation)
      explains the $\mathrm{H_I}$ MF better than that with the \nugc-SS simulation
      and the result is nearly consistent with the recent observational estimates \citep[][green triangles]{Jones18}
      The modelling of the SFR might be important since the low mass end slope
      is sensitive to $\alpha_\mathrm{star}$ and $V_\mathrm{star}$.
      The modelling of the gas stripping and cooling of satellite galaxies
      should also be important.
      However, we do not use sub-halo merger trees in this work, and do not consider
      gas cooling for satellite galaxies.
      Since such less massive galaxies do not have an impact on the main results of this paper,
      we leave this issue for future work.

\begin{figure}
        \begin{center}
          \includegraphics[width=\hsize]{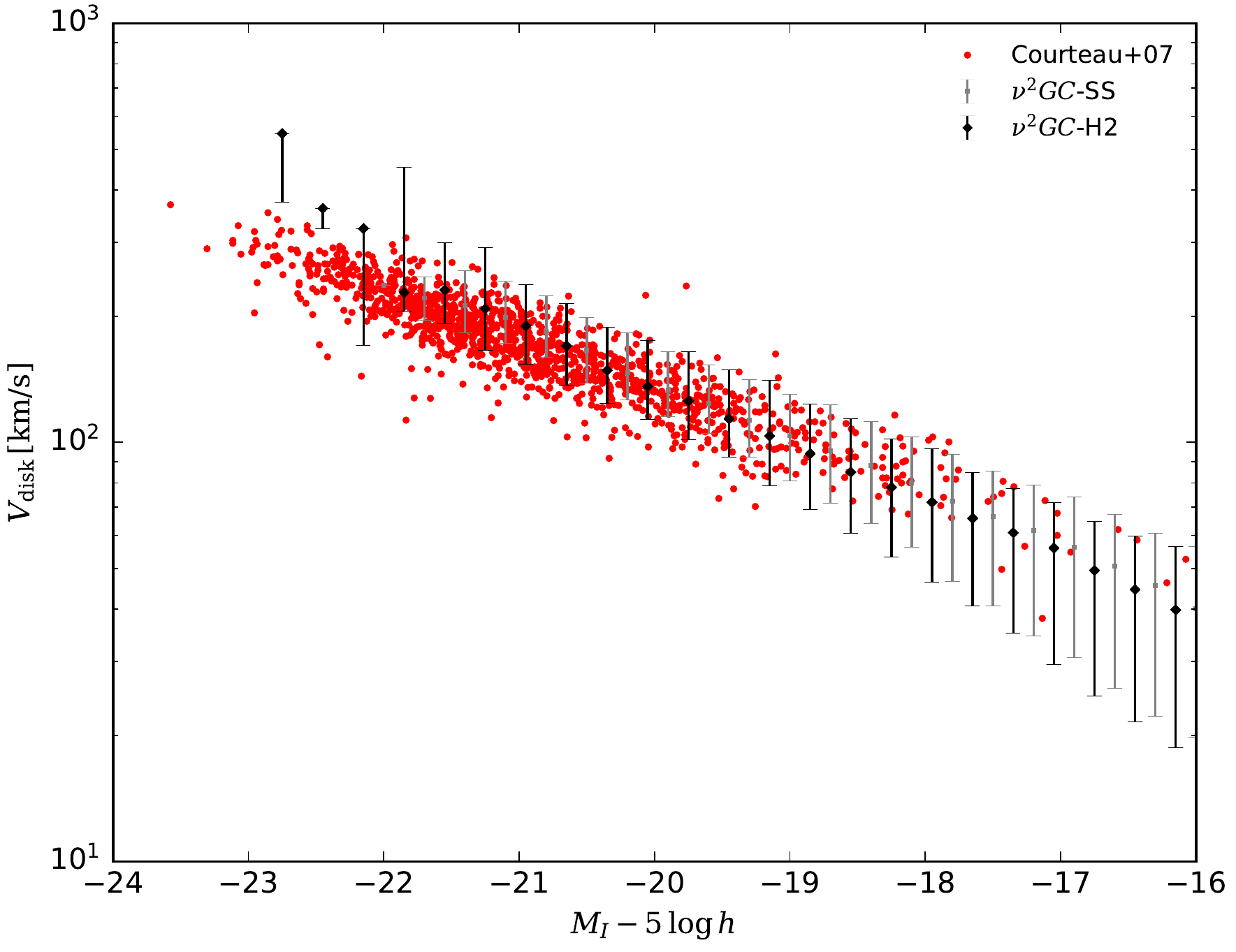}
        \end{center}
        \caption{Rotation velocities of spiral galaxies as a function of $I-$ band magnitude.
        (Tully-Fisher relation). The black line shows the median value obtained by the model
        and the error bars show the 10th and 90th percentiles
        from the \nugc-SS \nugc-H2 simulations.
        Red points show the observational
        data obtained from \protect\cite{Courteau07}.}
        \label{fig:TF}
      \end{figure}

\begin{figure}
        \begin{center}
          \includegraphics[width=\hsize]{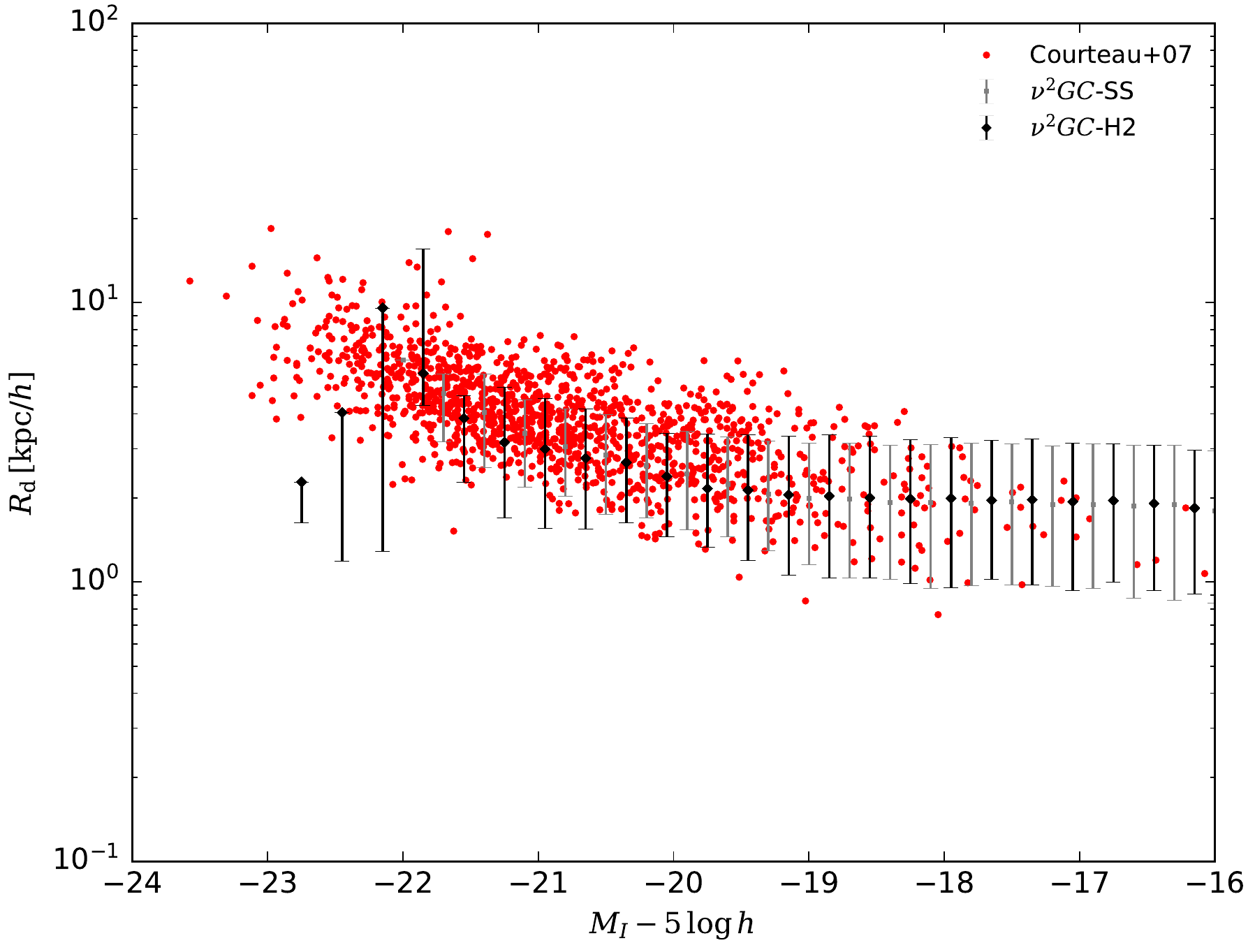}
        \end{center}
        \caption{Effective radius of spiral galaxies as a function of $I-$ band magnitude.
        The black line shows the median value obtained by the model
        and the error bars show the 10th and 90th percentiles.
        We employ the \nugc-SS \nugc-H2 simulations.
        Red points show the observational
        data obtained from \protect\cite{Courteau07}.
        We convert the scale length obtained by \protect\cite{Courteau07} to
        the effective radius with $R_\mathrm{d} = 1.68 r_\mathrm{ds}$.}
        \label{fig:discsize}
      \end{figure}

      We compare the predicted effective radius and rotation velocity of spiral galaxies
      at $z \sim 0$ with observations.
      We employ the \nugc-SS \nugc-H2 simulations to obtain the result.
      We use the data obtained from \cite{Courteau07} who estimated the disc scale lengths
      from $I-$ band image and the disc rotation velocities
      from $H_\alpha$ or $\mathrm{H_I}$ line width.
      Figs.~\ref{fig:TF} and~\ref{fig:discsize} are the scaling relations
      between the rotation velocity and the $I-$ band magnitude
      (the so-called Tully-Fisher relation;~\citealt{TF77})
      and the effective radius and the $I-$ band magnitude, respectively.
      The data obtained from \cite{Courteau07} are presented as red points.
      The results of their model are shown as black lines with error bars
      which are the 10th to 90th percentiles.
      The model results are consistent with the observational results
      and the effect of the mass resolution of the simulations is negligible.

\begin{figure}
          \begin{center}
            \includegraphics[width=\hsize]{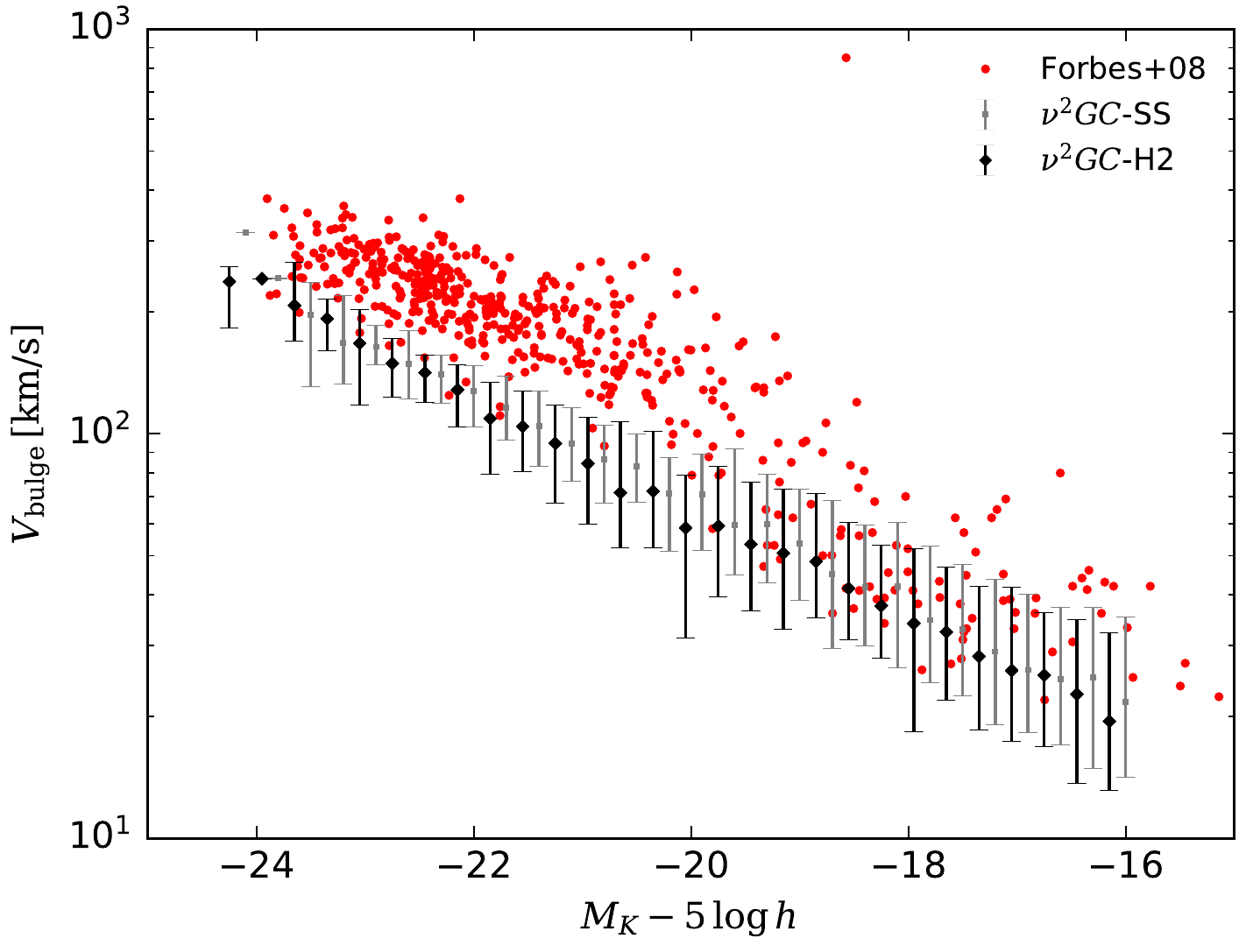}
          \end{center}
          \caption{Velocity dispersions of elliptical and S0 galaxies as a function of $K-$ band magnitude
          (Faber-Jackson relation). The black line shows the median value obtained by the model
          and the error bars show the 10th and 90th percentiles
          from the \nugc-SS and \nugc-H2 simulations.
          Red points show the observational
          data obtained from \protect\cite{Forbes08}.}
          \label{fig:FJ}
        \end{figure}

\begin{figure}
          \begin{center}
            \includegraphics[width=\hsize]{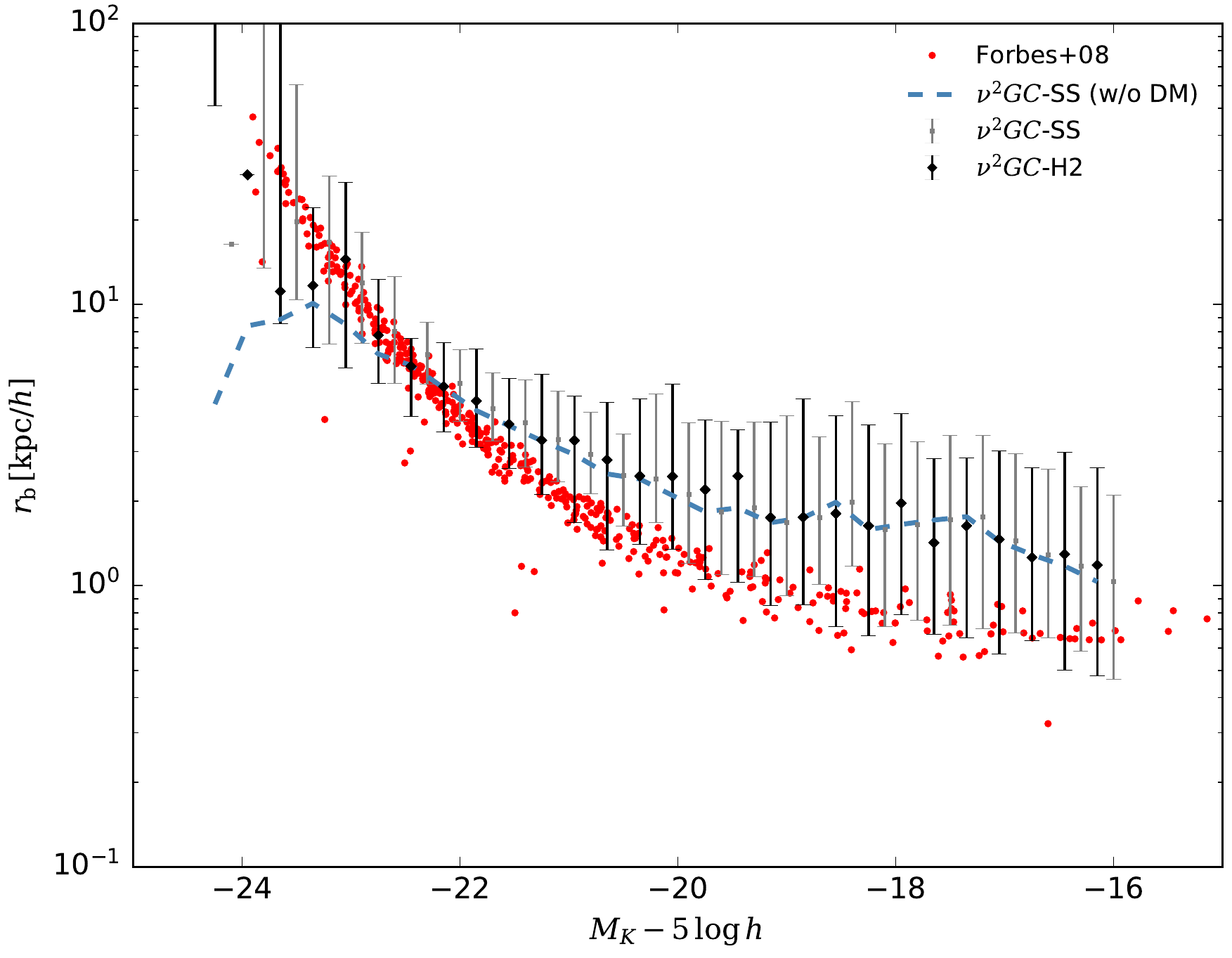}
          \end{center}
          \caption{Effective radius of elliptical and S0 galaxies as a function of $K-$ band magnitude.
          The black line shows the median value obtained by the model
          and the error bars show the 10th and 90th percentiles.
          The grey line with errorbars shows the median value obtained by the model considering $M_\mathrm{DM,1} = 0$
          from the \nugc-SS and \nugc-H2 simulations.
          Red points show the observational
          data obtained from \protect\cite{Forbes08}.}
          \label{fig:bulgesize}
        \end{figure}

\begin{figure}
          \begin{center}
            \includegraphics[width=\hsize]{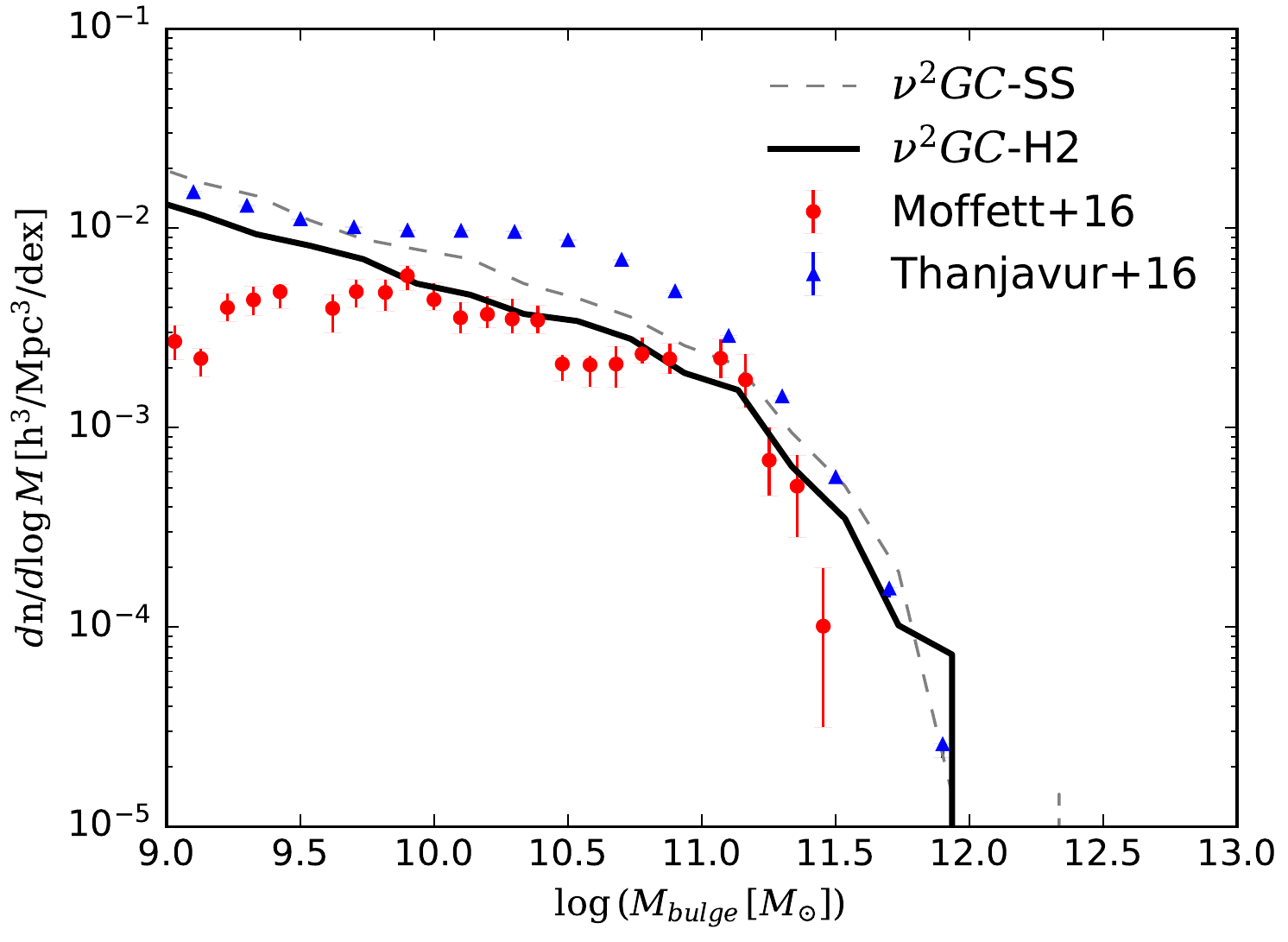}
          \end{center}
          \caption{Bulge mass function at $z \sim 0$ obtained with \nugc-SS and -H2 simulations.
            The black solid line denotes the result obtained from the model.
            Red filled circles and blue filled triangles present observed MFs
            obtained from \protect\cite{Moffett16} and \protect\cite{Thanjavur16}, respectively.}
          \label{fig:bulgeMF}
        \end{figure}

        Next, we compare the predicted effective radius and velocity dispersion of elliptical and S0
        galaxies at $z~\sim~0$ with observations since these values are used for calculating
        the dynamical time of bulges.
        Here we also employ the \nugc-SS and \nugc-H2 simulations,
        although the effect of the mass resolution of the simulations is negligible.
        We use the data obtained from \cite{Forbes08} who calculate the half-light radii are
        from 2MASS $K-$band 20th isophotal by using an empirical relation
        based on S\'ersic light profiles \citep{Forbes08}.
        Figs.~\ref{fig:FJ} and~\ref{fig:bulgesize} are the scaling relations
        between the bulge velocity dispersion and the $K$-band magnitude
        (the so-called Faber-Jackson relation;~\citealt{FJ76})
        and the effective radius and the $K-$ band magnitude, respectively.
        The data obtained from \cite{Forbes08} are shown in red points.
        The results of the fiducial model with \nugc-SS/-H2 are described as grey squares/black diamonds
        with error bars indicating 10th and 90th percentiles.
        For comparison, we overplot the model results with $M_\mathrm{DM,1} = 0$ as grey
        diamonds with error bars.
        We find that the effective radius of bulges with $M_K~-~5\log~h~<-23$
        becomes smaller when we set $M_\mathrm{DM,1} = 0$.
        The results obtained from the fiducial model have some discrepancies
        with the observational results, especially for the velocity dispersion
        while the bulge MF at $z \sim 0$ is consistent with observed bulge MF
        obtained from \cite{Moffett16}, and \cite{Thanjavur16}, as shown in Fig.~\ref{fig:bulgeMF}.

        The velocity dispersion obtained from the fiducial model becomes smaller
        with massive galaxies than those obtained from observations.
        There might be two possible reason for the inconsistency.
        First, due to the underestimate of gas mass especially in the small galaxies.
        We find that the model overproduces gas-poor galaxies, whose $r$-band magnitude
        are less than $\sim -18.5$.
        The dissipation process plays important roles for calculation of the velocity dispersion
        (Sec.~\ref{sec:bulgeprop}). Since the dissipated energy becomes larger with mergers of more gas-rich galaxies,
        underestimation of the cold gas mass would cause the underestimation of the velocity dispersion.
        Another possibility to reproduce Faber-Jackson relation might be
        related with the estimation of the gravitational potential of galactic discs.
        Galaxies which experience bulge growths should contain a galactic disc.
        The potential energy of the remained disc is estimated
        assuming that the rotation velocity of the disc remain unchanged (Eq. \ref{eq:Edisc}).
        When the discs have a shallower potential, the bulge should display a larger velocity dispersion.

\begin{figure}
          \begin{center}
            \includegraphics[width=\hsize]{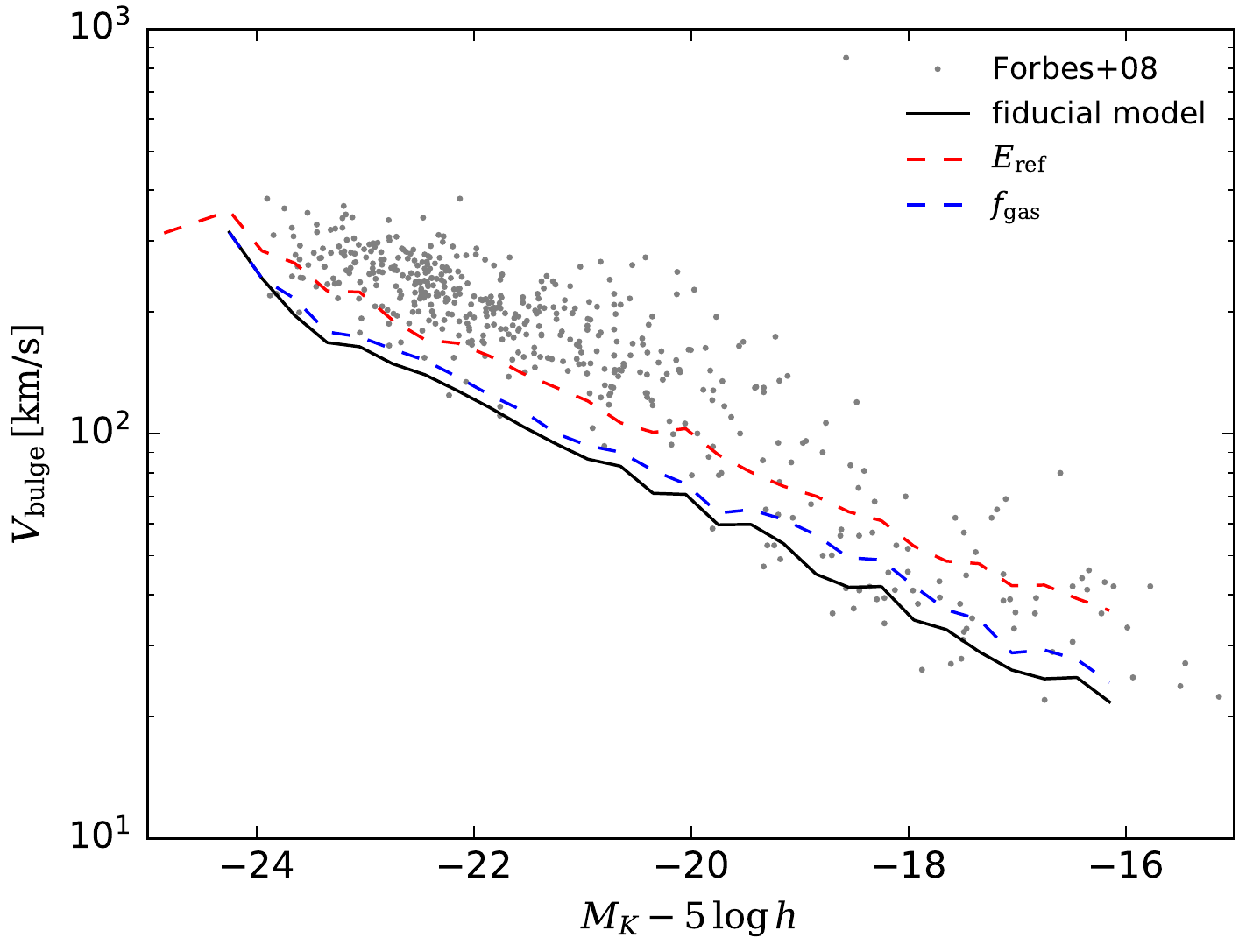}
          \end{center}
          \caption{Velocity dispersions of elliptical and S0 galaxies as a function of $K-$ band magnitude
          (Faber-Jackson relation). The black solid, red dashed, and blue dashed  lines show
          the median value obtained by the fiducial model (\protect\nugc~-SS), that by the artificially fixed gas fraction
          (Eq. \protect\ref{eq:fgas_test}), and that by the artificially fixed energy which remains in the galactic disc,
          respectively.
          Grey points show the observational
          data obtained from \protect\cite{Forbes08}.}
          \label{fig:FJ_test}
        \end{figure}

        To check these two effects,
        we test arbitrary models with the gas fraction $f_\mathrm{gas,test}$ of the galaxy
        and that with $0.3$ times smaller $E_\mathrm{0,disc}$ value.
        The new gas fraction, $f_\mathrm{gas,test}$ is described as:
        \begin{equation}
          f_\mathrm{gas,test} = f_\mathrm{gas} \times \left(\frac{M_{1d}}{10^{11} M_\odot}\right)^{-0.2},
          \label{eq:fgas_test}
        \end{equation}
        where $f_\mathrm{gas}$ and $M_\mathrm{1d}$ are the same definition in Sec.~\ref{Merger} and~\ref{sec:bulgeprop}.
        As an example, we consider a galaxy with $M_K - 5\log h \sim -20$.
        The re-estimated gas fraction, $f_\mathrm{gas,test}$ is $\sim~1.3$ times larger than the fiducial value.
        We use $f_\mathrm{gas,test}$ instead of $f_\mathrm{gas}$ in Eq.~\ref{eq:gasfrac},
        and re-estimate velocity dispersion.
        Fig.~\ref{fig:FJ_test} shows Faber-Jackson relation obtained from these simple tests.
        The model result is roughly consistent with observational one.
        We conclude that the discrepancy of bulge velocity dispersion with observational estimates
        would become smaller when we can reproduce observed colour-magnitude relation and HI MF
        of less massive galaxies.

    \subsection{Galaxy evolution}
    \label{App:Galev}
\begin{figure}
        \begin{center}
          \includegraphics[width=\hsize]{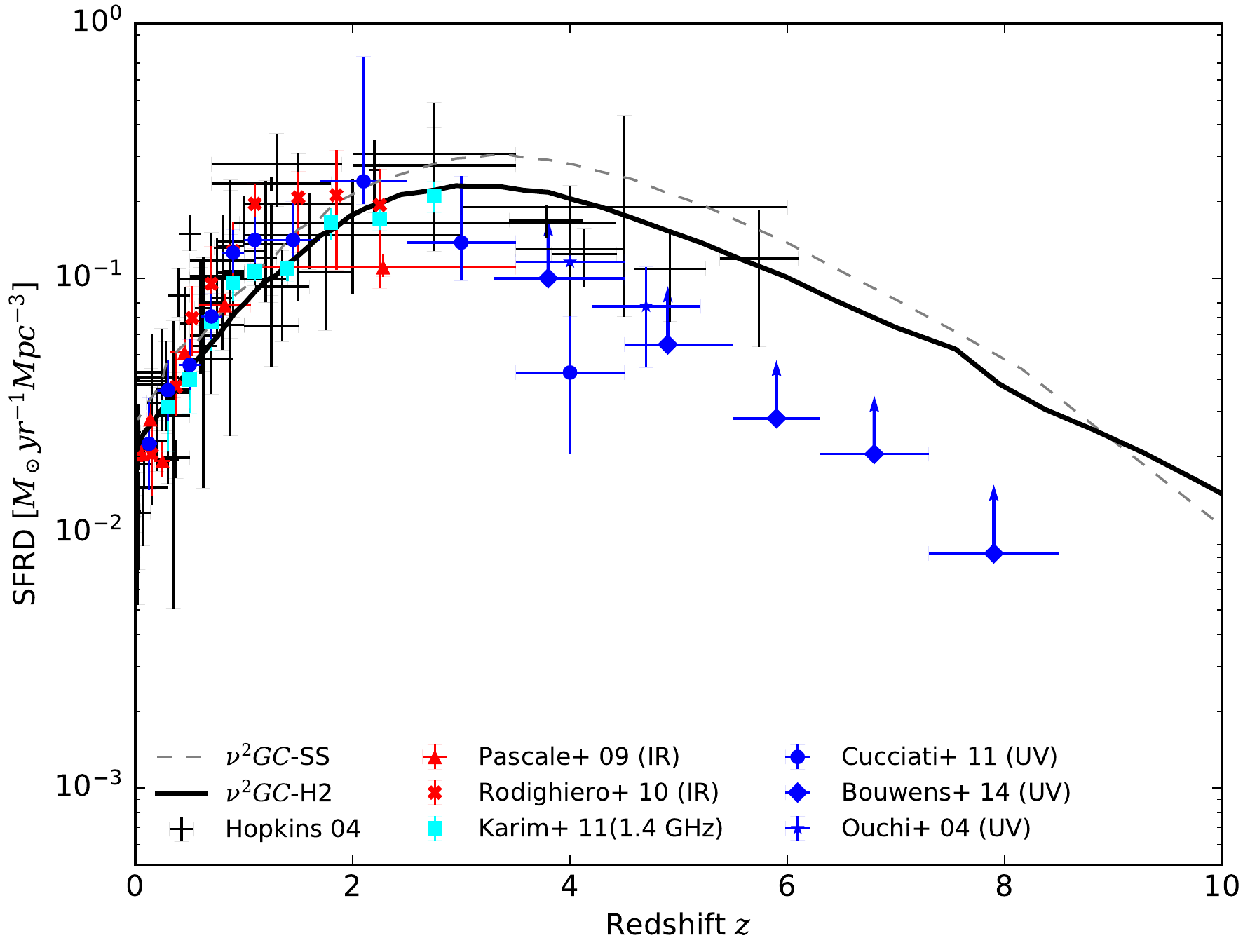}
        \end{center}
        \caption{Cosmic SFR density as a function of redshift.
                 The black solid line is the model results
                 obtained with the \nugc-SS and \nugc-H2 simulations.
                 Red filled triangles and stars
                 and cyan filled squares are obtained from dust continuum emission
                 \protect\citep[][respectively]{Pascale09, Rodighiero10,Karim11}.
                 Blue filled circles, filled diamonds, and stars are from UV continuum emission
                 \protect\citep[][respectively]{Cucciati12,Bouwens14,Ouchi04}.
                 Black crosses are obtained from \protect\cite{HopkinsA04}, which is a compilation of
                 various other observational results.}
        \label{fig:csfh}
      \end{figure}

      We firstly show the cosmic SFR density as a function of redshift in Fig.~\ref{fig:csfh}.
      The black solid line is the model result obtained with the \nugc-SS simulation and points are the results obtained from observations
      in IR-bands \citep{Pascale09,Rodighiero10}, radio 1.4 GHz \citep{Karim11}, UV-bands
      \citep{Cucciati12, Bouwens14,Ouchi04}, and a compilation of various observations \citep[][and therein]{HopkinsA04}.
      We find that the cosmic SFR density obtained by the fiducial model
      is consistent with the data over wide redshift range.

      Next, we present the evolution of $K-$ and $B-$ band LFs and stellar MFs of galaxies
      obtained by the fiducial model
      with the \nugc-M and -H2 simulations
      to show the result of bright and rare populations of galaxies.
      The LFs and MFs presented here are volume-weighted.
      The details of the calculation of LFs and MFs from the simulation are described in Appendix \ref{sec:LFMF}.

      Fig.~\ref{fig:GLFK_ev} shows the model $K-$ band LFs (black solid lines) compared with observational results
      \citep{Bell03,Huang03,Pozzetti03,Drory03,Caputi06,Saracco06,Devereux2009,Cirasuolo10,Driver12}.
      Model LFs reproduce observational results well for $z < 3.5$
      including faint-end slopes.
      The model of \citetalias{nu2gc} also explains observed $K-$ band LFs for $z < 2.0$ well
      (Fig. 21 of \citetalias{nu2gc}),
      although it over estimates number density of less luminous galaxies ($M_K > -22$).

      Fig.~\ref{fig:GLFB_ev} compares the model $B-$ band LFs (black lines)
      with observational results
      \citep{Norberg02,Gabasch04,Ilbert05,Giallongo05,Jones06}.
      The dust-attenuated model LFs are shown by the solid lines (for dust correction, see Sec.~\ref{LumMor})
      and LFs without dust attenuation are shown by the dashed lines.
      We note that the data obtained from \cite{Norberg02} and \cite{Jones06} at $z < 0.25$ are not dust attenuation-corrected.
      Therefore, their results allow a fair comparison with the LF of the dust-attenuated model.
      The dust attenuation-corrected model LFs at $z > 0.8$ seem to be inconsistent with observational estimates.
      The observational data of \cite{Giallongo05} are dust attenuation-corrected by assuming SMC and Calzetti extinction curves.
      Considering the correction for the dust attenuation,
      the model reproduces observed $B$-band LFs at $z < 3.5$ reasonably well.
      The data of \cite{Ilbert05} and \cite{Gabasch04} are not dust attenuation corrected.
      Since the bright-end of LFs of \cite{Giallongo05}, \cite{Ilbert05}, and \cite{Gabasch04} are similar
      and the dust attenuation in $B$-band should have less impact than those suggested from the fiducial model,
      we conclude that some  modifications of the dust attenuation are needed, which we leave for future studies.

      \begin{figure*}
        \begin{center}
          \includegraphics[width=\hsize]{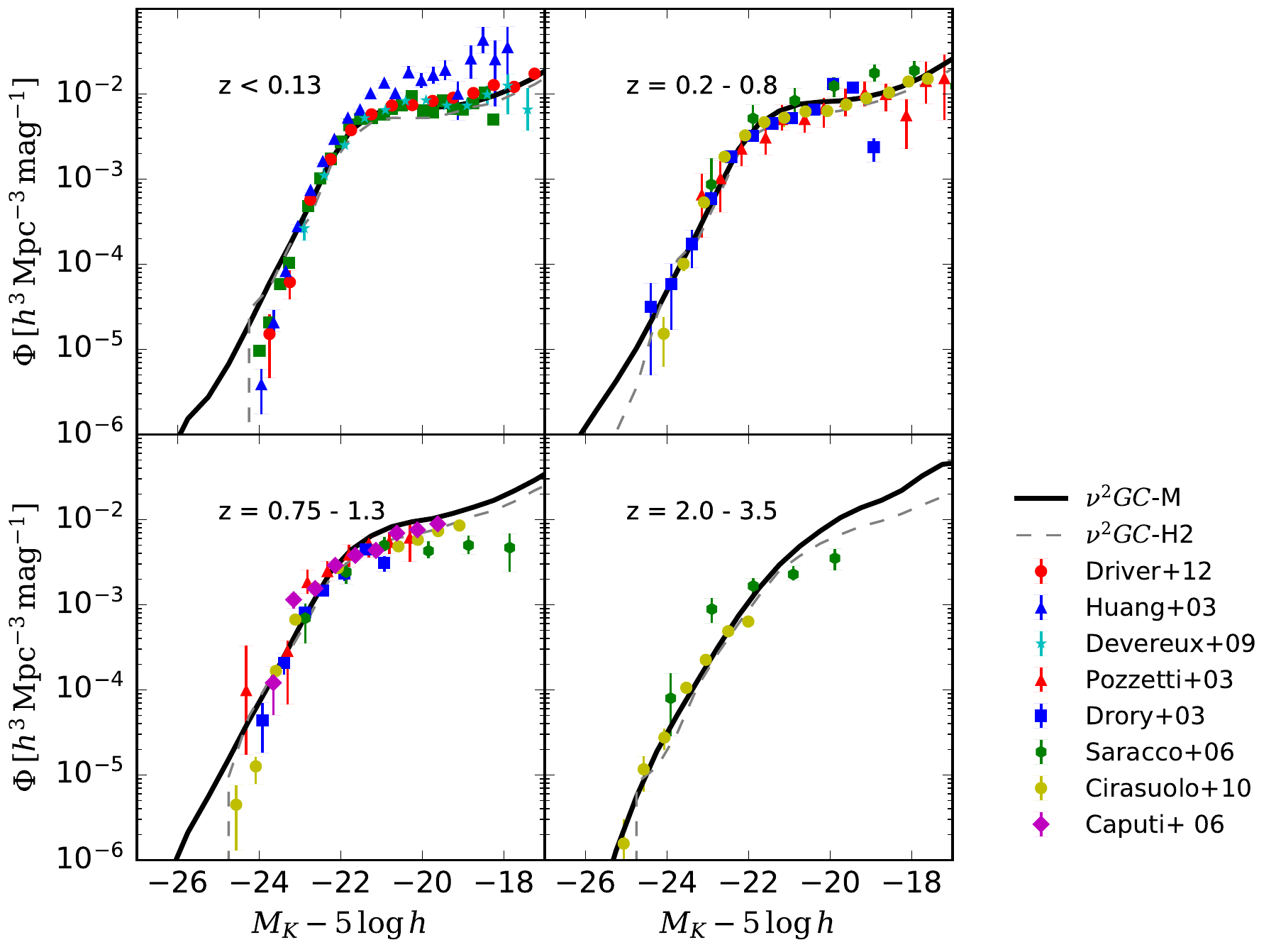}
        \end{center}
        \caption{$K-$ band LFs of galaxies at $z < 0.13$, $z = 0.2 - 0.8$,
          $z = 0.75 - 1.3$, and $z = 2.0 - 3.5$.
          The model LFs (volume-weighted) by the \nugc-M simulation
          appear as black solid lines.
          Observational estimates are taken from
          \protect\cite{Bell03}, \protect\cite{Huang03}, \protect\cite{Pozzetti03}, \protect\cite{Drory03},
          \protect\cite{Caputi06}, \protect\cite{Saracco06}, \protect\cite{Devereux2009}, \protect\cite{Cirasuolo10}, and \protect\cite{Driver12}.}
        \label{fig:GLFK_ev}
      \end{figure*}

      \begin{figure*}
        \begin{center}
          \includegraphics[width=\hsize]{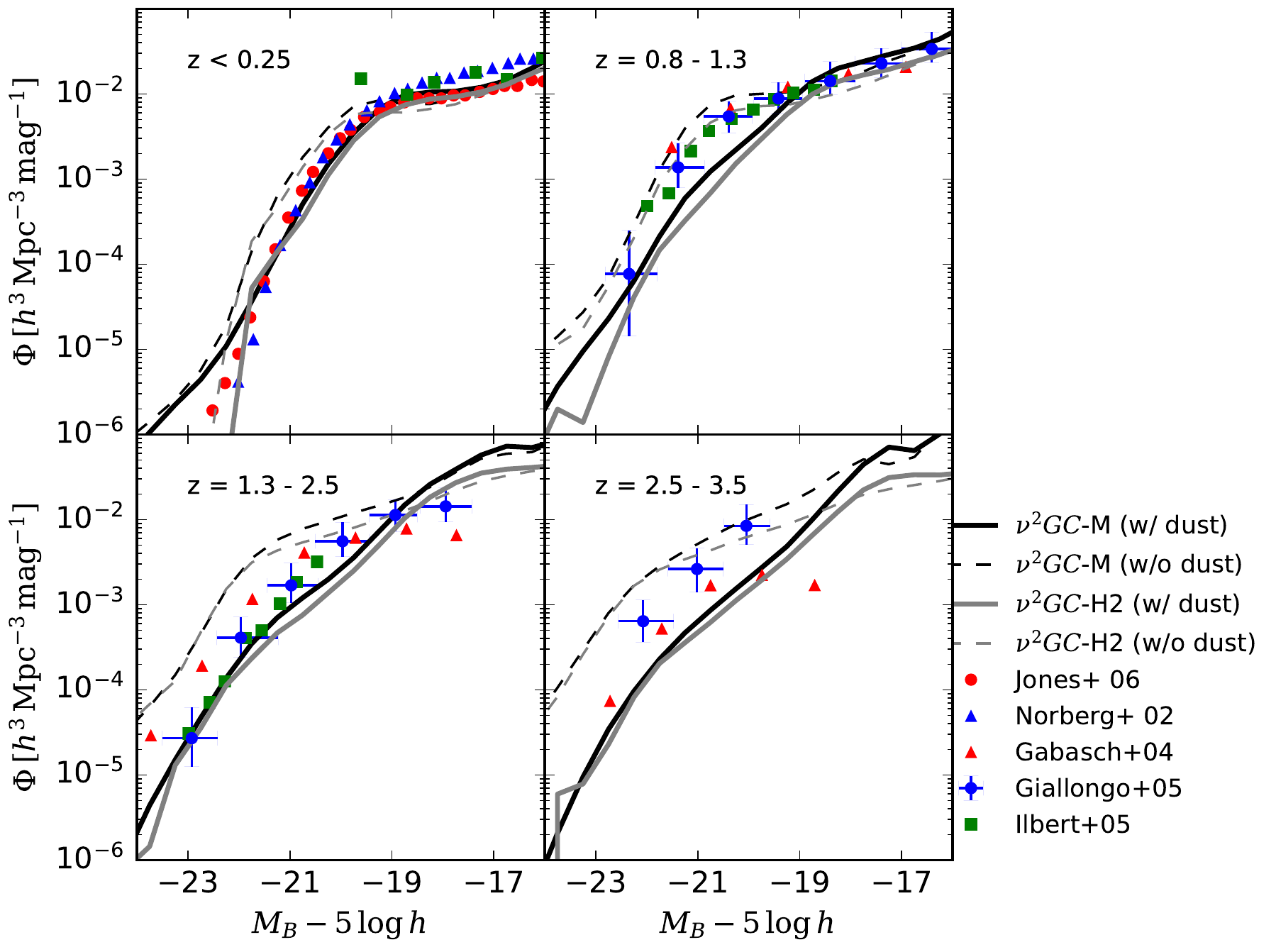}
        \end{center}
        \caption{$B-$ band LFs of galaxies at $z < 0.25$, $z = 0.8 - 1.3$,
          $z = 1.3 - 2.5$, and $2.5 - 3.5$.
          The model LFs (volume-weighted) obtained with the \nugc-M and -H2 simulations
          appear in black solid and grey dashed lines (with dust attenuation)
          and black dashed lines (without dust attenuation).
          Observational results are obtained from
          \protect\cite{Norberg02,Gabasch04,Ilbert05,Giallongo05,Jones06}.}
        \label{fig:GLFB_ev}
      \end{figure*}

\begin{figure*}
        \begin{center}
          \includegraphics[width=\hsize]{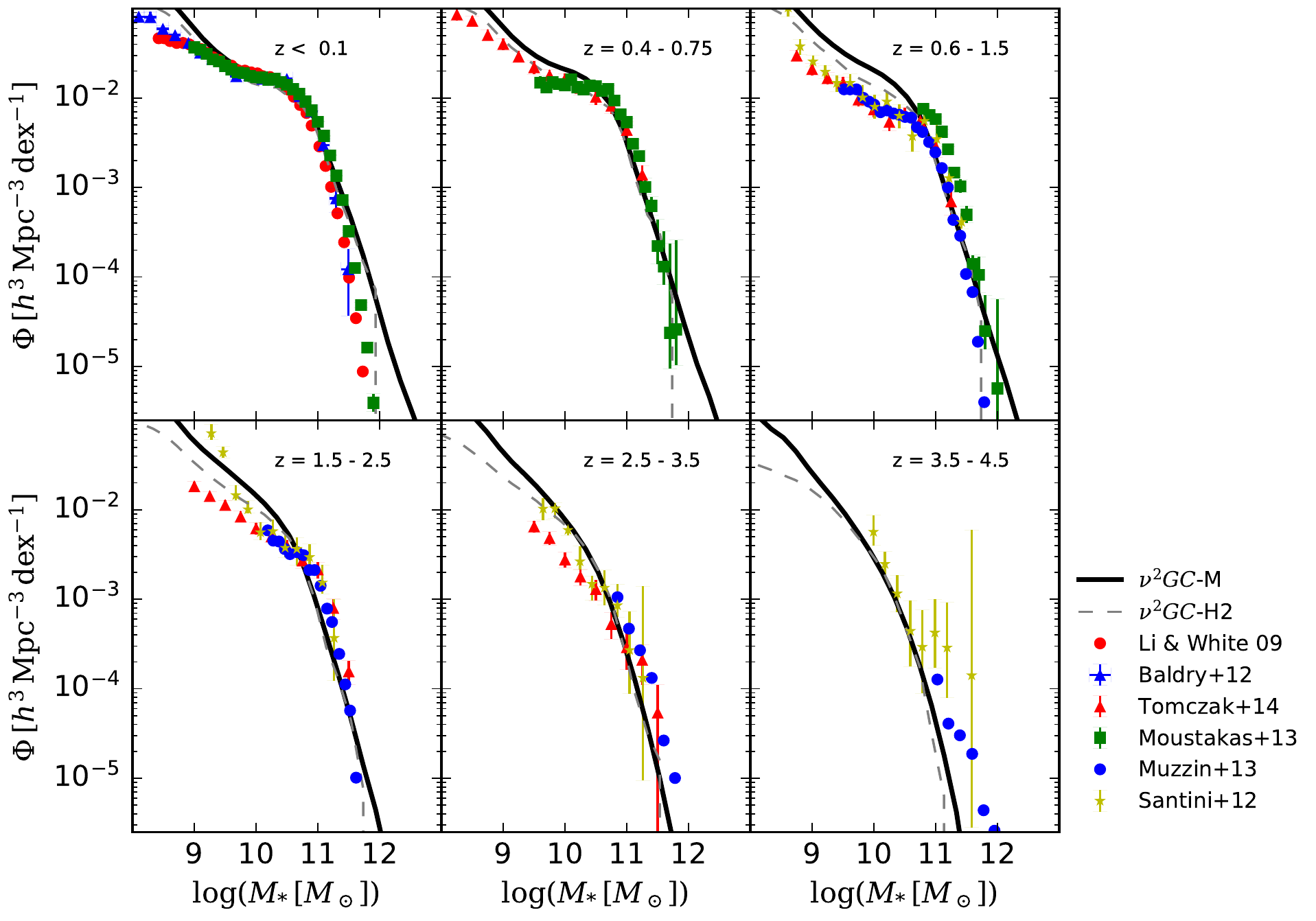}
        \end{center}
        \caption{Stellar MFs at $z < 0.1$, $z = 0.4 - 0.75$, $z = 0.6 - 1.5$,
        $z = 1.5 - 2.5$, $z = 2.5 - 3.5$, and $z = 3.5 - 4.5$.
        The model MFs (volume-weighted) obtained with the \nugc-M
        and -H2 simulations
        are shown in black solid and grey dashed lines.
        Observational results are obtained from
        \protect\cite{LW09}, \protect\cite{Baldry12}, \protect\cite{Santini2012},
        \protect\cite{Muzzin13}, \protect\cite{Moustakas13}, \protect\cite{Tomczak14}.}
        \label{fig:SMF_ev}
      \end{figure*}

      Fig.~\ref{fig:SMF_ev} shows the stellar MFs from $z~\sim~0$ to $z~\sim~4.5$.
      We adopt Chabrier IMF \citep{Chabrier03} as described in Sec. \ref{SF}.
      We compare our results (black lines) with observational estimates by
      \cite{LW09}, \cite{Baldry12}, \cite{Santini2012}, \cite{Muzzin13},
      \cite{Moustakas13}, and \cite{Tomczak14},
      who employ either a Chabrier IMF \citep{Chabrier03} or Kroupa IMF
      \citep{Kroupa01}.
      \footnote{Since the stellar mass difference between Chabrier and Kroupa IMF
      is only $\sim~0.04$ dex \citep{Muzzin13}, we assume a negligible difference in our results.}
      While the model can reproduce the massive end of the stellar MFs at $z < 3.5$,
      we find that the model underestimates the number of massive galaxies at $z~>~3.5$ (bottom right panel).
      This similar feature is seen in other SA models \citep[e.g.][]{Hirschmann12,Lacey16}.
      The derivation of stellar masses from observations is commonly performed by the
      broad-band SED fitting with galaxy templates assuming a single dust attenuation law.
      Alternatively, \cite{Mitchell13} suggest that the discrepancy between SA models and observations
      in the stellar MFs at high redshifts stems from the uncertainties in the dust attenuation curve.
      For less massive galaxies, we also find that we overproduce their number density
      at $0.4 < z < 2.5$, which is the similar trend to other SA models \citep[e.g.][]{Weinmann12}.
      Some previous studies with SA models investigate this problem.
      \cite{Henriques13} show that the ejected gas should be reincorporated into the system
      on a timescale which depends on the halo mass; the smaller halo should have
      the larger timescale, and the gas returns to the system more slowly.
      The importance of the timescale to reproduce SMFs are also proposed by \cite{White15}.
      They also suggest the mass-loading factor which strongly depends on the redshift
      also plays a role in reproducing SMFs.
      \cite{White15} imply a detailed comparison with observations are required
      to differentiate these two effects.
      \cite{Hirschmann16} consider the decrease of the gas infall rate by ``pre-heating''
      and find that their model can reproduce not only the low mass end of SMFs but also the
      metal enrichment of galaxies.
      We need to consider such effects in the \nugc, which we leave it for future studies
      to decrease of the degree of freedom.
      As \cite{White15} suggest, the values of parameters which are required for reproducing SMFs
      strongly depends on the treatment of the reservoir of reheated and/or ejected gas in each SA models.
      The value of these parameters, therefore, have almost no constraints now.

      For checking the mass resolution effect, we overplot the results with the
      \nugc-H2 simulation as grey dashed lines in Figs. \ref{fig:GLFK_ev} to \ref{fig:SMF_ev},
      although the \nugc-H2 simulation has $8^3$ times smaller box size than the \nugc-M simulation.
      We find the effect of the resolution is negligible.

\begin{figure*}
      \begin{center}
        \includegraphics[width=\hsize]{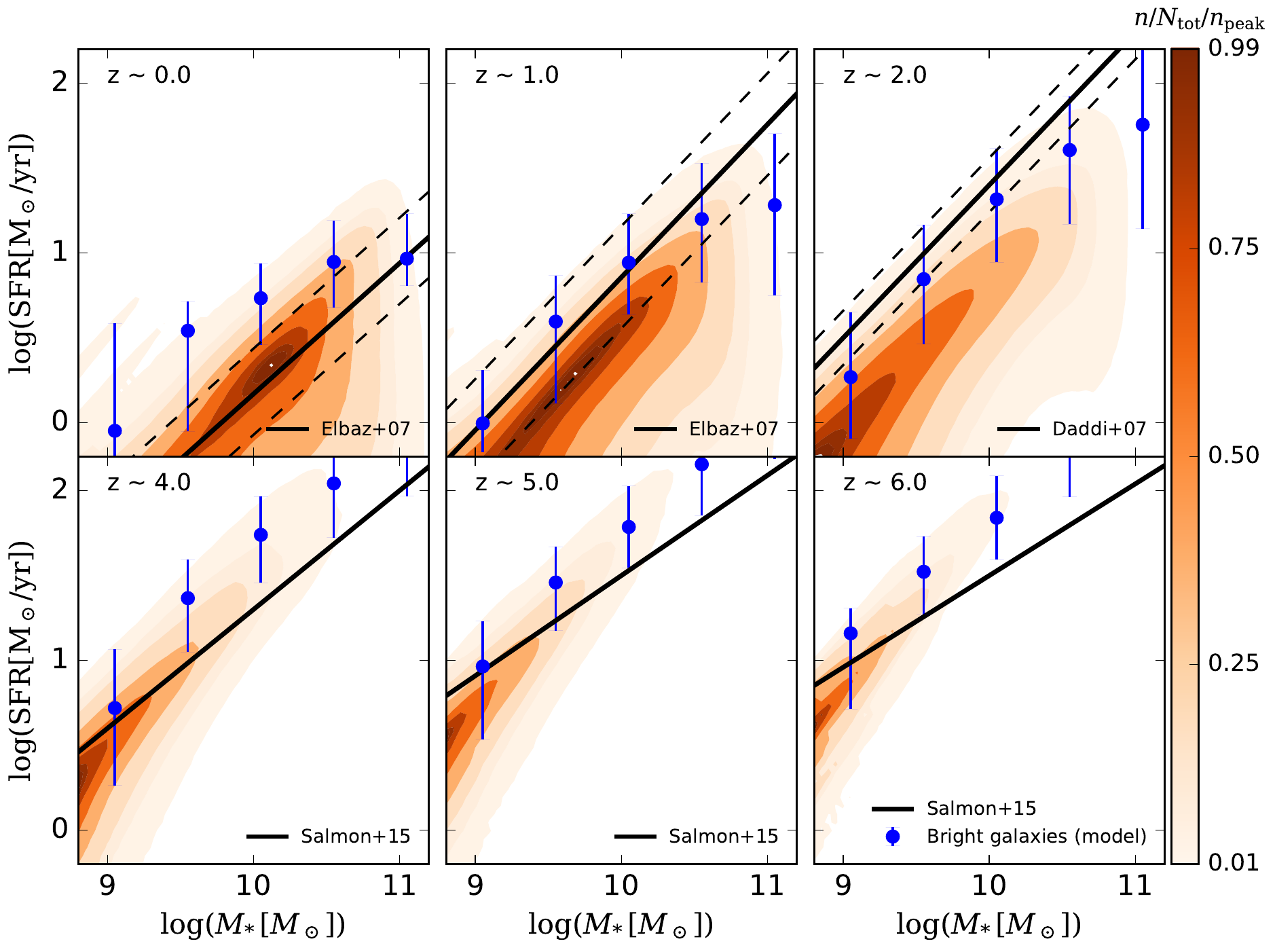}
      \end{center}
      \caption{The relation between total stellar mass and SFR at $z < 6.0$.
        The model results (obtained with the \nugc-M simulation) including all galaxies and
               those including only luminous galaxies ($M_{FUV} < -19.0$) are shown by
               the orange colour map and the blue points with errorbars (10th and 90th percentiles), respectively.
               For comparison, we overplot the results obtained from observations
               at $z \sim 0$ and $1$ \protect\citep{Elbaz07}, $z \sim 2$ \protect\citep{Daddi07},
               and $z \sim 4, 5,$ and $6$ \protect\citep{Salmon15Feb}.}
      \label{fig:Ms-SFR}
    \end{figure*}

    We also present the relation between total stellar mass and SFR at $z < 6.0$
    obtained from the fiducial model with the \nugc-M simulation
    and compare it with that obtained from observations \citep{Elbaz07,Daddi07,Salmon15Feb}
    in Fig.~\ref{fig:Ms-SFR}.
    We select all galaxies (central $+$ satellite) without any luminosity or surface density limitations.
    The result is shown as the orange density map.
    In addition, blue points with errorbars show the relation for luminous galaxies with $M_{FUV} < -19.0$
    (where $M_{FUV}$ is the magnitude of the GALEX FUV band) obtained by the fiducial model,
    which are consistent with that of \cite{Salmon15Feb} at $z > 4.0$.
    The galaxies obtained by the fiducial model have larger SFRs than those obtained by observations
    when we take the selection effect into account at $z > 4$.
    Since the $M_\mathrm{*}$-SFR relation obtained by \cite{Salmon15Feb} with $\log(M_\mathrm{*}/M_\odot) > 10.3$ has a large dispersion,
    the slope of the $M_\mathrm{*}$-SFR relation would not be strictly constrained.
    We note that the number of luminous galaxies obtained by the fiducial model with the \nugc-M simulation
    is 135.1, 180.9, and 108.1 times larger than that of \cite{Salmon15Feb} at $z \sim 4, 5,$ and $6$, respectively.
    The galaxies with $\log(M_\mathrm{*}/M_\odot) > 10.5$ and $M_{FUV} < -19.0$ at $z \sim 2$
    have smaller SFR than those predicted by the observational fitting.
    This could be a result of the AGN feedback effect (see also Sec. \ref{sec:discussion}).
    At $z \sim 2$, gas cooling of most of such massive galaxies are quenched by the AGN feedback.
    The cold gas mass, thus, becomes smaller, resulting in lower SFRs.

    \cite{Izumi18Apr} compare this relation obtained from the fiducial model
    employing the \nugc-L simulation with the data of four observed AGN host galaxies at $z \sim 6$.
    These four AGNs, which are optically low-luminosity quasars ($M_{UV} < -25$), are originally detected with Subaru Hyper Sprime Cam (HSC) \citep{Matsuoka17}
    and are observed with Atacama Large Millimeter/Submillimeter Array (ALMA) to
    investigate their host galaxies' properties.
    They find that the sample galaxies are on or \textit{below} the so-called ``main sequence'' at $z \sim 6$,
    which are very rare population in the fiducial model of \nugc.
    Luminous quasars ($M_{UV} < -25$) at $z \sim 6$, on the other hand,
    have host galaxies with higher SFR than the ``main sequence''.
    The fiducial model of \nugc~can reproduce such a bursty population.
    As shown in Fig. 8 in \cite{Izumi18Apr} and Fig.~\ref{fig:Ms-SFR},
    the distribution of the SFR seems to have several sub-sequences.
    These sub-sequences should be artificial which result from time and mass resolution of the simulations
    and/or the discrete treatment of the time evolution of the hot gas density profiles and
    cooled gas mass.
    As we show in Sec.~\ref{GasCooling}, the radial profiles of hot gas haloes
    remain unchanged until the DM halo mass doubles.
    It means that no hot gas distributes in $r < r_\mathrm{cool}$ until the DM halo mass doubles.
    Since the minimum halo mass of \nugc-M and -SS simulations is $8.79\times 10^{9} M_\odot$,
    the radial profile of the hot gas halo of galaxies with $M_\mathrm{*} < 10^9 M_\odot$
    is not updated from the formation time.
    A part of such galaxies, therefore, would contain an unphysically smaller amount of the cold gas.

  \section{The Timescale Dependency on BH and Accreted Gas Mass}
  \label{App:Tloss}
      We firstly show that the accretion timescale from the accretion disc to the SMBH
      has a negative (positive) dependency on the mass of the accreted gas (SMBH),
      following the viscous timescale in the accretion discs.
      We classify the accretion discs by their accretion rate following \cite{KFM08_text}.
      Then, we analytically calculate the radial velocity of the gas, $|v_\mathrm{r}|$, and
      the outer radius of the accretion disc which is determined as the boundary
      between self gravitating and non-self gravitating disc, $r_\mathrm{sg}$.
      The details appear in \cite{KPH04}.
      Here we define the Schwarzschild radius, $r_\mathrm{Sch}$, as $2GM_\mathrm{BH}/c^2$,
      the distance from the BH normalised by $r_\mathrm{Sch}$, $\hat{r}$,
      the viscous parameter, $\alpha$, and a non-dimensional variable,
      $f = 1 - \sqrt{3r_\mathrm{Sch}/r}$.
      The accretion rate is simply described as $\Delta M_\mathrm{acc} / t_\mathrm{vis}$ for this calculation,
      where $t_\mathrm{vis}$ is the viscous timescale determined as $t_\mathrm{vis} = r_\mathrm{sg} / |v_\mathrm{r}|$.
      The accretion rate normalised by the Eddington mass accretion rate, $\dot{m}$
      (the Eddington mass accretion rate: $ L_\mathrm{Edd} / c^2$), is employed.
      The disc is classified according to the dominant opacity and pressure sources as follows.
      \begin{enumerate}
        \item The outer region in which the main opacity source is (free-free) absorption
          and the gas is the dominant pressure source. Then
          \begin{equation*}
            |v_\mathrm{r}| \propto \alpha^{4/5}M_\mathrm{BH}^{-1/5}\dot{m}^{3/10}\hat{r}^{-1/4}f^{-7/10}
          \end{equation*}
          and
          \begin{equation*}
            r_\mathrm{sg}/r_\mathrm{Sch} \propto \alpha^{28/45}M_\mathrm{BH}^{-52/45}\dot{m}^{-22/45}.
          \end{equation*}
          We obtain $t_\mathrm{vis}~\propto~M_\mathrm{BH}^{15/2}~\Delta~M_\mathrm{acc}^{-41/4}$.

        \item The middle region in which the main opacity source is electron scattering
          and the gas is the dominant pressure source. Then
          \begin{equation*}
            |v_\mathrm{r}| \propto \alpha^{4/5}M_\mathrm{BH}^{-1/5}\dot{m}^{2/5}\hat{r}^{-2/5}f^{-3/5}
          \end{equation*}
          and
          \begin{equation*}
            r_\mathrm{sg}/r_\mathrm{Sch} \propto \alpha^{14/27}M_\mathrm{BH}^{-26/27}\dot{m}^{-8/27}.
          \end{equation*}
          We obtain $t_\mathrm{vis}~\propto~M_\mathrm{BH}^{18/5}~\Delta~M_\mathrm{acc}^{-22/5}$.

        \item The inner region in which the main opacity source is electron scattering
          and the radiation is the dominant pressure source. Then
          \begin{equation*}
            |v_\mathrm{r}| \propto \alpha M_\mathrm{BH}^{0}\dot{m}^{2}\hat{r}^{-5/2}f^{1}
          \end{equation*}
          and
          \begin{equation*}
            r_\mathrm{sg}/r_\mathrm{Sch} \propto \alpha^{2/9}M_\mathrm{BH}^{-2/9}\dot{m}^{4/9}.
          \end{equation*}
          We obtain $t_\mathrm{vis}~\propto~M_\mathrm{BH}^{6/5}~\Delta~M_\mathrm{acc}^{-4/5}$.
      \end{enumerate}

      Considering these conditions, we conclude that the viscous timescale has a positive correlation to the BH mass
      and negative correlation to the accreted gas mass at all radii.

      Next, we consider the Circumnuclear disc (CND).
      We consider the CND model of \cite{KW08}, as an example,
      although the physical mechanisms of
      how the CND maintains its structure is still under discussion.
      In \cite{KW08}, SNe occurred in the CND induces the tidal torque
      which enhances the gas accretion rate from the CND to the SMBH.
      When the CND becomes unstable considering from the Toomre criterion \citep{Toomre64}, then the star formation occurs and
      the accretion rate increases.
      Since the CND becomes stable for the massive SMBH,
      $\gamma_\mathrm{BH}$ should be positive.
      On the other hand, since the SFR becomes more significant
      for the more gas-rich galaxies,
      $\gamma_\mathrm{gas}$ should be negative.
      We cannot obtain constraints on the values of $\gamma_\mathrm{BH}$ and $\gamma_\mathrm{gas}$
      since the model of CND is too complicated to construct a single phenomenological model
      of the accretion timescale (i.e. the outer radius of the CND depends on the SMBH mass,
      mass density of CND itself and their host galaxy; see Sec. 2.3 in \citealt{KW08}).
      With the simple assumptions (based on \citealt{KW08}), we estimate $\gamma_\mathrm{BH} \sim -0.5$ and
      $\gamma_\mathrm{gas} \sim 1.0$, assuming a constant star formation efficiency,
      constant surface densities of the host galaxy and CND, the outer radius of the CND which is proportional to
      $M_\mathrm{BH}^{0.5}$.

  \section{The calculation of luminosity and mass function}
  \label{sec:LFMF}
    We describe the calculation of the volume-weighted LFs and MFs from the model output.
    We obtain LFs and MFs from the model at discrete output redshifts.
    On the other hand, LFs and MFs are estimated from observations
    in continuous redshift ranges.
    We thus should estimate model LFs and MFs in the same redshift ranges
    as observations by averaging model LFs and MFs.
    We will now describe the derivation of the model LFs.
    The calculation of MFs is the same as that of LFs,
    with the magnitude replaced by the logarithmic stellar mass.

    The average model LFs have the constant co-moving volume ($dV$),
    while the solid angle ($d\Omega$) is constant for observations.
    The luminosity function, $\phi(z, M)$, in which $z$ and $M$ are
    the redshift and magnitude, respectively, is described as follows:
    \begin{equation}
      \phi(z, M) = \frac{dN(z, M)}{dV},
    \end{equation}
    where $N(z, M)$ is the number density of objects over the whole sky at $z$ with a magnitude, $M$.
    The differential volume (co-moving), $dV$, is written
    with the differential solid angle, $d\Omega$, as
    \begin{equation}
      dV = \frac{cr^2(z)}{H(z)} dz d\Omega.
    \end{equation}

    We calculate the model LF at a magnitude ($M$)
    which is averaged in a redshift range ($z_0 < z < z_n$),
    $\bar{\phi} (M)$, as follows:
    \begin{align}
      \bar{\phi}(M) = \frac{\sum^{n}_{i=0} W_i \phi_i (z_i)}{\sum^{n}_{i=0} W_i}, \\
      W_i = \frac{r^2(z_i) dz_i}{H(z_i)}, \\
      dz_i = (z_{i+1} - z_{i-1}) / 2,
    \end{align}
    where $i$ means the corresponding output number,
    $r(z)$ and $H(z)$ are the line-of-sight distance and Hubble parameter, respectively.
    At the larger redshift, the weight becomes larger.
    Then we can obtain averaged LFs/MFs at a constant solid angle.

    \section{The Difference of Observable Fraction with Hopkins et al. 2007}
    \label{App:ObsFrac}
      Here we show the difference of observable fractions obtained from
      \cite{Hopkins07J} and this paper (Eq.~\ref{eq:ObsFrac}).
      \cite{Hopkins07J} derives an observable fraction as follows.
      They obtain intrinsic bolometric correction which is a similar shape to that of \cite{Marconi04}.
      By employing the observed hydrogen column density distribution \citep{Ueda03},
      they calculate the photoelectric absorption in $X-$ ray.
      For optical and mid-IR bands, they adopt a canonical gas-to-dust ratio and SMC-like dust attenuation curve
      \citep{Pei92} to obtain the probability of observing AGNs in optical/mid-IR bands.
      By the bolometric correction and the correction of the photoelectric absorption and the dust attenuation,
      they obtain intrinsic bolometric AGN LFs.
      Using this bolometric AGN LF, they estimate the probability of observing AGNs with an intrinsic luminosity of hard-/soft- $X$-ray
      and optical $B$-band.
      They fit the probability as a function of the bolometric luminosity, $L_\mathrm{bol}$,
      which is the observable fraction of AGNs:
      \begin{equation*}
        f(L_\mathrm{bol}) = f_{46}\left(\frac{L_\mathrm{bol}}{10^{46}~\mathrm{erg~s^{-1}}}\right)^\beta,
      \end{equation*}
      where $(f_{46}, \beta)$ is $(1.243,0.066)$ in hard $X-$ ray (2-10 keV), $(0.260,0.082)$
      in $B-$ band (4400 \AA).

      The method for the estimation of the observable fraction in this paper
      is slightly different from that of \cite{Hopkins07J}.
      We convert hard $X$-ray (2-10 keV) LFs obtained from \cite{Aird15A}
      to UV (1450\AA) LFs by using a bolometric correction \citep{Marconi04}
      and $M_{UV} = M_B + 0.85$ \citep{KSM01}.
      The LFs obtained from these processes are regarded as the intrinsic UV LFs
      since hard $X$-ray (2-10 keV) LFs of \cite{Aird15A} are absorption-corrected.
      By comparing these intrinsic UV LFs with LFs obtained from observations,
      we obtain the parameters of observable fractions as
      $(A_0,~A_1,~\beta_0,~\beta_1)~=~(0.16,~0.07,~-0.05,~0.00)$
      (Eq. \ref{eq:ObsFrac}).

\begin{figure*}
        \begin{center}
          \includegraphics[width=\hsize]{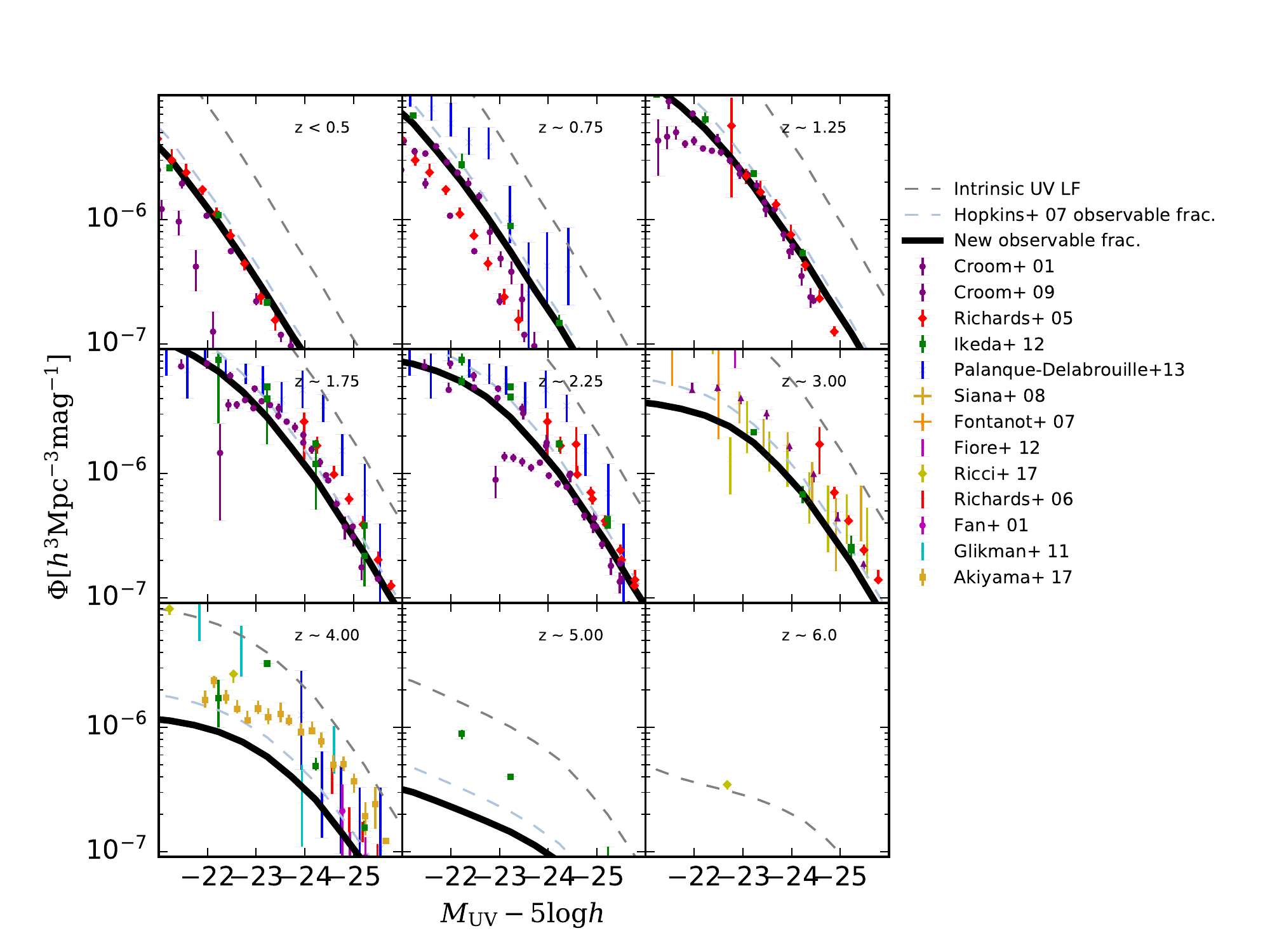}
        \end{center}
        \caption{AGN LFs in UV- band(1450 \AA) in $0.0~<~z~<~6.5$.
              Grey dashed line is the intrinsic UV LFs. Blue dashed and black solid lines
              are UV LFs considering observable fractions obtained from
              \protect\cite{Hopkins07J} and this paper (Sec. \protect\ref{sec:ObsFrac}), respectively.
              Observational results are obtained from
              \protect\cite{Croom01}, \protect\cite{Croom09Nov}, \protect\cite{Fan01},
              \protect\cite{Richards2005July}, \protect\cite{Richards06June}, \protect\cite{Fontanot07},
              \protect\cite{Siana08}, \protect\cite{Glikman11},\protect\cite{Fiore12},
              \protect\cite{Ikeda12}, \protect\cite{Palanque-Delabrouille13}, \protect\cite{Ricci17},
              and \protect\cite{Akiyama18}.}
        \label{fig:ObsFracComp}
      \end{figure*}

      We show the differences of observable fractions obtained by \cite{Hopkins07J} and
      by our new method in Fig.~\ref{fig:ObsFracComp}.
      The grey dotted line indicates intrinsic UV LFs and blue dashed and black solid lines
      show LFs considering observable fraction obtained from \cite{Hopkins07J} and this paper,
      respectively.
      We assume that the observable fraction obtained by \cite{Hopkins07J} is the same in $B$ and $UV$ bands.
      We find that in such a simple assumption, the observable fraction obtained in this paper
      is roughly consistent with those obtained by \cite{Hopkins07J},
      although they have a small ($\sim 20$ \%, at most) difference.

      We note that UV LFs with observable fractions obtained from both \cite{Hopkins07J} and our calculation
      are inconsistent with observations at $z > 5.0$ since the fitting function of hard $X-$ ray LFs obtained from \cite{Aird15A}
      can explain the observational results only at $z < 5.0$.
      We also note that the scatter of the conversion from the hard $X$-ray to $UV$ luminosity
      are not considered for deriving the observable fraction.
      \cite{Akiyama18} suggest that this scatter has significant effect on the shape of the LFs
      (see Fig. 21 in \citet{Akiyama18}).
      We need to consider the effect although we leave it for future studies.

  \bsp	
  \label{lastpage}
\end{document}